\documentclass[a4paper, 11pt, aas_macros]{FSKH_623_Report} 

\title{Modelling the multi-wavelength emission and polarisation signatures of the novel white-dwarf pulsar system AR Sco}                                                                      
\author{Louis du Plessis}                                         
\date{\today}  

\usepackage{amsmath}
\usepackage{amssymb}
\usepackage{graphicx}
\usepackage{url}
\usepackage{xfrac}
\usepackage{esvect}
\usepackage{varwidth} 
\usepackage{rotating}
\usepackage{multicol}
\usepackage{pdfpages}
\usepackage{aas_macros} 
\usepackage{titlesec}

\numberwithin{equation}{chapter}
\renewcommand{\theequation}{\thechapter.\arabic{equation}}
\renewcommand{\thesection}{\thechapter.\arabic{section}}  

\defcitealias{Harding2021}{AH21}
\defcitealias{Harding2015}{AH15}
\defcitealias{Cerutti2016}{CS16}
\defcitealias{Barnard2022}{BH22} 
\defcitealias{Takata2017}{TK17}
\defcitealias{Takata2019}{TK19}                                
\begin{document}

\maketitle
\includepdf[pages=-,scale=1.0,offset=25.5mm -25.85mm,noautoscale]{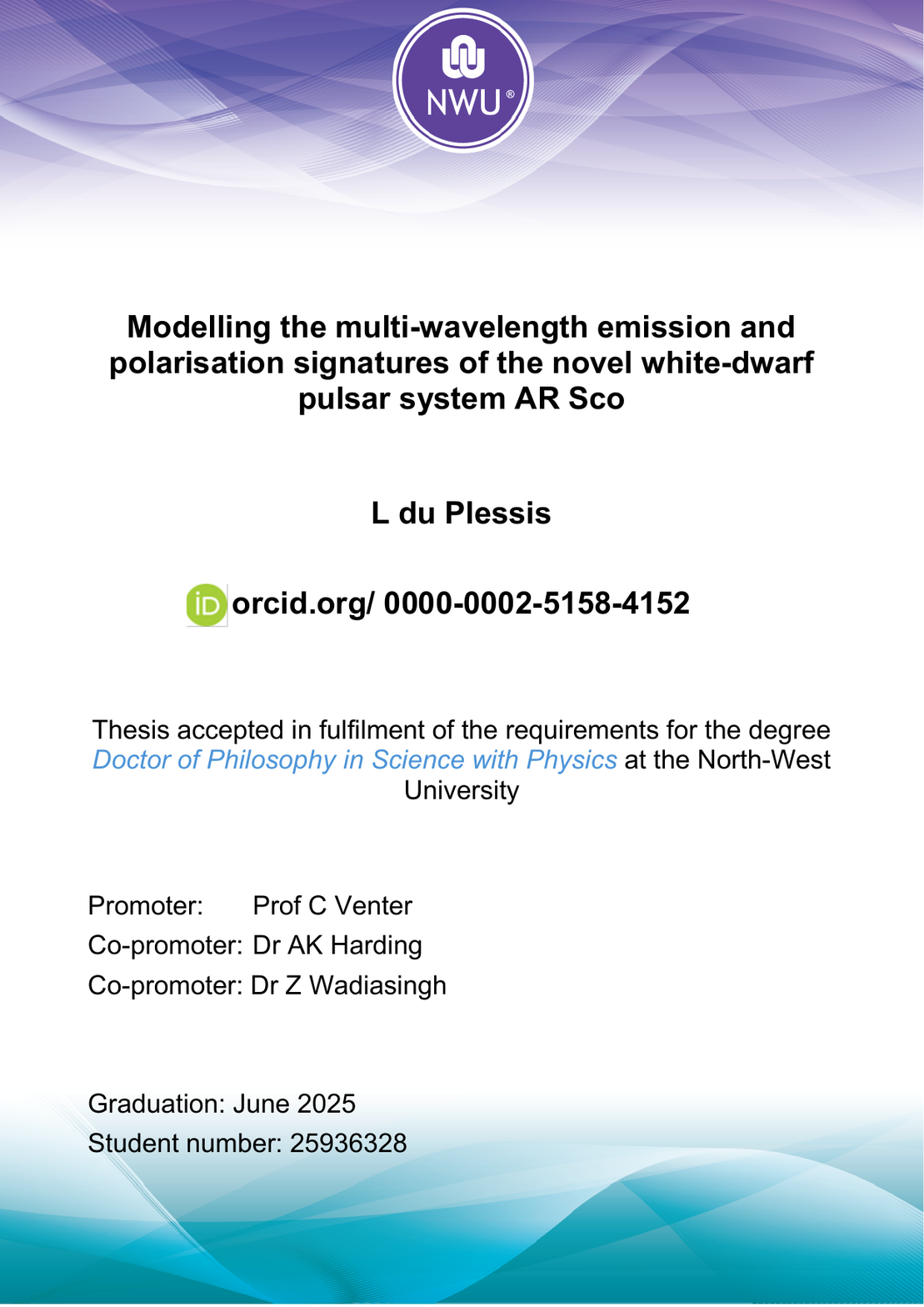}

\pagenumbering{roman}                                             

\titleformat{\section}{\fontsize{22}{16}\bfseries\filcenter}{\thesection}{1em}{}

\section*{Abstract}
The serendipitous reclassification of AR Scorpii (AR Sco) from a delta Scuti variable star to a white dwarf binary system has initiated an in-depth exploration of this novel system. These observations led to discovering pulsar-like emission features from the source, earning it the title of `first white dwarf pulsar'. Over the last decade, there have been many multi-wavelength and polarimetric observations on this source yet the proposed models remain unable to concurrently predict the emission and polarisation features. The main aim of this work was to develop a general emission code to concurrently model the emission maps, light curves, and spectra at various orbital phases for AR Sco. As a precursor to developing this code, I followed up on our previous work by fitting our geometric rotating vector model to the orbital-resolved polarisation position data. I found the magnetic inclination angle and observer angle to vary by $\sim 10^{\circ}$ and $\sim 30^{\circ}$, respectively, over the binary orbit. Further periodicity analysis of the spin- and beat-coupled linear fluxes showed that the spin-coupled emission dominates between orbital phases $0.1 - 0.6$ and the beat-coupled emission dominates between orbital phases $0.6 - 1.1$. These results emphasise the need to probe various injection scenarios proposed for AR Sco. For the development of the emission code, I solved the general equations of motion with included classical radiation reaction forces (RRF) by implementing the Dormand-Prince 8(7) numerical integrator with adaptive time-step methods to exploit the higher accuracy of the scheme. This yielded improved accuracy and computational time vs. the commonly used Vay symplectic integrator, particularly for the high $B$-fields, $E_{\perp}$-fields, and RRF needed for pulsar and pulsar-like magnetospheres. In my investigation of the RRF's effect on the particle pitch angle, I reaffirmed the notion that the RRF does not significantly affect the pitch angle in the super-relativistic regime as well as the $\mathbf{E}\times\mathbf{B}$-drift maintaining a large pitch angle deviation of the trajectory with respect to the local $B$-field. Additionally, I demonstrated the novel result of the particles entering and conforming to the radiation reaction limit regime of Aristotelian Electrodynamics (AE) for uniform electromagnetic fields. This served as an excellent validation for my RRF calculations. As calibration for the radiation calculations, emission maps, and spectra I compared my model output with results from the code of the pulsar emission model of Harding and collaborators for a pulsar with $10\%$ the $B$-field strength of Vela and how my results converged to theirs. I also showed the novel results of the particles conforming to the AE regime but for more realistic force-free fields. This once again validates my results but also validates the AE assumptions since it is notably a solution to the equilibrium electrodynamics. Furthermore, I assessed two existing synchro-curvature radiation calculation methods, assessing whether they are appropriate for use in the high $E_{\perp}$-fields present in pulsar magnetospheres. These results highlight the importance of following the particle along its $\mathbf{E}\times\mathbf{B}$ drifted trajectory (AE trajectory) and calculating the applicable radiation along this curve. One is thus interested in the angle of the particle as it gyrates around this curve instead of the traditional pitch angle. Having established confidence in my particle dynamics and emission calculations, I next showed my exploratory modelling of the magnetic mirror scenario proposed for AR Sco. I demonstrate I could fit the observational spectral energy distribution including the recent \textit{NICER} pulsed X-ray spectrum well, constraining the white dwarf $B$-field to $B_{\rm S} = (2.5 - 3.0) \times 10^{8} \, \rm{G}$ and the electron power-law index to $p\sim 2.9$, falling within the observational estimate of between $0.8 - 1.4$ for the Optical/UV photon spectral index. Finally, I showed the effect the $B$-field, $E_{\perp}$-field, initial pitch angle, and initial particle Lorentz factor have on the mirror points, RRF, emission maps, and spectra. This demonstrates how crucial it is to include the general particle dynamics to accurately model the micro-physics present in magnetic mirror models.       
         
\vspace*{\fill}

\textbf{\textit{Keywords:}} AR Sco, White Dwarf Pulsar, General Particle Dynamics, Radiation Reaction Forces, Aristotelian Electrodynamics, Pulsar Emission Modelling, Numerical Modelling

\titleformat{\section}{\fontsize{14}{16}\bfseries}{\thesection}{1em}{}

\tableofcontents    
\listoffigures                                              
\pagebreak
\pagenumbering{arabic}
\chapter{Introduction}
This thesis is presented in article format, with each article and Chapter~\ref{sec:ARSCO} including its own relevant introduction. In the following, I will present an overarching introduction to give context concerning the topic, problem statement, and aims of the thesis. I will therefore repeat some of the necessary topics from the other introductions but keep the content more concise.       

\section{Binary White Dwarf Pulsars}
With the recent discovery of the AR Scorpii (AR Sco) sibling J$1912 - 4410$ by \citet{Pelisoli2023} the class of binary `white dwarf pulsar' has finally been established, making it the second confirmed source in this class after the prototype AR Sco. AR Sco was originally misclassified as a variable delta Scuti star, but was fortuitously re-classified by \citet{Marsh2016} as a binary system containing a white dwarf (WD) and an M-dwarf companion with pulsed multi-wavelength emission. The non-thermal emission was found to be primarily pulsed at the WD spin period of $P_{\rm S} = 117 \, \rm{s}$, ranging from the radio to X-ray band \citep{Marsh2016}. \citet{Buckley2017} observed the optical polarisation to be up to $40\%$, as well as a $90\%$ pulse fraction. Given the narrow emission lines, absence of flickering \citep{Garnavich2019}, and no signatures of an accretion column \citep{Takata2018}, there is no indication of significant accretion in the system. Due to the WD spin-down power being sufficient to power the non-thermal emission, \citet{Buckley2017} proposed the name of `first WD pulsar' for the source. Interestingly, some of the emission is observed to be coupled to the binary period $(P_{\rm S} = 118.2 \, \rm{s})$ of the system, indicating that emission is linked to the binary interaction between the WD and companion star \citep{Potter2018, Takata2018}. This differentiates these systems from standard canonical neutron star pulsars since the companion seems to be fundamental in producing the non-thermal emission. The underlying particle injection and emission mechanism have not yet been elucidated for AR Sco with the majority of models suggesting that the magnetic interaction between the WD and companion injects and accelerates particles in the magnetosphere of the WD where the particles radiate synchrotron radiation (SR) as they travel along the $B$-fields lines of the WD \citep{Takata2017, Potter2018b, Bednarek2018, DuPlessis2019, Lyutikov2020, Singh2020}. The current most popular model among these is the magnetic mirror model proposed by \citet{Takata2017}. This model proposes that particles are injected at the heated surface of the companion due to the $B$-field interaction. These particles travel towards the WD surface until they encounter a magnetic mirror close to the WD surface where they radiate most of their energy. These particles then become trapped in the WD magnetosphere as they are bounced between the magnetic mirrors at the WD magnetic poles until they are reabsorbed by the companion or escape the magnetic confinement at the poles. We will discuss all the observations and proposed models for AR Sco in detail in Chapter~\ref{sec:Paper1} and the more recent models in Chapter~\ref{sec:Paper2}.

When assuming that the non-thermal emission is powered by the WD spin down, a surface $B$-field of $\geqslant 200 \, \rm{MG}$ was inferred \citep{Buckley2017, Potter2018b}. However, there has been no direct detection of the WD $B$-field, only an upper limit of $\leqslant 100 \, \rm{MG}$ due to the lack of Zeeman splitting of the Ly $\alpha$ line \citep{Garnavich2020}. This feeds into the current biggest problem faced by AR Sco models, namely that the spin rate of the WD suggests a low $B$-field $\leqslant 10 \, \rm{MG}$ to be able to have been spun up to its current rate \citep{Lyutikov2020, Pelisoli2023}, yet the high spin down of the WD suggests a high $B$-field required for the synchronising torque \citep{Katz2017, Pelisoli2023}. Using the high inferred $B$-field for AR Sco and assuming mass transfer via Roche-Lobe overflow, \citet{Lyutikov2020} shows that the mass transfer rate required to spin up the WD is $\dot{M}= 10^{-4} \, \rm{M}_{\odot}\rm{yr}^{-1}$, which is $10^{5}$ times higher than in similar cataclysmic variables. Conversely, no model has shown the ability using low $B$-fields to concurrently fit the non-thermal synchrotron spectrum observed for this source. The \citet{Lyutikov2020} model also proposes a high mass transfer rate where there is no observational evidence of an accretion column or the system being in a propeller state. A model reconciling the high $B$-field and spin rate has recently been proposed by \citet{Schreiber2021} suggesting that the WD is formed without a $B$-field, meaning the WD can be spun up unimpededly and becomes magnetic via a rotational and crystallisation dynamo. They propose that AR Sco was originally spun up via accretion until the spin rate and $B$-field were too large, leading the transfer of angular momentum to cause the system to become detached. Eventually, via magnetic braking and gravitational wave losses, the system reattached leading to the high spin rate and high spin down with a high $B$-field. As the WD is spun down, the system will enter a Roche Lobe overflow state where the companion accretes material onto the WD, causing it to be spun up. This means these systems switch between the current spin-down state and accretion state, analogous to transitional pulsars that switch between accretion-powered and spin-powered states. An alternative model solving this problem has been proposed by \citet{Beskrovnaya2024} assuming that during the spin-up phase, the magnetosphere radius is smaller than the Alfven radius, thus the accretion can penetrate the WD magnetosphere via Bohm diffusion. Interestingly the AR Sco sibling seems to exhibit possible indications of flaring \citep{Pelisoli2024}, which could support the model proposed by \citet{Schreiber2021} but is contrary to AR Sco's absence of visible flaring, suggesting they are in different evolutionary states. 
       
\section{Global Pulsar Magnetosphere Modelling}
Over the years, various pulsar emission models have been proposed to explain the multi-wavelength signatures observed from these objects. Local acceleration gap models such as the slot gap \citep{Arons83, Muslimov03} and the outer gap \citep{Chen1984, Romani95} could produce reasonable light curves and spectra but fell short of explaining pair production, where particles are accelerated, global current patterns, and the multi-wavelength emission. These topics are still somewhat uncertain, even with the significant advancements in modern pulsar emission models, but they have highlighted the importance of solving the Maxwell equations self-consistently, including the particle dynamics and the pulsed emission. One of the important advancements for these models is the force-free electrodynamics (FFE) framework \citep{Contopoulos1999, Komissarov2002, Spitkovsky2006, Kalapotharakos2009} where the plasma is solely governed by the Lorentz force, meaning inertia and gas pressure are ignored. Also, no gaps would form in this plasma-filled magnetosphere, so in principle there can be no particle acceleration and subsequent radiation. An important addition to the FFE framework were the dissipative force-free formulations \citep{Gruzinov2008, Kalapotharakos2012, Li2012} that allowed for gaps or dissipative regions to form, giving hints of the distribution of the accelerating electric fields, but these models still struggled to address the exact emission regions, particle acceleration, and pair formation, being dependent on the assumption of the spatial distribution of the conductivity. From these works and modern pulsar magnetosphere simulation solutions, the common belief is that most pulsar magnetospheres where enough pair production occurs are nearly force-free and the dominant site of dissipation of energy is in the equatorial current sheet outside the light cylinder (where the co-rotation speed equals the speed of light).

Kinetic models were subsequently developed to address the required kinetic-scale plasma physics, particle dynamics, and radiation reaction from first principles. Particle-in-cell (PIC) models do a great job of integrating the micro-physics into model relativistic global pulsar magnetospheres \citep{Chen2014, Philippov2014, Cerutti2015, Belyaev2015, Philippov2015}, but cannot resolve the required pair micro-physics near the stellar surface. PIC models have also been used to explain the high-energy pulsar emission \citep{Cerutti2016, Kalapotharakos2018, Brambilla2018, Philippov2018, Kalapotharakos2023}. However, PIC codes are very computationally demanding since they solve the full particle gyration with included classical radiation-reaction forces (RRF). Moreover, a major limiting factor in these simulations is the large-scale separation between the gyro-period compared to the stellar rotation period, requiring large computational power to run these. PIC models approach this limitation by scaling up the gyro-period to computationally realistic scales by lowering the electromagnetic field values, particle Lorentz factors $\gamma$, and RRF by orders of magnitude. The problem is that these parameters are many orders of magnitude different than what is realised in real pulsars, and one may therefore be unable to probe the true pulsar environment by such simulations. Simply re-scaling the parameters and forces to higher values after the simulation is questionable, since different considerations come into play when operating in the RRF limit at high $\gamma$ and field strengths \citep{Petri2023}.

An alternate approach to modelling the pulsar magnetosphere uses the concept of Aristotelian Electrodynamics (AE), first proposed by \citet{Finkbeiner1989} and later popularised by \citet{Gruzinov2012}. AE proposes that the particle is quickly accelerated by the parallel $E$-field to a critical particle Lorentz factor $\gamma$, where the Lorentz force and RRF are in equilibrium as the particle follows the principal null direction (i.e., moving at the speed of light $c$). The advantage is that in this case one can make approximations for the super-relativistic particle trajectories and avoid integrating the full equations of motion. \citet{Petri2023} blends the FFE and AE approaches, allowing him to avoid integrating the equations of motion and simply use a particle pusher. In that model, he balances the Lorentz force with a radiative force that is linear in velocity, reducing to the AE result in the limit of $v=c$. A similar approach is that of \citet{Yangyang2022}, using the ideas of AE and letting the particles follow the principal null direction and equating the spatial component of the light-like moving particle to the radiation-reaction-limited velocity. AE assumes that curvature radiation (CR) dominates the particle losses, but one can also balance the gained power with the synchro-curvature power radiated as in \citet{Vigano2015} for a more general value of the critical $\gamma$. AE is therefore an equilibrium solution following the principal null direction, thus the particle has to be quickly accelerated to the critical $\gamma$ for this limit to be applicable. Additionally, the RRF can exceed the Lorentz force in the observer frame, as discussed by other authors \citep{landau1975, Cerutti2012, Vranic2016, Yangyang2022}, meaning these equations apply only once the particle enters equilibrium. An advantage of AE is that it traces out the particle trajectory as it heads out to infinity due to the influence of the fields, incorporating the $\mathbf{E}\times \mathbf{B}$-drift, which is important for synchro-curvature radiation (SCR), synchrotron radiation (SR), and CR calculations. The $\mathbf{E}\times \mathbf{B}$-drifting effect on the trajectory and radiation needs to be accounted for in pulsar-like sources, due to the $E_{\perp}$-field being a significant fraction of the $B$-field in and beyond the pulsar magnetosphere (demarcated by the light cylinder). Hence, this consideration is important since the standard SR calculations are derived in the absence of an $E$-field \citep{Blumenthal1970}, so it is technically not applicable in this case. 

\section{Problem Statement and Research Aims} 
The wealth of multi-wavelength observations for AR Sco at high cadence enable measurement of the system parameters versus orbital phase instead of averaging over large ranges of the orbital phase. Current emission and geometric models for AR Sco are unable to accurately and jointly model the light curves, spectra, and polarisation signatures of the source AR Sco. Thus, it is crucial to develop an emission model that is able to concurrently reproduce all of these signatures. The first step is to determine the magnetic field geometry and system parameters such as the magnetic inclination angle and observer angle. Additionally, this constrains the $B$-field structure. Since geometric models cannot produce energy-dependent light curves or spectra, we need to introduce more physics to construct a sophisticated emission model (involving general particle dynamics and radiation physics) in order to probe the system properties. This is analogous to the recent development history of pulsar models, where geometric models informed the development of more sophisticated emission models. As is clear from the particle injection and proposed models for AR Sco, one needs to solve the general particle or plasma dynamics for highly relativistic particles in large $B$-fields and induced $E_{\perp}$-fields to account for the mirroring and $\mathbf{E}\times\mathbf{B}$-effects experienced by these particles. This means more sophisticated modelling is required akin to the PIC modelling used for global pulsar magnetosphere simulations.

Our aims with this study are to firstly fit our geometric rotating vector model (RVM) to the orbital-phase resolved polarisation position angle (PPA) observations of \cite{Potter2018}. This will allow us to constrain the WD magnetic inclination angle and the observer angle at various orbital phases over the whole orbit. Additionally, we will do a deeper analysis of the linear polarisation of the source using Lomb-Scargle periodograms to probe the spin and beat-coupled emission components at the various orbital phases. This will give us more insight and information to develop an emission model for the source.            

For the development of our emission model, we aim to predict the emission maps, spectra, and light curves at various orbital phases which have not been concurrently or adequately modelled for AR Sco to date. To achieve this we aim to solve the general particle dynamics and incorporating a general radiation loss term, without making assumptions of large particle Lorentz factors or small pitch angles as well as accounting for all the particle drift effects. This is thus novel to previous models proposed for the source. There is a very high computational demand when resolving the full particle gyration in high $B$-field regimes with small particle $\gamma$-values due to the small particle gyro-radius and needing to resolve the full particle gyration. Therefore, we will implement a high-order numerical solver with adaptive timestep methods to improve both accuracy and computational time above that of current pulsar PIC methods. We will also test if our particle converges to the radiation-reaction limit regime of AE assumed to operate in pulsar modelling \citep{Gruzinov2012, Yangyang2022, Petri2023}. To calibrate our radiation, phase corrections, emission maps, and spectral calculations we will use the same methods as the pulsar emission models of \citet{Harding2015, Harding2021, Barnard2022} and reproduce a pulsar scenario ($10\%$ field strength of Vela) to compare our results to. We will also assess the effect the large $E_{\perp}$-field present in pulsars and pulsar-like sources has on the particle trajectories and emission while identifying a suitable SCR method that accounts for the $\mathbf{E}\times\mathbf{B}$-drift effects. The effect of the $E_{\perp}$-field on the particles is also highly relevant for AR Sco, which was neglected by previous models, especially for the proposed magnetic mirror scenario.     

Finally, we will use our code to model the magnetic mirror scenario for AR Sco proposed by \citet{Takata2017}. We will investigate the model parameters suggested by \citet{Takata2017, Takata2019} as well as assess various WD $B$-field strengths, magnetic inclination angles ($\alpha$), particle power law indices, and observer angles ($\zeta$). Therefore, we will fit our model predictions to the available multi-wavelength spectral energy distribution (SED) data for AR Sco as well as compare our emission maps to the observational orbital phase-resolved emission maps from \citet{Potter2018}. Due to solving the general particle dynamics, we will also investigate the micro-physics operating in the magnetic mirror scenario.

\section{Thesis Outline} 
The structure of this thesis is presented in the following order:

{\bf Chapter}~\ref{sec:background} is a brief and broader literature overview of additional topics relevant to pulsar physics that are not covered in each of the article chapters. The specific background for each article is presented in each of their respective introductions and methods sections.   

{\bf Chapter}~\ref{sec:Paper1} presents the first article \citep{DuPlessis2022} and gives a thorough observational introduction to AR Sco as well as an overview of the models proposed for AR Sco up to the point of publication of the article. The article then expands on the work of \cite{DuPlessis2019} to fit our implementation of the RVM to the orbital phase-resolved polarisation data as well as extended periodicity analysis of the linear polarisation data.   

{\bf Chapter}~\ref{sec:Paper2} presents the second article \citep{DuPlessis2024} and serves as the methods article for how I solved our model's general particle dynamics, including RRF using higher-order numerical solvers and adaptive step size methods. Additionally, I demonstrate how our results conform to the radiation-reaction-limited results of AE for uniform $B$- and $E$-fields. 

{\bf Chapter}~\ref{sec:Paper3} presents the calibration article (submission-ready draft), where I show how I reproduced the trajectories, emission maps, and spectra of the pulsar emission models of \citet{Harding2015, Harding2021} using lower field values as well as the comparison to their model results. Additionally, I explain how I identified an SCR calculation method for our modelling purposes as well as assess the results' convergence to the AE regime using FF fields.  

{\bf Chapter}~\ref{sec:ARSCO} I discuss the magnetic mirror model for AR Sco proposed by \citet{Takata2017, Takata2019} in detail and how we reproduced their scenario setup using our model. I then show the SED and emission maps using their best-fit parameters, better fitting parameters for our model, and assessing some of the micro-physics involved in this scenario.  

{\bf Chapter}~\ref{sec:Conclusion} presents concluding remarks regarding this study, future work to be done on AR Sco using the code I have developed, and future improvements to and applications of the code.  

\section{Publications}
The following publications resulted from this work. \\
\textbf{Journal Publications:}
\begin{itemize}
\item Du Plessis, L., C. Venter, Z. Wadiasingh, A.K. Harding, D.A.H Buckley, S.B. Potter and P.J Meintjes, \textit{Probing the non-thermal emission geometry of AR Sco via optical phase-resolved polarimetry}, MNRAS, 510 (2), 2998-3010, doi:10.1093/mnras/stab3595, 2022.
\hspace{2pt}
\item Du Plessis, L., C. Venter, A.K. Harding, Z. Wadiasingh, C. Kalapotharakos and P. Els, \textit{Towards modelling AR Sco: Generalised particle dynamics and strong radiation-reaction regimes}, MNRAS, 532 (4), 4408–4428, doi:10.1093/mnras/stae1791, 2024.
\end{itemize}

\textbf{Conference Proceedings Publications:}
\begin{itemize}
\item Du Plessis, L., C. Venter, Z. Wadiasingh, and A. K. Harding, 2023, In proceedings of annual High Energy Astrophysics in Southern Africa (HEASA2022) conference,  \textit{Modelling the Multi-Wavelength Non-thermal Emission of AR Sco}, \\ https://ui.adsabs.harvard.edu/abs/2023heas.confE..25D.
\end{itemize}

\textbf{Journal Articles in Preparation:}
\begin{itemize}
\item Du Plessis, L., C. Venter, A.K. Harding, Z. Wadiasingh and C. Kalapotharakos,  \textit{Towards Modelling AR Sco: Calibration -- Reproducing High-Energy Pulsar Emission and Testing Convergence to Aristotelian Electrodynamics}, (MNRAS, in prep), 2025.
\item Du Plessis, L., C. Venter, A.K. Harding, Z. Wadiasingh and C. Kalapotharakos,  \textit{Towards Modelling AR Sco: Magnetic Mirror Modelling Results}, (MNRAS, in prep), 2025.
\end{itemize}

%%%%%%%%Background%%%%%%%
\chapter{Background} \label{sec:background}
This chapter serves as a brief additional background of topics not covered in the articles or Chapter~\ref{sec:ARSCO}. Since the most relevant background is already covered in each of the articles and methods section of Chapter~\ref{sec:ARSCO}, I aim to avoid repetition where possible. I have used the broad background from my Masters mini-thesis \citep{DuPlessisThesis}, removing, altering, and focusing the content to supplement the articles and Chapter~\ref{sec:ARSCO}. For details on white dwarf and neutron star formation processes and properties namely mass limits, densities, luminosity relations, spin relations, and B-field relations see Chapter 2.1 of my Masters mini-thesis \citep{DuPlessisThesis}.       

\section{Binary Systems}
Binary star systems are commonly found throughout the Universe in a variety of classes. I will mainly focus on the two classes that are most relevant to this thesis. Eclipsing binaries are orbiting stars that periodically eclipse one another, causing visible signatures in the observed luminosity. Through observations, one can calculate the orbital period, radii, and effective temperature of the stars constituting the system. Spectroscopic binaries are stars with discernible spectra, where Doppler shifting causes the spectral lines to be shifted from their respective rest-frame wavelengths. This allows one to calculate the radial velocities of the stars, constraining their masses \citep{carroll2017introduction}. Using these observables, one can apply Kepler's law for two circularly-orbiting bodies given by
\begin{equation} \label{eq 2.5}
\frac{P_{\rm orb}^{2}}{a^{3}} = \frac{4\pi^{2}}{G(M_{\rm{1}}+M_{\rm{2}})},
\end{equation}           
where $P_{\rm orb}$ is the orbital period, $a$ is the orbital separation, $G$ is the gravitational constant, and $M_{\rm{1}}$ and $M_{\rm{2}}$ are the masses of the stellar bodies. Conventionally, $M_{\rm{1}}$ is used to donate the mass of the compact object in a system if a compact object is indeed present. Using Equation~(\ref{eq 2.5}) and the orbital inclination angle\footnote{The angle between the plane of the orbit and the plane of the sky.} $i$ a mass function can be derived to constrain the mass of the compact object \citep{hartle2003}
\begin{equation}
\frac{M_{2}^{3}}{(M_{1}+M_{2})^{2}}\sin^{3}i = \frac{P_{\rm orb}V_{1}^{3}}{2\pi G},
\end{equation}  
where $V_{1}$ is the orbital velocity of the first body. This equation can be written in terms of the mass ratio $M_{1}/M_{2}$ for further simplification. Estimating $i$ and $M_{2}$ from the spectrum, one can constrain $M_{1}$.

\begin{figure}[h!]
\centering
\includegraphics[width=0.5\textwidth]{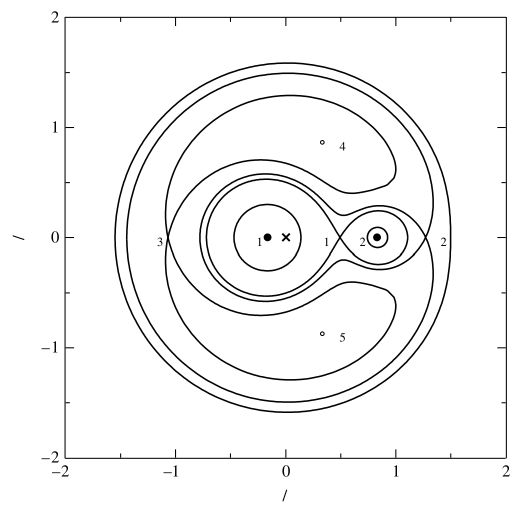} 
\caption[Roche Lobes Illustration]{Equipotential surfaces of a binary system where the stellar bodies are shown by the black dots marked $1$ and $2$ and centre of mass as an $x$ \citep{carroll2017introduction}. The inner Lagrangian point is located at the intersecting point numbered $1$.}
\label{Roche_Lobe}
\end{figure}
A close binary system can now be defined as a system where the two stars in the system share a common envelope. If we assume the binary system has a circular orbit, then using Newton's gravitational attraction force, equipotential surfaces can be calculated for each star\footnote{Equipotential surfaces are surfaces constructed by connecting points with equal gravitational potential shown in Figure \ref{Roche_Lobe}.}. Close to each star, these equipotential surfaces are spherical and become deformed as the distance from the star increases. The point where the two stars' equipotential surfaces intersect between the two stars is called the inner Lagrangian point where these two respective surfaces define the critical Roche lobes. The inner Lagrangian point is also where mass passes through from the donor star to the compact object. The size of the Roche lobe is measured by its average radius $r$, such that the volume inside the Roche lobe is equal to the volume of a sphere with radius $r$ \citep{1971ARA&A...9..183P}. Using the mass ratio $M_{\rm{1}}/M_{\rm{2}} = q$, the average radius is given by
\begin{equation} \label{eq 2.2.6}
\frac{r_{\rm{1}}}{a} = 0.38 + 0.2\log\left(q \right) \ \ \	\rm{for} \ 0.3< \textit{q} <20, or
\end{equation}
\begin{equation} \label{eq 2.2.7}
\frac{r_{\rm{1}}}{a} = 0.46224\left(\frac{M_{\rm{1}}}{M_{\rm{1}}+M_{\rm{2}}} \right)^{\frac{1}{3}} \ \ \ \rm{for} \ 0<\textit{q}<0.8, 
\end{equation}
where $a$ is the orbital separation. Importantly when calculating the radius $r_{\rm{1}}$ with Equation (\ref{eq 2.2.6}) it yields a larger value than using Equation (\ref{eq 2.2.7}) when $q>0.523$ \citep{1971ARA&A...9..183P}. Hence, accretion via the inner Lagrangian postulates that for a close binary system, there is a critical surface with average radius $r_{\rm{cr}}$ where if the star has a radius greater than $r_{\rm{cr}}$, matter will flow through the inner Lagrangian from the donor star and accrete onto the star \citep{1971ARA&A...9..183P}.  

\section{Pulsars} \label{sec2}
\subsection{Standard Braking Model}
Pulsars are generally considered to be extremely fast-rotating and highly magnetised neutron stars (NS). Their large rotating $B$-fields subsequently induce a large $E$-field that co-rotates the particles in the plasma-filled magnetosphere. As a conceptual idea, a pulsar may be described as a rotator with the source having a magnetic pole that is offset by an angle $\alpha$ (magnetic inclination angle) with respect to the rotation axis. A spin-down model is therefore used to describe the slowing rotational period of the pulsar as it loses its rotational energy due to braking forces in the case where it is not spun up via accretion \citep{lorimer2005handbook}. The rotational kinetic energy of the NS is thus converted into particle acceleration. Traditionally, the rotational energy loss rate $\dot{E}_{\rm rot}$ is typically equated to the vacuum magnetic dipole radiation loss rate \citep{Ostriker1969}, yielding
\begin{equation} \label{eq 2.8}
\frac{d}{dt}\left(\frac{1}{2}I_{\rm m}\Omega^{2} \right)=I_{\rm m}\Omega\dot{\Omega} = -\frac{2\mu^{2}\Omega^{4}}{3c^{3}}\sin^{2}\alpha, 
\end{equation}
where $I_{\rm m}$ is the moment of inertia, $P_{\rm s}=2\pi/ \Omega$ is the spin period, $\Omega$ the angular velocity, $\dot{\Omega}=4\pi^{2}I_{m}\dot{P}_{\rm s}/P^{3}_{\rm s}$ the rate of change in angular velocity, $\mu=BR^{3}/2$ is the magnetic moment, and $R$ the radial distance. Solving Equation (\ref{eq 2.8}) for $B$, the $B$-field at the surface of the star can be calculated as
\begin{equation} \label{eq 2.9}
B_{\rm S} \equiv \sqrt{\frac{3c^{3}}{8\pi^{2}} \frac{I_{m}}{R^{6}\sin^{2}\alpha} P_{\rm s}\dot{P}_{\rm s}}.
\end{equation}
Using $I=10^{45}\rm{g\,cm^{2}}$, $R=10^{6}\, \rm{cm}$, and $\alpha=90^{\circ}$, reduces Equation (\ref{eq 2.9}) to
\begin{equation}
B_{\rm S} = 6.4\times10^{19} \sqrt{P_{\rm s}\dot{P}_{\rm s}}~\rm{G}
\end{equation}
at the poles. If the spin down is characterised as $\dot{\Omega}=-k\Omega^{n}$, where $k$ is a constant, then the braking index $n$ is obtained by solving for $n$ using $\ddot{\Omega}/\dot{\Omega}$. To calculate the age of a pulsar one integrates $\dot{\Omega}=-k\Omega^{n}$ assuming $n>1$ and $\Omega_{0}\gg \Omega$, with $\Omega_{0}$ the initial angular velocity. This yields $ \tau =P_{\rm s}/\left(n-1\right)\dot{P}_{\rm s} $, thus substituting $ n=3 $ for magnetic dipole radiation one obtains $ \tau =P_{\rm s}/2\dot{P}_{\rm s} $ as the characteristic age of a pulsar \citep{Ostriker1969}.

\begin{figure}[h!]
\centering
\includegraphics[scale=10]{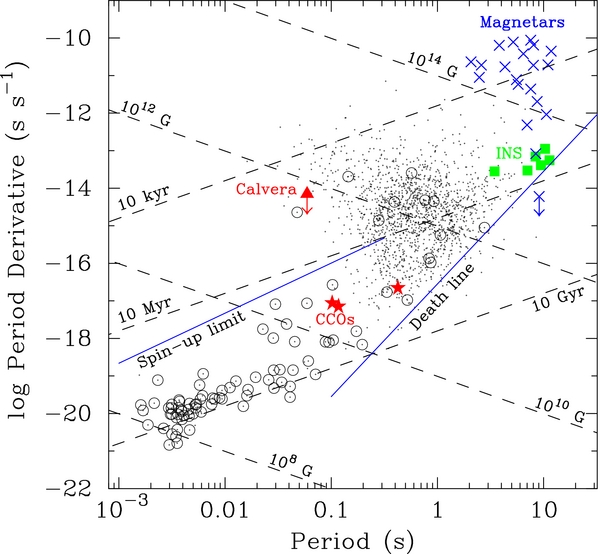} 
\caption[$P_{\rm s}\dot{P}_{\rm s}$ diagram]{$P_{\rm s}\dot{P}_{\rm s}$ diagram where the positive-slope dashed lines indicate the age of the pulsar and the negative-slope dashed lines indicate the surface $B$-field strength \citep{Gotthelf2013}.}
\label{fig:2.2}
\end{figure}

Since the $B$-field and the age of a pulsar are both dependent on $P_{\rm s}$ and $\dot{P}_{\rm s}$ one can construct a $ P_{\rm s}\dot{P}_{\rm s} $ graph as shown in Figure \ref{fig:2.2}. In the figure the positive-slope dashed lines indicate the characteristic age and the negative-slope dashed lines the surface $B$-field strength of a pulsar.

\subsection{GJ Magnetosphere} \label{Back_GJ}
After \cite{Gold1968} proposed some of the first conceptual pulsar models, \cite{GJ69} started developing the foundations for pulsar electrodynamics. The initial assumption was that an NS is a perfectly conducting sphere with an aligned magnetic field and spin axis ($\mathbf{\mu}\parallel\mathbf{\Omega}$). Outside of the star a dipole $B$-field structure ($r=k'\sin^{2}\theta$, where $k'$ indicates the specific field line) is assumed, with the interior $B$-field being $\mathbf{B}_{in}=B_{\rm S}\vec{e}_{z}\parallel\mathbf{\mu}$ \citep{Padmanabhan2001}. Since inside the conductor $\mathbf{E}\cdot\mathbf{B}=0$, \cite{GJ69} showed that the Maxwell equations yield  
\begin{equation}
\begin{split}
\mathbf{E}_{\rm in} & + \frac{\left(\mathbf{\Omega}\times\mathbf{r}\right)}{c}\times\mathbf{B}_{in} = 0 \\
\mathbf{E}_{\rm in} & = -\frac{B_{0}\Omega r\sin\theta}{c}\left(\sin\theta\mathbf{e}_{\rm r}+ \cos\theta\mathbf{e}_{\theta}\right),
\end{split}
\end{equation}
where $r$, $\theta$, and $\phi$ are the base spherical coordinates and the $\phi$ dependence is removed due to the axisymmetry of a sphere. In the stellar interior the condition $\mathbf{\nabla}\mathbf{E}_{in}=0$ is met, therefore the Laplace equation can be solved to find the potential, where $R$ is the radius of the NS and $r$ the radial distance from the surface,
\begin{equation}\label{phi_neut}
\Phi = -\frac{B_{\rm S}\Omega R^{5}}{2cr^{3}}\sin^{2}\theta.
\end{equation}
If one assumes a vacuum magnetosphere conditions, the Lorentz invariant $\mathbf{E}\cdot\mathbf{B}$ can be calculated assuming a dipole $B$-field and using the potential in Equation~(\ref{phi_neut}),
\begin{equation}
\mathbf{E}\cdot\mathbf{B}=-\left(\frac{\Omega R}{c}\right)\left(\frac{R}{r}\right)^{7}B_{\rm S}^{2}\cos^{3}\theta\neq0.
\end{equation} 
The $E$-field component parallel to the local $B$-field can then be calculated as
\begin{equation}
\frac{\mathbf{E}\cdot\mathbf{B}}{\mid \mathbf{B} \mid} \approx -\left(\frac{\Omega R}{c}\right)B_{0}\cos^{3}\theta \sim 2\times10^{8}B_{12}P^{-1}~\rm statvolt \, cm^{-1},
\end{equation}  
where $B_{12}=B_{0}/10^{12}~\rm G$. This is $\sim 10^{8}$ times the gravitational binding force of a proton, meaning that particles are ripped from the surface of the NS, filling the magnetosphere. It was therefore realised that an NS cannot be surrounded by a vacuum \citep{GJ69}. Considering that a particle's speed cannot exceed the speed of light in a vacuum, a cylindrical radius can be defined using the dipolar $B$-field structure, 
\begin{equation} \label{R_LC}
R_{\rm LC}=\frac{c}{\Omega}=\frac{cP}{2\pi},
\end{equation}
where particles travelling along these $B$-field lines at this radius are co-rotated at approximately the speed of light. This defines the light cylinder radius and also defines the last closed $B$-field line forming the magnetosphere (the latter being tangent to this radius). Using the magnetic dipole equation ($r=k\sin^{2}\theta$), Equation (\ref{R_LC}), and the small-angle approximation\footnote{$\sin^{-1}\theta\sim\theta$} the polar cap angle $\theta_{\rm pc}$ can be calculated as \citep{rybicki2008radiative}
\begin{equation} \label{theta_pc}
\theta_{\rm{pc}} \simeq \sqrt{\frac{\Omega R}{c}}. 
\end{equation}
Using this angle, a corresponding polar cap radius $R_{\rm PC}=R\sin\theta\sim R\sqrt{\Omega R / c}$ can subsequently be derived. A potential difference can hence be derived between the magnetic pole and $R_{\rm PC}$ by substituting $R_{\rm PC}$ into Equation (\ref{phi_neut}), yielding
\begin{equation}
-\Delta\Phi = \frac{1}{2}B_{0}R\left(\frac{\Omega R}{c}\right)^{2}.
\end{equation}   

The particles pulled from the NS surface fill the magnetosphere, configuring themselves such that there is no Lorentz force acting on them, hence creating a force-free (FF) magnetosphere. Gauss's equation can then be solved to find the Goldreich-Julian charge density required to fill the magnetosphere
\begin{equation} \label{eq 2.14}
\rho_{\rm{GJ}}=\frac{\mathbf{\nabla} \cdot \mathbf{E}}{4\pi}=\frac{\mathbf{\Omega} \cdot \mathbf{B}}{2\pi c} \frac{1}{1-(\Omega r/c)\beta_{\rm t}\sin\theta} \simeq -\frac{\mathbf{\Omega} \cdot \mathbf{B}}{2\pi c},
\end{equation} 
where $\beta_{\rm t}$ is the temporal component of the $\beta$ four-vector. We then obtain an electron number density of
\begin{equation} \label{eq 2.15}
n_{\rm e} \simeq -\frac{\mathbf{\Omega} \cdot \mathbf{B}}{2\pi ec},
\end{equation} 
where $e$ is the electron charge. 

% The closed magnetic field lines are rotating with the NS, where particles are forced to co-rotate along these closed magnetic field lines of the WD and stay confined to the magnetosphere. The open magnetic field lines are forced open, since particles travelling along the magnetic field lines cannot exceed the speed of light in vacuum as they are co-rotating. Particles liberated from the polar caps are thus accelerated along the open magnetic field lines away from the NS, and emit radiation.

\begin{figure}[h!]
\centering
\includegraphics[scale=1]{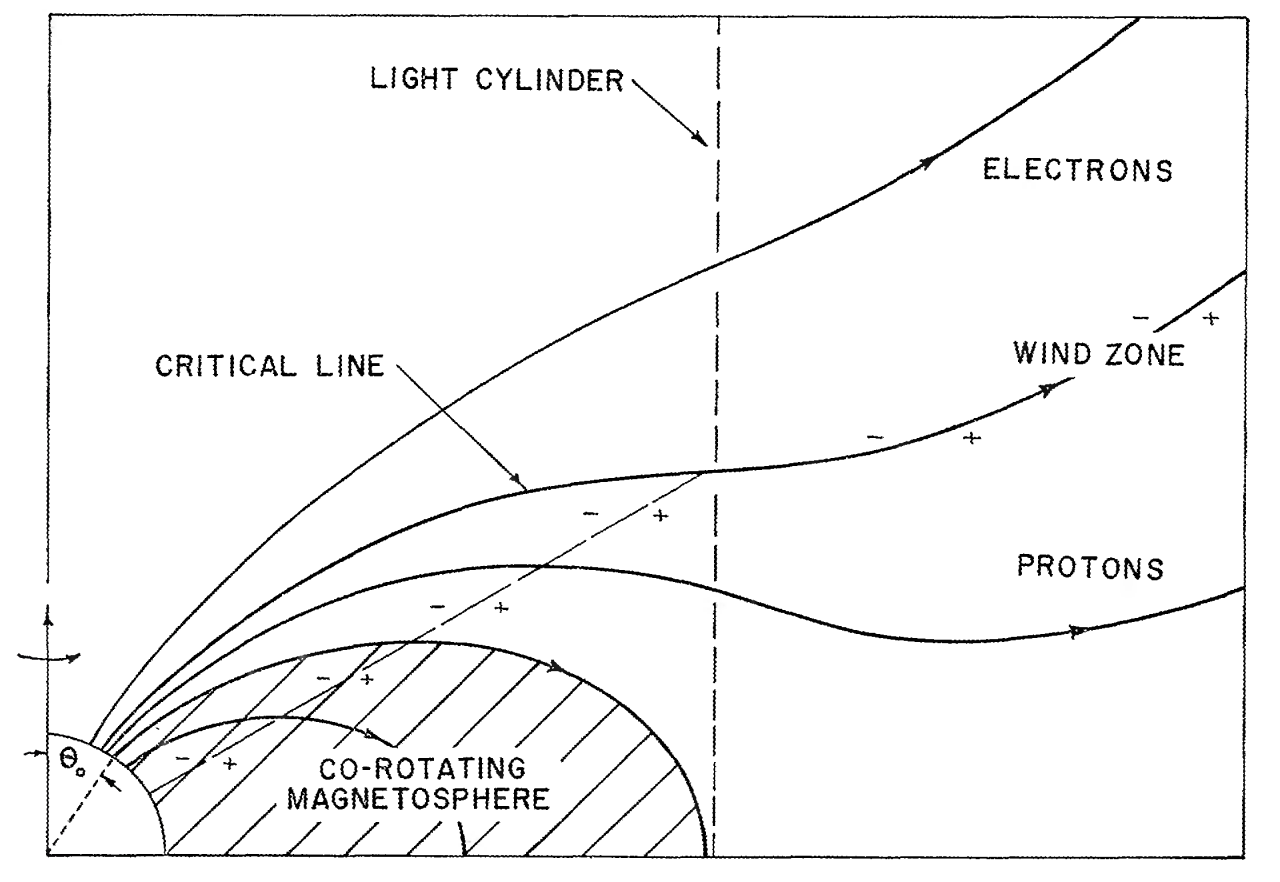}
\caption[Geometric Pulsar Magnetosphere Illustration]{A pulsar magnetosphere showing the light cylinder and the slant dashed line indicating where $\mathbf{\Omega}\cdot\mathbf{B}=0$ ($\rho_{\rm GJ}=0$) from \citet{GJ69}.}
\label{fig:2.3}
\end{figure}

\section{Radiation Mechanisms} \label{sec3}
\subsection{Synchrotron Radiation}
From classical electrodynamics, it can be shown that radiation from accelerating charges can be described by the Lamour equation yielding the total radiated power per charge. In the absence of an $E$-field a charged particle can be described to gyrate around a $B$-field with an angular frequency $\omega_{\rm g}=eB/\gamma mc$ known as the relativistic angular gyrofrequency. This can be derived from the equation of motion by setting $\bf{v}\cdot\bf{a}=0$, namely using the centripetal force. One obtains cyclotron radiation when these charges gyrate at non-relativistic speeds, leading to a spectrum of harmonics. In the relativistic case, this is known as SR, where a geometric analysis can be used to show that the emitted frequencies are boosted to $\nu_{\omega}\sim \gamma^{3}\nu_{\rm g}$, with $\nu_{\rm g} = \omega_{\rm g}/2\pi$ \citep{longair1983high}. For classical SR there is a non-zero velocity component parallel to the local magnetic field and a perpendicular acceleration component, assuming $a_{\parallel}=0$. Using the Lamour formula and substituting the frame-invariant scalar product of the four-vector acceleration $\mathbf{a}\cdot\mathbf{a}$ yields
\begin{equation}
\begin{split}
P &= \frac{2e^{2}\mathbf{a}\cdot\mathbf{a}}{3c^{3}}\\
&= \frac{2e^{2}\gamma^{4}}{3c^{3}}\left(a_{\perp}^{2}+\gamma^{2} a_{\parallel}^{2}\right)\\
&= \frac{2e^{2}\gamma^{4}}{3c^{3}}\omega_{\rm g}v_{\perp}^{2} \\
&= \frac{2e^{2}\gamma^{4}}{3c^{3}}\left(\frac{eB}{\gamma mc}\right)^{2}\left(v\sin\theta_{\rm p}\right)^{2}.
\end{split}
\end{equation}
where $a_{\parallel}=0$, $a_{\perp}=\omega_{\rm g}v_{\perp}$, and $\theta_{\rm p}$ is the pitch angle \citep{Jackson1962}. Reformulation yields
\begin{equation}
P = 2\sigma_{\rm T}c\beta^{2}\gamma^{2}U_{\rm B}\sin^{2}\theta_{\rm p},
\end{equation}
where the Thomson cross section is given by $\sigma_{\rm T}=8\pi e^{4}/3m^{2}c^{4}$ and the magnetic energy density $U_{\rm B}=B^{2}/8\pi$. Taking the average over an isotropic pitch angle distribution \citep{Padmanabhan2001} one obtains 
\begin{equation} \label{E_dot_SR}
P = \frac{4}{3}\sigma_{\rm T}c\beta^{2}\gamma^{2}U_{\rm B}.
\end{equation}
Due to the relativistic motion of the particles the emission cone is highly beamed towards the observer, with $\sin\theta\sim 1/\gamma$. A critical frequency is defined \citep{Blumenthal1970} by
\begin{equation}
\omega_{\rm c} = \frac{3}{2}\gamma^{3}\omega_{\rm g}\sin\theta_{\rm p},
\end{equation}
which arises from the fact that the acceleration rate cannot exceed the gyrofrequency, otherwise the particle would not be bound to the $B$-field, thus yielding the limit $t_{\rm acc}\geq t_{\rm SR}$.   
The single particle power per unit frequency \citep{Blumenthal1970} is calculated  using
\begin{equation} \label{P_SR}
P(\omega) = \frac{\sqrt{3}e^{3}B\sin\theta_{\rm p}}{2\pi mc^{2}}F(x),
\end{equation}
where $F(x)= x\int^{\infty}_{x}K_{5/3}\left(\epsilon\right)d\epsilon$, $K_{5/3}$ is the modified Bessel function of the order $5/3$ and $x=\omega/\omega_{\rm c}$. The Bessel function has the following asymptotic limits \citep{longair1983high}
\begin{equation}
\begin{split}
F(x) &\sim \frac{4\pi}{\sqrt{3}\Gamma\left(\frac{1}{3}\right)}\left(\frac{x}{2}\right)^{1/3},\qquad x\ll 1 \\
F(x) &\sim \left(\frac{\pi}{2}\right)^{1/2}\exp^{-x}x^{1/2},\qquad x\gg 1 
\end{split}
\end{equation}  
where $\Gamma(x)$ is the Gamma function, the generalisation of the factorial. Using a change in variable $x=\lambda y$ one can calculate the Gamma function with $\Gamma(\lambda) = \int^{\infty}_{0} y^{\lambda -1}\exp^{-y}$ for $\lambda > 0$. For an SR spectra consisting of multiple particles, one assumes a particle energy distribution, in this case a simple power law, with a number density of $N(E)dE$ with the particle energy distribution from $E$ to $E + dE$. This can be expressed as the relation \citep{rybicki2008radiative} 
\begin{equation}\label{par_dist}
N(E)dE = \kappa E^{-p}dE,
\end{equation}  
where $p$ is the power law index of the particles and $\kappa$ a normalisation constant. We can now use Equation (\ref{P_SR}), the power per unit frequency, and calculate the total power radiated as a function of frequency 
\begin{equation} \label{Fv_SR}
\begin{split}
P_{\rm tot}\left(\omega
\right)&= \int^{E_{2}}_{E_{1}}P\left(\omega
\right)N\left(E\right)dE \\
&\propto \int^{\gamma_{2}}_{\gamma_{1}}F\left(\frac{\omega}{\omega_{\rm c}}\right)\gamma^{-p}d\gamma \\
&\propto \omega^{-\left(p-1\right)/2},
\end{split}
\end{equation}
where the spectral index is $\alpha'=\left(p-1\right)/2$. If $F_{\nu}$ is taken to be the flux, then $F_{\nu}\propto P_{SR}(\nu)\propto \nu^{-\alpha'}$. This simplifies to $\nu F_{\nu}\propto \nu^{-(p-3)/2}$, which is used to plot the SED, since $\nu F_{\nu}=\left[\rm{erg/cm^{2}/s}\right]$ gives an energy flux per area. We use $\nu F_{\nu}$ since $dF_{\nu}/d(ln(\nu)) = \nu F_{\nu}/d\nu$, which is the energy flux per logarithmic frequency interval. Additionally, the SR radiation-reaction-limited Lorentz factor can be calculated by equating the particle acceleration to the SR loss rate, $ecE_{\parallel}=\dot{E}_{\rm SR}=4\sigma_{\rm T}u_Bc\gamma^2/3$.  Solving for $\gamma$ yields
\begin{equation} \label{SRR}
\gamma_{\rm SRR} = \left(\frac{E_{\parallel}6\pi e}{\sigma_{\rm T}B_{\rm p}^{2}}\right)^{1/2}.
\end{equation}   

In the scenario of synchrotron self-absorption, where a relativistic charged particle in the $B$-field can absorb a photon or if stimulated emission occurs, the spectral index can change. If one assumes the absorbed photon energy is much smaller than the charged particle energy it interacted with, the absorption coefficient can be calculated with a dependence of $\nu^{-(p+4)/2}$ \citep{Bottcher2012}. This indicates that the opacity of a non-thermal source increases with smaller frequencies until the source becomes optically thick. The frequency where this occurs is known as the synchrotron self-absorption frequency and taking this effect into account causes $\alpha'=5/2$ in that regime \citep{rybicki2008radiative}. This regime usually is located in the radio or low-frequency optical band.       

Let us define $G(x)$ similar to how $F(x)$ was defined \citep{rybicki2008radiative} 
\begin{equation}
G\left(x\right)=xK_{3/2}\left(x\right),
\end{equation}
where $K_{3/2}$ is the modified Bessel function of order $3/2$. The power radiated can then be written in terms of the two linear polarisation components, with respect to the local $B$-field \citep{longair1983high}
\begin{equation}
\begin{split}
P_{\perp}\left(\omega\right) &= \frac{\sqrt{3}e^{3}B\sin\theta_{\rm p}}{4\pi mc^{2}}\left(F(x)+G(x)\right) \\
P_{\parallel}\left(\omega\right) &= \frac{\sqrt{3}e^{3}B\sin\theta_{\rm p}}{4\pi mc^{2}}\left(F(x)-G(x)\right).
\end{split}
\end{equation}
The ratio of power yielded for the two polarisation components for one electron can be shown to be $P_{\perp}/P_{\parallel}=7$ \citep{longair1983high}. Similarly, it is shown that the fractional polarisation is
\begin{equation}
\Pi = \frac{P_{\perp}(\omega)-P_{\parallel}(\omega)}{P_{\perp}(\omega)+P_{\parallel}(\omega)} = \frac{G(x)}{F(x)}.
\end{equation} 
Using a power-law distribution, the fractional polarisation reduces to 
\begin{equation}
\Pi = \frac{p+1}{p+7/3}.
\end{equation}

\subsection{Curvature Radiation}
In the previous segment, it was noted that SR is associated with a change in perpendicular momentum or energy. CR is associated with a longitudinal energy change, namely a charged particle following a curved path. Particles are liberated from the pulsar surface and accelerated along the open magnetic field lines, making CR an important radiation process for pulsars' primary particles \citep{Sturrock1971}. Particles need to be accelerated above a Lorentz factor of $\sim 10^{7}$ for efficient CR contributions. Assuming a sufficiently large radius of curvature ($\rho_{\rm c}$) to use classical electrodynamics, \cite{Jackson1962} defines a critical frequency beyond which the radiation is exponentially suppressed as
\begin{equation}
\omega_{\rm c} = \frac{3}{2}\gamma^{3}\left(\frac{c}{\rho_{\rm c}}\right),
\end{equation} 
The critical energy can be calculated by 
\begin{equation}
E_{\rm c} = \hbar \omega_{\rm c} = \frac{3\hbar c\gamma^{3}}{2\rho_{c}},
\end{equation}
which can be obtained through similar means as for SR in the previous subsection. 
The $\rho_{\rm c}$ close to the stellar surface for a dipolar field is approximated in \cite{Harding1978} as 
\begin{equation}
\rho_{\rm c} \approx \frac{4R\theta_{\rm pc}}{3\theta^{2}},
\end{equation} 
where $\theta_{\rm pc}$ is the polar cap angle as defined in Equation (\ref{theta_pc}) and $\theta$ the polar angle with respect to the magnetic axis. At large distances, the curvature radius can be approximated as the light cylinder radius $\rho_{\rm c} \approx R_{\rm LC}$, although this becomes even larger for force-free magnetosphere \citep{Kalapotharakos2014}. The single particle CR power spectrum is given in \cite{Jackson1962} and \cite{erber1966} as  
\begin{equation} \label{P_CR}
\left(\frac{dP_{\rm CR}}{dE}\right) = \sqrt{3}\alpha_{\rm f}\frac{\gamma c}{2\pi \rho_{\rm c}}F\left(\frac{\omega}{\omega_{\rm c}}\right),
\end{equation}
where $F$ is defined similarly as before, $\alpha_{\rm f} = e^{2}/\hbar c$ is the fine structure constant, and $\hbar = h/2\pi$ is the normalised Planck's constant. Integrating over all frequencies yields \citep{Jackson1962}
\begin{equation} \label{E_dot_CR}
P_{\rm tot} =\frac{2e^{2}c}{3\rho_{\rm c}^{2}}\beta^{4}\gamma^{4}. 
\end{equation}
For detailed derivations of these equations, see \cite{Jackson1962}. By substituting standard pulsar radii and magnetic fields into Equation (\ref{E_dot_CR}), CR photons are found to have GeV energies and can produce electron-positron pair cascades in young pulsars. Equation (\ref{Fv_SR}) can similarly be used for CR by substituting $P_{\rm SR}$ for $P_{\rm CR}$. Once again one can obtain a curvature radiation-reaction-limited Lorentz factor by equating the acceleration rate to the CR loss rate, $ecE_{\parallel}=\dot{E}_{\rm CR}=2e^2c\gamma^4/3\rho_c^2$ \cite{Harding2005b}. Solving for $\gamma$ yields 
\begin{equation}\label{CRR}
\gamma_{\rm{CRR}} \sim (1.5E_{\parallel}/e)^{3/4}\rho_{\rm{c}}^{1/2} \sim 5\times 10^{7}.
\end{equation}   
    
\subsection{Pair Production}
The process of pair production operates in environments with efficient radiation processes and high $B$ and $E$-fields \citep{Harding1982}. Pair production can occur in one of two regimes, namely single-photon pair production requiring high $B$-fields and two-photon pair production requiring high photon densities. In the case of single-photon pair production, the photons can interact with a strong $B$-field to produce an electron-positron pair. If these photons have large enough energies, the probability of the photons travelling a distance $d$ in the $B$-field to create pairs can be calculated for a uniform $B$-field as \cite{erber1966}
\begin{equation}
n_{\rm p} = n_{\rm ph}\left(1 - e^{-\alpha_{\rm at}(\chi)d}\right). 
\end{equation}
Here $n_{\rm ph}$ is the photon number density, $n_{\rm p}$ the pairs number density, $\alpha_{\rm at}(\chi)$ the photon attenuation coefficient, and 
$\chi = 0.5\left(h\nu/m_{\rm e}c^{2}\right)\left(B_{\perp}/B_{\rm cr}\right)$ the Erber parameter. The critical $B$-field $B_{\rm cr} = m_{\rm e}^{2}c^{3}/e\hbar$ is a useful fiducial value used in \cite{Harding1983} where the particle gyrational energy is equal to its rest mass. The attenuation coefficient, which is the photon absorption per length for a photon propagating perpendicular to a uniform $B$-field \citep{erber1966}, is given as
\begin{equation}
\alpha_{\rm at}(\chi) = \frac{1}{2}\left(\frac{\alpha_{\rm f}}{\lambda_{\rm c}}\right)\left(\frac{B_{\perp}}{B_{\rm cr}}\right)T\left(\chi\right),
\end{equation} 
where $\lambda_{\rm c} =h/mc$ is the Compton wavelength and $T(\chi)$ a modified Bessel function. The asymptotic limits of $T(\chi)$ are 
\begin{equation}
\begin{split}
T(\chi) &= 0.46e^{-\frac{4}{3\chi}},\qquad \chi\ll 1 \\
T(\chi) &= 0.60\chi^{-1/3},\qquad \chi\gg 1 
\end{split}
\end{equation}
These limits can be inserted into the attenuation equation to be written in the following form \citep{Luo2000}.
\begin{equation}
\alpha_{\rm at}(\chi) = 0.46\left(\frac{\alpha_{\rm f}}{\lambda_{\rm c}}\right)\left(\frac{\chi}{\epsilon_{\rm ph}}\right),
\end{equation}
where $\epsilon_{\rm ph}=E_{\rm ph}/m_{\rm e}c^{2}$ is the photon energy in normalised units. From this equation one can see that if $B_{\perp}$ increases, the attenuation coefficient increases, meaning the pair production becomes more efficient. An approximated formulation for the conditions necessary for single-photon pair production is given by \cite{Sturrock1971} as
\begin{equation}
\epsilon_{\rm ph}B_{\perp}\geq 10^{11.9}.
\end{equation}   

Two-photon pair production occurs when two photons with sufficiently high photon energies, namely $E_{\rm ph}>m_{\rm e}c^2$ interact, creating an electron-positron pair. The cross-sectional area for this collision can be calculated in terms of the photon energy in the centre of momentum frame as shown by \cite{Svensson82}.  

Notably in \citet{DuPlessis2019} we estimated the pair creation conditions and inverse Compton scattering finding SR and CR (with high enough particle $\gamma$) to be the dominant radiation mechanisms.  
 
\section{Polarisation}
If the $E$-field of an electromagnetic wave is oscillating in one plane, the wave is described to be linearly polarised. The direction of oscillation and the direction of propagation define the plane of polarisation. Consider an $E$-field vector composed of two linearly polarised components, namely
\begin{equation}
\mathbf{E}= (E_{1}\mathbf{\hat{x}}+E_{2}\mathbf{\hat{y}})e^{\rm{-i\omega t}}, 
\end{equation} 
where $\mathbf{\hat{x}}$ and $\mathbf{\hat{y}}$ are the unit vectors, $\omega$ is the angular frequency and $t$ is the time \citep{rybicki2008radiative}. The complex amplitudes can then be written as $E_{1}=\xi_{1}e^{i\phi_{1}}$ and $E_{2}=\xi_{2}e^{i\phi_{2}}$. This leads to the $E$-field $x$ and $y$ components
\begin{equation} \label{eq 2.10}
E_{\rm{x}}=\xi_{1}\cos(\omega t-\phi_{1}), \ \ \ E_{\rm{y}}=\xi_{2}\cos(\omega t-\phi_{2}),
\end{equation}
where $\phi_{1}$ and $\phi_{2}$ are the phase shifts. 
\begin{figure}[h!]
\centering
\includegraphics[scale=1]{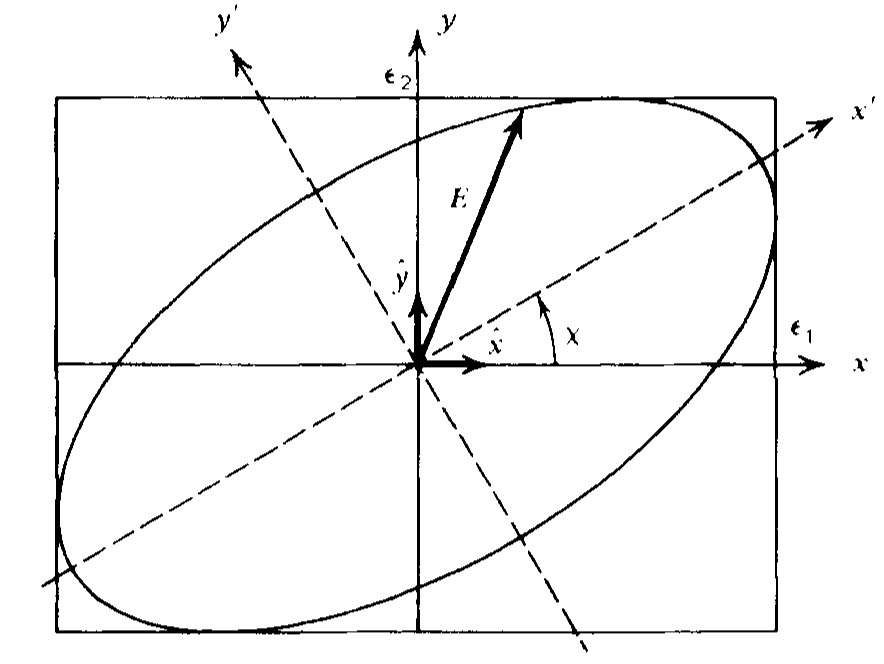} 
\caption[Rotated Polarisation Ellipse]{The principal axis of the polarisation ellipse where the $E$-field components are rotated by an angle $\chi$ \citep{rybicki2008radiative}.}
\label{fig:1.1}
\end{figure}
If the two components differ, using a rotated (primed) axis we may write in the case of elliptical polarisation: 
\begin{equation} \label{eq 2.11}
E_{\rm{x}}'=\xi_{0}\cos\beta\cos\omega t, \ \ \ E_{\rm{y}}'=-\xi_{0}\sin\beta\cos\omega t,
\end{equation} 
where $\beta$ is the ratio between the $x'$ and the $y'$ component (or the phase shift between these two components) and $\xi_{0}$ is the amplitude. Rotating Equations (\ref{eq 2.11}) by $\chi$ yields
\begin{equation}
\begin{split}
& E_{\rm{x}}=\xi_{0}(\cos\beta\cos\chi\cos\omega t+\sin\beta\sin\chi\sin\omega t), \\  
& E_{\rm{y}}=\xi_{0}(\cos\beta\sin\chi\cos\omega t-\sin\beta\cos\chi\sin\omega t).
\end{split}
\end{equation} 
Equating these two equations with Equation (\ref{eq 2.10}) produces four equations given in \cite{rybicki2008radiative}, where solving for $\xi_{0}$, $\beta$, and $\chi$ leads to the Stokes parameters.
\begin{equation}
\begin{split}
& I \equiv \xi_{1}^{2} + \xi_{2}^{2} = \xi_{0}^{2} \\
& Q \equiv \xi_{0}^{2}\cos2\beta\cos2\chi \\
& U \equiv \xi_{0}^{2}\cos2\beta\sin2\chi \\
& V \equiv \xi_{0}^{2}\cos2\beta, \\
\end{split}
\end{equation}
where $I$ is the intensity, since $I^{2}=Q^{2}+U^{2}+V^{2}$, and $V$ is the circular polarisation. The equations above can also be written in terms of linear polarisation components or right- and left-handed components. The Stokes parameters given as linear polarisation components are:
\begin{equation} \label{Stokes_linear}
\begin{split} 
& I = \left\langle E^{2}_{\rm x}\right\rangle + \left\langle E^{2}_{\rm y}\right\rangle  \\
& Q = \left\langle E^{2}_{\rm x}\right\rangle - \left\langle E^{2}_{\rm y}\right\rangle \\
& U = 2\left\langle E_{\rm x}E_{\rm y}\cos\delta\right\rangle \\
& V = 2\left\langle E_{\rm x}E_{\rm y}\sin\delta\right\rangle, \\ 
\end{split}
\end{equation}
where $\delta=\phi_{1}-\phi_{2}$ is the phase difference and the brackets represent the time-averaged value \citep{Trippe2014}.
Using the Stokes parameters, we can calculate the elliptical polarisation parameters
\begin{equation} \label{PPA_Stokes}
\begin{split}
& \xi_{0}=\sqrt{I} \\
& \sin2\beta=\dfrac{V}{I} \\
& \tan2\chi=\dfrac{U}{Q}, 
\end{split}
\end{equation}
where the parameters $Q$ and $U$ are used to measure the orientation of the ellipse with respect to the $x$-axis. It is important to note that the degree of linear polarisation $(m_{\rm L}=\sqrt{Q^2 +U^2}/I)$ and $\chi$ the polarisation angle, gives the length and orientation of a vector centred in the vector space of $Q$ and $U$, respectively \citep{Trippe2014}. 
 
\section{The Rotating Vector Model}
\begin{figure}[!h]
\begin{minipage}{20pc}
\includegraphics[width=20pc]{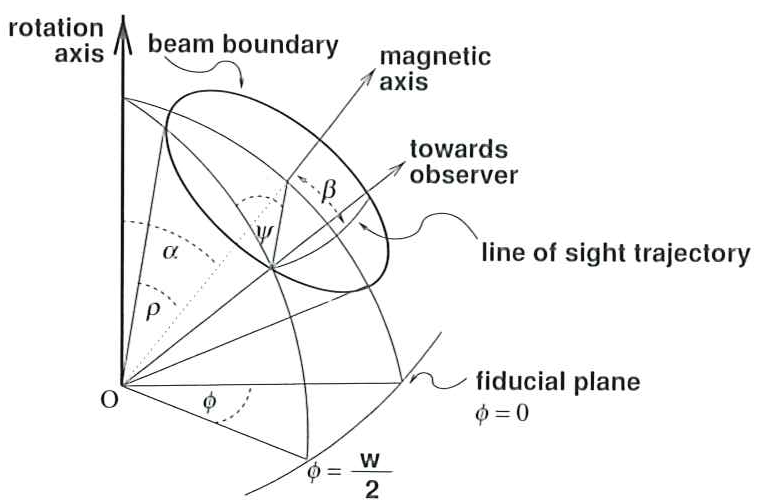}
\caption[Geometric Radio Emission Beam of a Pulsar]{ Schematic of the radio emission beam of a pulsar, indicating all the geometric angles associated with the emission beam \citep{lorimer2005handbook}.}
\label{fig:3.3}
\end{minipage}\hspace{1pc}%
\begin{minipage}{18pc} 
\includegraphics[width=18pc]{Figures/Background/fig1.png} 
\caption[Polarisation Signature from Pulsar Radio Beam]{ The emission beam and the path of the line of sight across the beam. The bottom image shows the observed PPA as it changes along this path \citep{lorimer2005handbook}.}
\label{fig:3.4}
\end{minipage} 
\end{figure}

Pulsars can be schematically thought of as rotating NSs with two radio emission cones located near the magnetic poles. Figure \ref{fig:3.3} defines a magnetic inclination angle $\alpha$ of the magnetic dipole moment $\mu$ with respect to the rotation axis $\bold{\Omega}$. The beam geometry of a pulsar shown in Figure~\ref{fig:3.3} also defines an observer angle $\zeta$, which is the angle between the rotation axis and the observer's line of sight. The impact angle is $\beta =\zeta - \alpha$, where $\psi$ is the polarisation position angle (PPA), $\phi$ is the sweeping or azimuthal angle, $W$ is the pulse width, and $\rho$ is the half-opening angle of the beam. The RVM is a geometrical model normally used for radio pulsars, where the PPA is predicted by the following equation \citep{Radhakrishnan}
\begin{equation} \label{eq 3.1} 
\tan(\psi-\psi_{0})=\frac{\sin\alpha\sin(\phi-\phi_{0})}{\sin\zeta\cos\alpha-\cos\zeta\sin\alpha\cos(\phi-\phi_{0})}.
\end{equation}
The parameters $\phi_{0}$ and $\psi_{0}$ are used to define a fiducial plane. The derivation of Equation (\ref{eq 3.1}) was done in my Masters mini-thesis \citep{DuPlessisThesis}.

The RVM makes the following assumptions: a zero-emission height, all the emission is tangent to the local $B$-field, the pulsar's co-rotational speed at the emission altitude is non-relativistic, the emission beams are circular, the $B$-field is well approximated by a static vacuum dipole $B$-field, and the plane of polarisation is perpendicular to the local $B$-field. It is important to note that the RVM is a geometric model, meaning that one can not produce spectra or light curves using the RVM alone.
  
Figure \ref{fig:3.4} shows that at the edges of the emission beam, the PPA changes slowly with $\phi$, but as it approaches the centre, ($\phi=\phi_{0}$), it increases more rapidly, forming the well-known canonical S-shape. The steepest gradient of Equation (\ref{eq 3.1}) is given by

\begin{equation} \label{eq 3.2} 
\left(\frac{d\psi}{d\phi}\right)_{\max} = \frac{\sin(\alpha)}{\sin(\beta)}, 
\end{equation}
which is useful to help determine the continuity of the model at specific parameter choices. An example of this is where $\alpha=\zeta$ meaning $\beta=0$, leading to Equation (\ref{eq 3.2}) blowing up due to the division by zero.  
%%%%%%%%Paper 1%%%%%%%%%%
\chapter{Introduction to AR Sco and Geometric Modelling} \label{sec:Paper1}
\section{Paper Context}
This work was published as a follow-up on the \citet{DuPlessis2019} paper published in my Master's thesis. The intent was to use our geometrical RVM from the previous work and fit it to the optical orbital phase-resolved polarisation position data from \citet{Potter2018}. This would allow us to constrain $\alpha$ and $\zeta$ over the orbit of the system, assessing if there are any changes in these parameters vs orbital phase. The paper also served to summarise the existing models proposed for AR Sco (up to the date of publication) highlighting some of their assumptions. 

\section{Author Contribution}
As the main author, I developed the code to use the Markov Chain Monte Carlo (MCMC) technique to fit the observational data with the help of C. Venter and Z. Wadiasingh. I developed the code to do the data analysis from the raw data supplied by D.A.H Buckley and S.B. Potter, reproduce their results from \citep{Potter2018} as a check, and extract the processed data into data files required for the fitting. I generated all the figures and produced the extra Lomb-Scargle analysis figures. Thus I developed and implemented the main pipeline. As the main author, I wrote the initial article draft with input from the co-authors. All the co-authors supplied valuable comments and interpretations from their respective speciality points of view.   

\includepdf[pages=-,scale=0.9,offset=27.0mm -23.0mm,noautoscale]{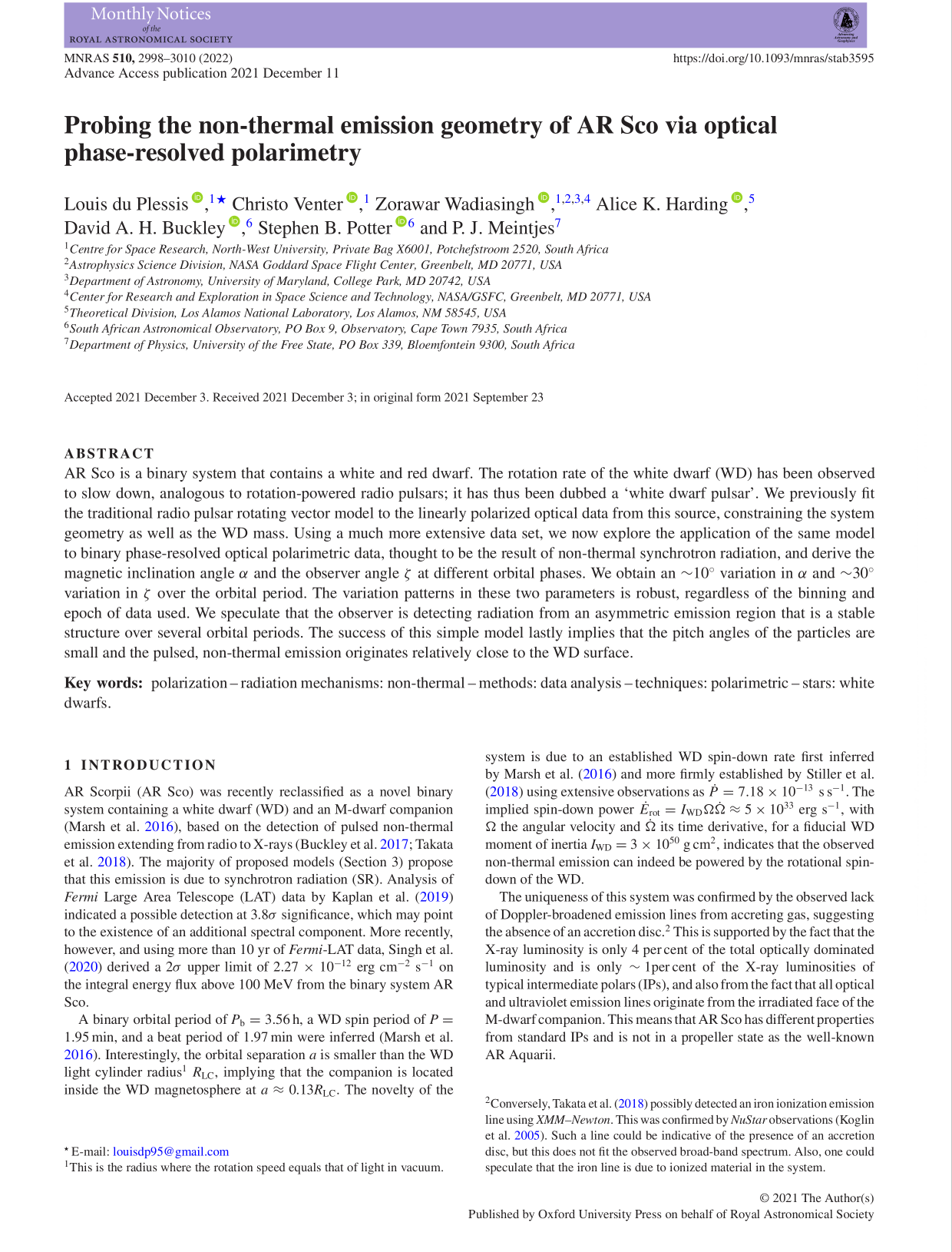}

\section{Observational Updates}
The detection of the AR Sco sibling by \citet{Pelisoli2023} as the second WD `pulsar' finally establishes these sources as a new class. Interestingly the new source seems to only have one peak in the light curves as well as showing possible indications of flaring in the system contrary to AR Sco. There is a somewhat new proposed model for AR Sco and this source by \citet{Pelisoli2024} which is discussed in Chapter~\ref{sec:Paper2}. The second major observational update on AR Sco is by \citet{Garnavich2023} showing the precession of the WD visible in the observations first proposed by \citet{Katz2017}. They estimate the precession to be between $40 - 100$ years. This thus adds more complexity to account for the precession of $\alpha$ in the modelling as well. 
%%%%%%%%%%%%%%%%%%%%%%%%%
%%%%%%%%Paper 2%%%%%%%%%%
\chapter{Towards Modelling AR Sco: Particle Dynamics} \label{sec:Paper2}
\section{Paper Context}
This chapter contains the published work \citet{DuPlessis2024} that serves as the methods paper for our emission model/code. The purpose of the paper firstly serves to show how I solved the equations of motion, implemented the RRF, implemented the adaptive time step methods, calculated our electromagnetic fields, and implemented our particle setup. Secondly, the paper shows how I implemented each numerical integrator, and assessed the accuracy, stability, and computational cost of each numerical integrator to identify the best one for our use case. Lastly, the paper indicates how our results converge to the radiation-reaction limit solutions for uniform $E$ and $B$-fields as well as illustrates the advantages of using our numerical approach over that of existing pulsar PIC codes.  
      
\section{Author Contribution}
As the main author I developed the code, implemented each numerical scheme, implemented the adaptive time step methods, implemented each test scenario, preformed the data analysis, plotted the figures and wrote the draft and revised versions of the article. All of the collaborators gave comments on the article and helped with insight to develop the code and interpret the results. The collaborators gave insight from their specialisation fields namely: C. Venter and A.K. Harding relating to their pulsar emission modelling, Z. Wadiasingh relating to his magnetar modelling, C.~Kalapotharakos relating to his own pulsar PIC modelling and P. Els relating to his kinetic modelling of particle scattering by turbulent B-fields in the Heliosphere.    

\includepdf[pages=-,scale=0.9,offset=27.0mm -23.0mm,noautoscale]{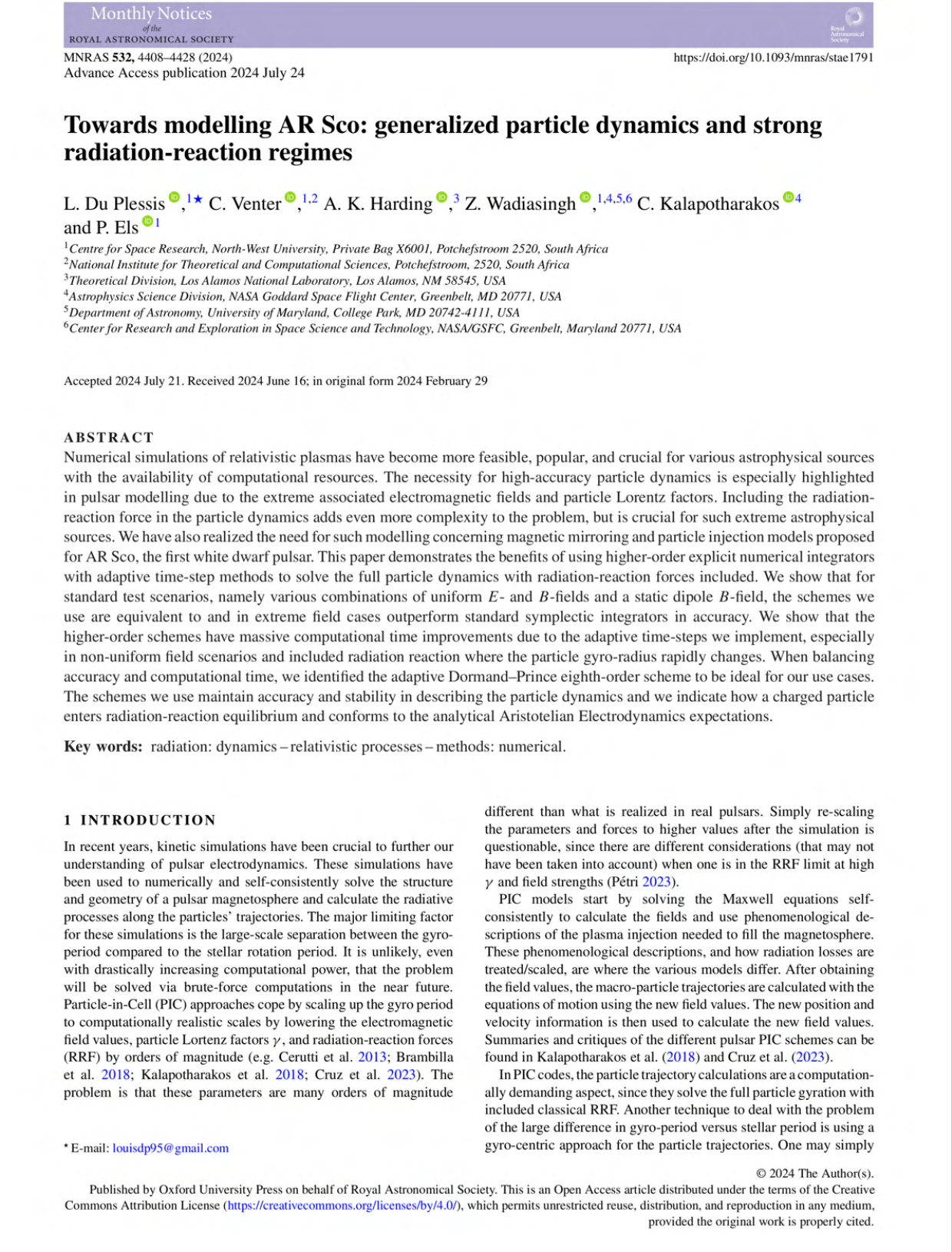}
%%%%%%%%%%%%%%%%%%%%%%%%%
%%%%%%%%Paper 3%%%%%%%%%%
\chapter{Towards Modelling AR Sco: Calibration} \label{sec:Paper3}
\section{Paper Context}
The work in this chapter is in the process of being submitted as the second paper in a series with the article in Chapter \ref{sec:Paper2} serving as the methods paper. The article is thus the draft to be submitted shortly after the submission of the thesis. In this work, I reproduce the emission maps from the \citetalias{Harding2015, Harding2021, Barnard2022} models and show how our results converge to the AE solution for an FF field. I also identify an applicable SCR method for our modelling use case. This work also serves as a comparison between a gyro-phase-resolved model and a gyro-centric model discussing the shortcomings and benefits of each. Further, the work is an analysis about the applicability of the radiation calculations in the high $E_{\perp}$-field scenarios of pulsars to be used since the standard SR and SCR equations are derived in the absence of $E$-fields. All of the abovementioned serves as a calibration for our model to have confidence in our simulated emission maps and spectra that we produce for AR Sco in Chapter \ref{sec:ARSCO} and for future modelling of similar sources as well as pulsars. 
    
\section{Author Contribution}
As main author, I implemented all of the required emission map, spectra, radiation calculations, and AE calculations. I extracted all the required parameter information from the \citetalias{Harding2015, Barnard2022} codes, implemented the comparison calculations, and performed all the data processing and visualisation. A.K Harding served as the expert on the \citetalias{Harding2015, Harding2021} code and results, C. Kalapotharakos served as the expert on the AE results and radiation calculations following the AE trajectory, C. Venter gave input with his experience working with the \citetalias{Harding2015, Harding2021, Barnard2022} codes and his pulsar modelling background, and Z. Wadiasingh gave input with his magnetar and FRB modelling background. I wrote the draft of the article, generated the plots, and implemented the comments. All of the collaborators gave comments and input into the article.    

\includepdf[pages=-,scale=0.9,offset=27.0mm -23.0mm,noautoscale]{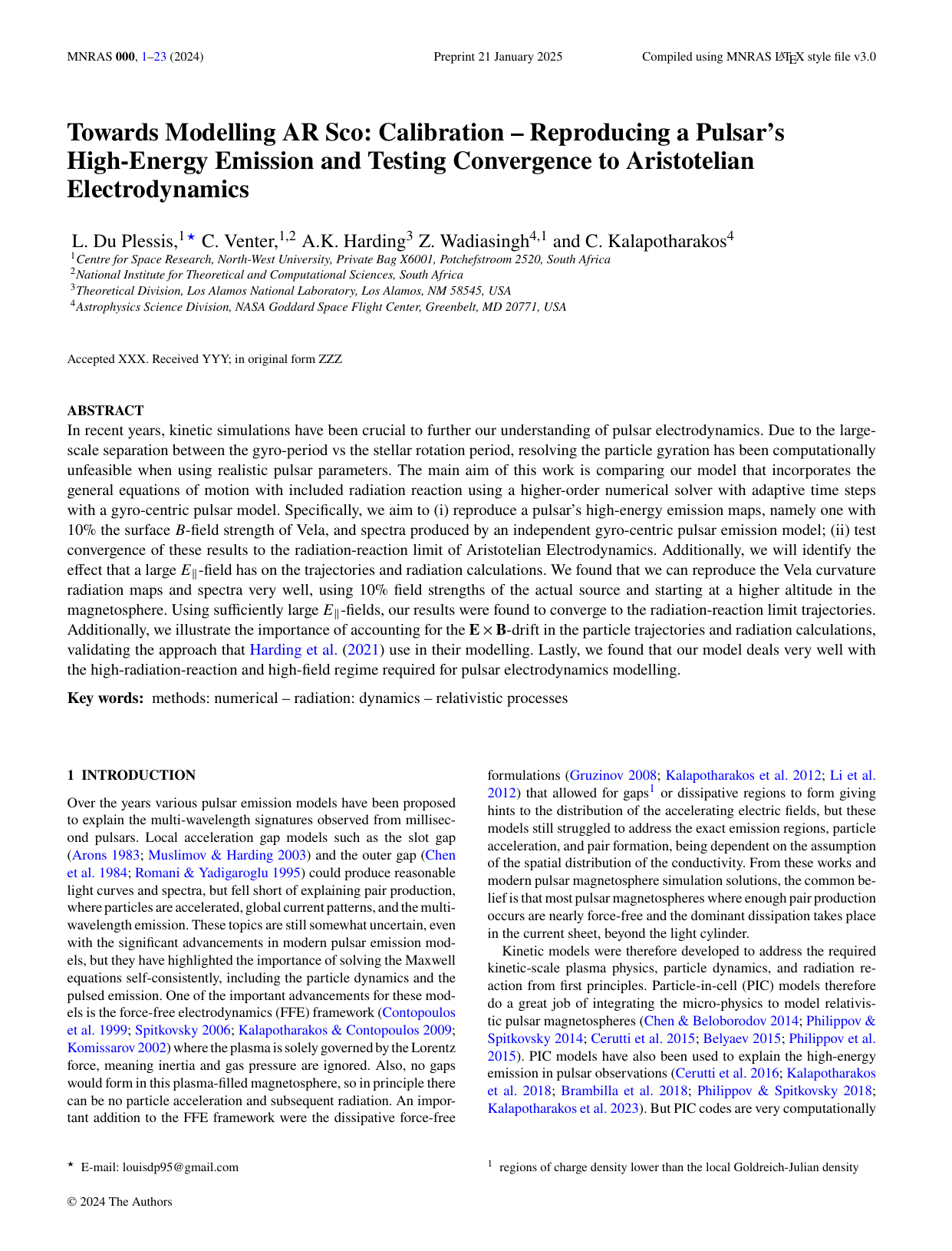}
%%%%%%%%%%%%%%%%%%%%%%%%%
%%%%%%%%AR Sco%%%%%%%%%%
\chapter{AR Sco Magnetic Mirror Results} \label{sec:ARSCO}
The source AR Sco has been introduced and discussed in detail along with summaries of the various models proposed for the source in Chapter~\ref{sec:Paper1}. Here I will specifically focus on the magnetic mirror model from \citetalias{Takata2017} we want to replicate and compare our results with. This model was discussed in Chapters \ref{sec:Paper1} and \ref{sec:Paper2} where in the latter I reproduced and compared the particle $\gamma$ value evolution of their model. Here I will mainly focus on how the model was set up and the considerations I had to make in our modelling due to many vague elements in the  \citetalias{Takata2017, Takata2019} model implementations. To briefly summarise the model: particles are injected from the companion star into the magnetosphere of the rotating WD where the particles travel along the field lines towards the WD surface, radiating energy along their trajectory. As the particles travel along the field lines they eventually encounter a magnetic mirror close to the magnetic pole of the WD where they radiate most of their energy due to the high $B$-fields close to the WD surface. The particle then turns around due to the magnetic mirror and follows the field line towards the opposite magnetic pole where it will be mirrored again, becoming trapped in the magnetosphere until it is reabsorbed by the companion or the WD itself as the particle escapes the magnetic confinement. This chapter is an exploratory study of the proposed magnetic mirror model to first constrain the $B$-field, particle index, particle energies, and pitch angles required to fit the observational SED. Due to the computationally heavy nature of these simulations, I will thus start by modelling one field line in this exploratory work.      

\section{Magnetic Mirror System Parameters} \label{sec:ARSCO_s1}
To set up the magnetic mirror scenario from \citetalias{Takata2017} I start by defining the system parameters, namely the radius of the WD as $R_{\rm WD} = 0.8R_{\odot}$, a binary separation of $a=8\times10^{10} \, \rm{cm}$, a WD spin period of $P_{\rm s} = 117$~s, and setting the orbital phase at $0.25$ where orbital phase $0.0$ would be at inferior conjunction as used by \citetalias{Takata2017}. I use orbital phase $0.25$ since the observation of \citet{Potter2018b} found the linear polarisation to be highest between orbital phase $0.2$ and $0.5$. I use the distance to the source of $d=117.8 \rm{pc}$ constrained by \citet{Stiller2018}, since I could not find a distance specified in \citetalias{Takata2017}. However, they found that they use $d=110 \rm{pc}$ in their \citetalias{Takata2019} model. The first ambiguity in their modelling is which WD magnetic moment $\mu_{\rm WD}$ they used, since all of their parameters are calculated using $\mu_{\rm WD}=10^{35} \, \rm{G}\, \rm{cm}^{3}$ $(B_{\rm{s}} \sim 6\times 10^{8} \, \rm{G})$, but for the spectra and $\gamma$ evolution, they used $\mu_{\rm WD}= 6.5\times 10^{34} \, \rm{G}\, \rm{cm}^{3}$ $(B_{\rm{s}} \sim 4\times 10^{8} \, \rm{G})$. I thus investigated both cases but the latter seems to be more in line with their results and I assume all their calculated values are scaled to the correct $\mu_{\rm WD}$ instead of the higher $\mu_{\rm WD}$ used in the article calculations. The next uncertainty is how they calculate their particle trajectories since they use Equations~(21) in Chapter \ref{sec:Paper2} to model their particle $\gamma$ and $p_{\perp}$ but these equations can not be used to calculate the particle trajectory since the particle position can not be recovered from these equations. Thus our best assumption is that they simply use some particle pusher to follow a $B$-field line and use this as the particle trajectory, since they also state they neglect any co-rotation or $\mathbf{E}\times \mathbf{B}$-effects. The second problem with Equations~(21) are that they can not be used to determine if a particle encounters a magnetic mirror, since the Equations are decoupled from the particle dynamics and have no $B$-field or $E$-field drift effects \footnote{Specifically, $\nabla \mathbf{B}$ and $\mathbf{E}\times\mathbf{B}$ drift.} included, as discussed in Chapter \ref{sec:Paper3}. Thus in \citetalias{Takata2017, Takata2019} the particles have to be manually turned around which is not specified that this is done or how this is done in their papers. I do not know if they use their theoretical mirror point radius that they derive by neglecting SR to turn the particle around. The other problem with Equations~(21) are the assumptions of small particle $\theta_{p}$ and no inclusion of an $\mathbf{E}\times\mathbf{B}$ effect, making them non-ideal to model the particle $\theta_{p}$\footnote{Modelling the angle between the particle velocity and local $B$-field instead of $\theta_{\rm VA}$.} as discussed in Chapter \ref{sec:Paper3}, as well as to model a magnetic mirroring effect\footnote{This requires the $\nabla B$ drift even without an $E$-field included} where one expects $\theta_{p}$ to range from $0^{\circ}$ to $180^{\circ}$\footnote{One finds motion in the same direction as the $B$-field $\theta_{p}$ being $[0^{\circ}; 90^{\circ})$, motion in the opposite direction of the $B$-field $\theta_{p}$ being $(90^{\circ}; 180^{\circ}]$, and $90^{\circ}$ the mirror point. This is discussed in Chapter~\ref{sec:Paper2}}. Additionally, they adopt the conclusion from \citet{Geng2016} that the $E_{\parallel}$-field is screened, thus I included this in our modelling as well. I decided to model two magnetosphere cases, namely using only the RD $B$-field, as is done by \citetalias{Takata2017, Takata2019}, and including the $E_{\perp}$-field for the RVD field from Equation~(7) in Chapter \ref{sec:Paper2}, since there is no justification to exclude the $E_{\perp}$-field\footnote{Only $E_{\parallel}$ is screened by the plasma buildup, not $E_{\perp}$; see \citet{Chen1984}.}. From their \citetalias{Takata2017} paper they identified a WD magnetic inclination angle of $\alpha=60^{\circ}$ thus I will use the same value but investigate $\alpha = 80^{\circ}$ as well since the observations discussed in Chapter \ref{sec:Paper1} point to the WD being an orthogonal rotator. I also adopted their observer angle of $\zeta=60^{\circ}$ to try to reproduce their spectra but will assess other observer angles as well. 

\section{Particle Setup and Spectral Normalisation} \label{sec:ARSCO_s2}
Due to there being no indication of how \citetalias{Takata2017, Takata2019} set up their particles in their $B$-field with a specified $\theta_{p}$ I use our method discussed in Chapter \ref{sec:Paper2} and start the particle at the companion midpoint as is the inferred starting point from their results. There is also no specification which $B$-field line they followed, how many they followed or any suggestions about the injection region. From their results, I infer they start at the companion midpoint and simply follow a single $B$-field line. Due to how computationally heavy our model is this was also the easiest particle setup to simulate but because I am not sampling multiple field lines this causes low-statistic results in our emission maps. \citetalias{Takata2017} estimate the typical particle Lorentz factor as $\gamma_{0} = 50$ via magnetic dissipation of the WD $B$-field heating and accelerating the particles, which they use as $\gamma_{\rm{min}} = \gamma_{0}$. From \citet{Geng2016}, $\gamma_{\rm{max}}= 3.4\times10^{6}$ is estimated by balancing the acceleration rate and the SR cooling rate which is also adopted by \citetalias{Takata2017}. To produce the SR spectrum for AR Sco they use a power law particle energy distribution given by
\begin{equation} \label{gam_dist}
f(\gamma) = K_{0}\gamma^{-p} \,\, ; \,\, \gamma_{\rm{min}} \leqslant \gamma \leqslant \gamma_{\rm{max}},
\end{equation}       
where $K_{0}$ is the normalisation constant and $p$ the power-law index. To calculate the normalisation constant for their spectrum they use the condition 
\begin{equation} \label{LB_norm}
\int^{\gamma_{\rm{max}}}_{\gamma_{\rm{min}}} \gamma mc^{2}f(\gamma)d\gamma = L_{\rm{B}}.
\end{equation} 
Here $L_{\rm{B}}$ is the power from the magnetic dissipation obtained from \citet{Buckley2017} and estimated by \citetalias{Takata2017} as
\begin{equation}
L_{\rm{B}} \sim 2.8\times 10^{32} \, \rm{erg} \, \rm{s}^{-1} \left(\frac{\mu_{\rm{WD}}}{10^{35} \, \rm{G}\, \rm{cm}^{3}}\right)^{2}.
\end{equation}

To obtain an additional condition they assume that due to charge conservation, the rate of electrons leaving the companion surface $(\dot{N}_{e})$ is equal or approximate to the rate of particles leaving the companion surface due to the magnetic dissipation $(\dot{N}_{p})$. This is estimated to be 
\begin{equation}
\dot{N}_{p} \sim 5\times 10^{35} \, \rm{s}^{-1} \left( \frac{L_{\rm{B}}}{10^{32} \, \rm{erg} \, \rm{s}^{-1}}\right).
\end{equation} 
They then use the condition 
\begin{equation} \label{Ne_norm}
\int^{\gamma_{\rm{max}}}_{\gamma_{\rm{min}}} f(\gamma)d\gamma = \dot{N}_{e}
\end{equation} 
to calculate the spectral normalisation. A problem with this approach is that Equation~(\ref{Ne_norm}) is not independent from Equation~(\ref{LB_norm}) since they use $L_{\rm{B}}$ to calculate $\dot{N}_{e}$. Further, to produce the SR spectrum one has to either assume a constant $\theta_{p}$ or for a more physically accurate model a $\theta_{p}$ distribution. In \citetalias{Takata2017} a uniform $\theta_{p}$ is assumed and a more physically accurate distribution is investigated in \citetalias{Takata2019}. For this exploratory work we are more interested in reproducing the \citetalias{Takata2017} spectrum and want to reproduce the \citetalias{Takata2019} model in a future paper, which will include these results as well. To calculate the normalisation constant for the uniform $\theta_{p}$ distribution, \citetalias{Takata2017} uses
\begin{equation} \label{Takata2017_pitch_dist}
\frac{d\dot{N}_{e}}{d\theta_{p}} = K_{1} \,\, ; \,\, 0\leqslant \theta_{p} \leqslant \pi/2,
\end{equation}  
where $K1$ is the normalisation constant. It is not clear if \citetalias{Takata2017} separately calculated the two normalisation constants, since one has to include both the particle energy distribution and the $\theta_{p}$ distribution into either Equation~(\ref{LB_norm}) or (\ref{Ne_norm}) and solve for the normalisation constants. This is shown in \citet{Yang2018} normalising the spectrum for both a particle energy and $\theta_{p}$ distribution with 
\begin{equation} \label{dist_Yang}
N_{e}(\theta_{p},\gamma)d\theta_{p}d\gamma = g(\theta_{p})f(\gamma)d\theta_{p}d\gamma,
\end{equation}
where $g$ and $f$ are the two distribution functions. They first normalise the $\theta_{p}$ distribution function and then solve for $K_{0}$ using Equation~(\ref{dist_Yang}) with $f$ and $g$ included, not separately solving $f$ or $g$ as is the case for Equation~(\ref{Ne_norm}) and (\ref{Takata2017_pitch_dist}). This problem is not elucidated in \citetalias{Takata2019} using other $\theta_{p}$ distributions since it would appear as though they are separately solving for the normalisation constants. I thus take the approach of \citet{Yang2018} and normalise the $\theta_{p}$ distribution then include the $\theta_{p}$ distribution function into Equation~(\ref{Ne_norm}) and solve for $K_{0}$\footnote{I scale $L_{\rm B}$ and $\dot{N}_{p}$ according to the $B$-field I use.}. I investigated using $p=2.5$ from \citetalias{Takata2017} but this caused our spectrum to be much too hard to fit the SED thus I show $p=3.0$ from \citetalias{Takata2019} in these results. The \citetalias{Takata2017} seem to model the thermal X-rays, not the non-thermal X-rays as is done in \citetalias{Takata2019}.   

For reproducibility purposes I use the same $\gamma_{\rm{min}}$ and $\gamma_{\rm{max}}$ as above, using logarithmic steps\footnote{For numerical integration using logarithmic steps, see \citet{Venter2010}.} and sampling five times per decade and using a $d\theta_{p} = 5.0^{\circ}$ ranging from $0^{\circ} - 90^{\circ}$. In the results, I refer to $0^{\circ} - 90^{\circ}$ but the particle is moving towards the WD, this is to avoid confusion having to refer to $180^{\circ} - 90^{\circ}$ to have the same angle with the $B$-field but have the correct direction. Due to having both a $\gamma$ and $\theta_{p}$ distribution, I had $475$ parameter combinations per field line which is very computationally demanding. The code is parallelised using Open MPI and was run on the FSK-GAMMA-1 computer cluster facility at the Centre for Space Research (North-West University) using between 30 and 100 CPUs depending on availability. Initially, it was run on a standard desktop with 6 CPUs. Luckily only the particles with low $\gamma$ values and small $\theta_{p}$ had very long runtimes. This is due to the particles with larger $\gamma$ values having a larger gyro-radius, meaning our adaptive methods can take larger time steps as discussed in Chapter \ref{sec:Paper2}. The particles with large $\theta_{p}$ values turn around quite quickly due to their magnetic mirror occurring much higher in the magnetosphere. To limit the computational time I allow the particle to mirror and stop the simulation once it is back at the companion, since Figure 10 in Chapter \ref{sec:Paper2} shows the particle loses most of its energy at the first pole. The figure does show the $\mathbf{E}\times \mathbf{B}$ case does not lose as much energy and mirrors further down in the magnetosphere. However, I found using larger $\alpha$ values than $\alpha=0^{\circ}$, which is used in Figure~10 in Chapter~\ref{sec:Paper2}, that the particles go much deeper into the magnetosphere, losing more energy before mirroring. This is due to the fact that the $\nabla \mathbf{B}$-drift scales as $1/B^{3}$, thus if the particle has to travel further down in the magnetosphere before reaching the pole it is mirrored earlier due to the extra contributions of the $\nabla B$-drift in the lower field region of the magnetosphere. Another condition I set is if $\gamma\leqslant 5$ I stop the particle. This is to avoid particles that do not mirror before they reach the stellar surface, at this point the particles have lost the vast majority of their energy and have reached the point where a plasma build-up is screening the $E_{\parallel}$ from the WD as proposed by \citet{Geng2016, Takata2017}. I found the point where the $\gamma$ was too low to be between $0.01R_{\rm{LC}} - 0.025R_{\rm{LC}}$ in our simulations. Therefore, I believe these particles will negligibly further contribute to the emission maps and spectra. Similar to \citetalias{Takata2017, Takata2019} I start the pulsar spin phase with the open field lines of the northern hemisphere pointing towards the companion. Importantly, I use our SCR calculations discussed in Chapter~\ref{sec:Paper3} to account for the $E_{\perp}$-field in this modelling but this scenario falls within the SR regime and the bulk particle $\gamma$ values are too low to produce significant CR.
For easy reference in the results the $\mathbf{E\times\mathbf{B}}$-drift velocity is given by 
\begin{equation} \label{v_EB}
\mathbf{v}_{\rm EB} = \frac{c\mathbf{E\times\mathbf{B}}}{B^{2}}.
\end{equation}       
The $\nabla \mathbf{B}$-drift is given by
\begin{equation} \label{v_grad}
\mathbf{v}_{grad} = \frac{r_{\rm L}v_{\perp}}{2}\frac{\mathbf{B}\times\nabla\mathbf{B}}{B^{2}},
\end{equation} 
where $r_{\rm L} = \gamma mc v_{\perp}/eB$ is the Larmor radius. Both these equations are from \citet{Chen1984} but written in cgs units.
 
\section{Results} 
Here I show the spectra and emission map results reproducing the magnetic mirror model proposed by \citetalias{Takata2017} for their best-fit case, as well as some alternate cases namely using different $B$-fields, $\alpha$ values, and excluding the $E_{\perp}$-field. In the SED plots, I have included the multi-wavelength observations from \citet{Marsh2016} with the blue triangles serving as upper limits. I have also included the \textit{XMM}-Newton thermal and non-thermal spectrum from \citet{Takata2018}, which is fitted in \citetalias{Takata2019}. Additionally, I include the \textit{NICER} pulsed data points for the SED from \citet{Takata2021} labelled as `NICER2020'. An important consideration for the modelling is that the observational SED is phase-averaged over the orbit and I model one snapshot of the orbit.   
  
\subsection{SEDs} \label{sec:ARSCO_s3.1}
\begin{figure}[!h]
\centering
\includegraphics[width=\textwidth]{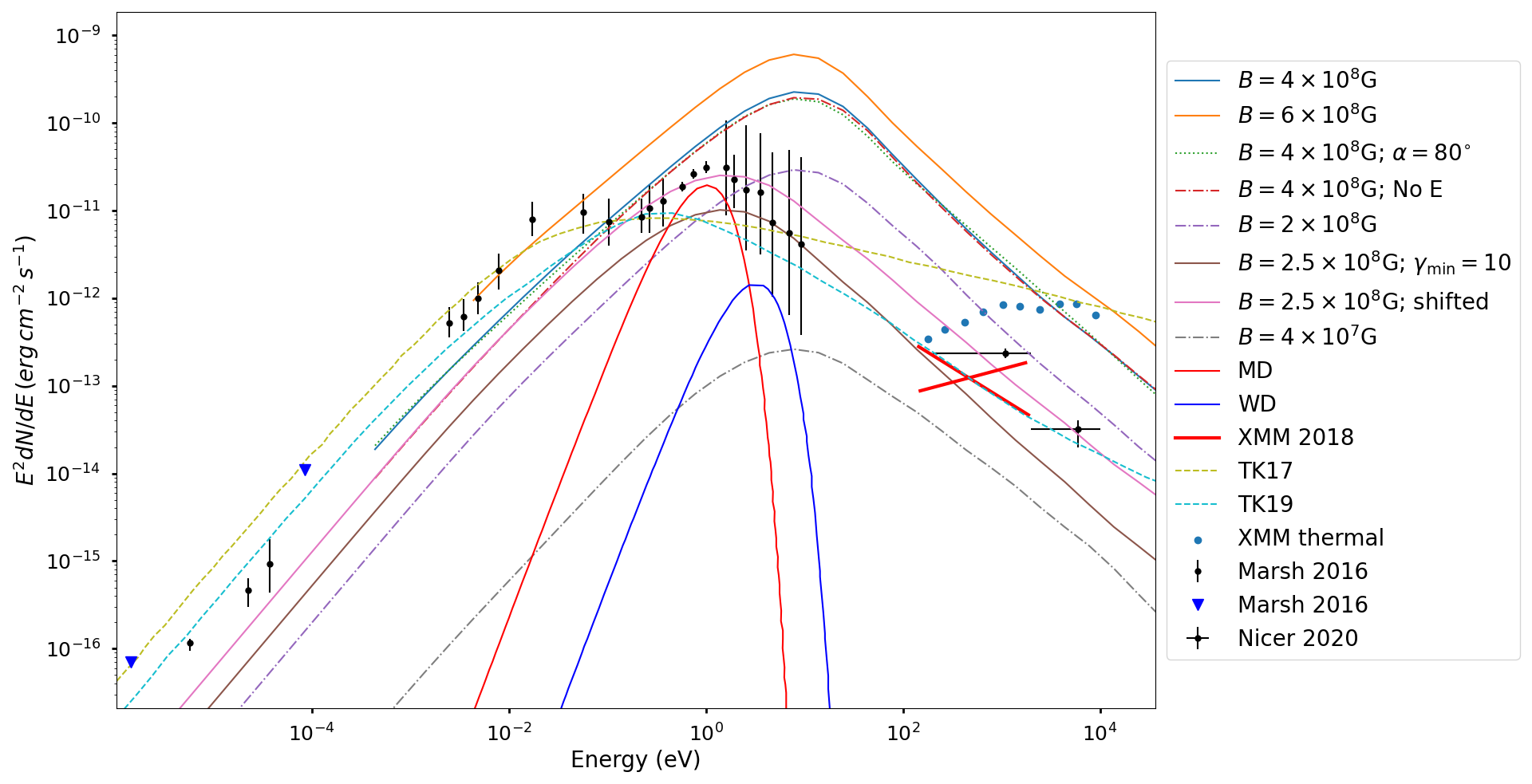} 
\caption[AR Sco Spectrum $p=3.0$]{AR Sco magnetic mirror model SED using a power-law index of $p = 3.0$, $\zeta = 60^{\circ}$, and plotted with the multi-wavelength observational data. The spectra include are, the $B_{\rm S}= 4.0\times 10^{8} \, \rm{G}$ and $\alpha=60^{\circ}$ case (\citetalias{Takata2019} ``best-fit parameters") plotted in pale blue, the $B_{\rm S}= 6.0\times 10^{8} \, \rm{G}$ and $\alpha=60^{\circ}$ case plotted in orange, the $B_{\rm S}= 4.0\times 10^{8} \, \rm{G}$ and $\alpha=80^{\circ}$ case plotted in green, the $B_{\rm S}= 4.0\times 10^{8} \, \rm{G}$ and $\alpha=60^{\circ}$ case with no included $E_{\perp}$ plotted in magenta, the $B_{\rm S}= 2.0\times 10^{8} \, \rm{G}$ and $\alpha=60^{\circ}$ case in purple, the $B_{\rm S}= 2.5\times 10^{8} \, \rm{G}$ and $\alpha=60^{\circ}$ case using $\gamma_{\rm min}=10$ plotted in brown, and the $B_{\rm S}= 4.0\times 10^{7} \, \rm{G}$ and $\alpha=60^{\circ}$ case plotted in grey. The observation data includes those of \citet{Marsh2016} labelled as `Marsh2018' with 2 upper limits given by the two blue triangles, the \textit{XMM}-Newton thermal X-ray component from \citet{Takata2018} given by the blue dots, the XMM-Newton pulsed component X-ray butterfly from \citet{Takata2018} given by the red cross, and the \textit{NICER} pulsed component X-ray data points from \citet{Takata2021} labelled as `NICER2020'. Additionally, included are the blackbody spectra for the WD in blue and the M-dwarf companion in red as well as the best-fit model from \citetalias{Takata2017} in yellow and the best-fit model from \citetalias{Takata2019} in cyan.}
\label{Spec_p3}
\end{figure}

For Figure~\ref{Spec_p3} I show the spectra produced from the different magnetic mirror cases using a power-law index of $p=3.0$ and $\zeta = 60^{\circ}$ as is used by \citetalias{Takata2019} to fit the non-thermal X-rays. I have plotted our spectrum results generated using $B_{\rm S}= 4.0\times 10^{8} \, \rm{G}$ and $\alpha=60^{\circ}$ (\citetalias{Takata2019} ``best-fit parameters") plotted in pale blue, using $B_{\rm S}= 6.0\times 10^{8} \, \rm{G}$ and $\alpha=60^{\circ}$ plotted in orange, using $B_{\rm S}= 4.0\times 10^{8} \, \rm{G}$ and $\alpha=80^{\circ}$ plotted in green, using $B_{\rm S}= 4.0\times 10^{8} \, \rm{G}$ and $\alpha=60^{\circ}$ with no included $E_{\perp}$ plotted in magenta, using $B_{\rm S}= 2.0\times 10^{8} \, \rm{G}$ and $\alpha=60^{\circ}$ plotted in purple, using $B_{\rm S}= 2.5\times 10^{8} \, \rm{G}$ and $\alpha=60^{\circ}$ with $\gamma_{\rm min}=10$ plotted in brown, and using $B_{\rm S}= 4.0\times 10^{7} \, \rm{G}$ and $\alpha=60^{\circ}$ plotted in grey. I have included the blackbody spectra for the WD and M-dwarf plotted in blue and red respectively. Additionally, I have added the spectrum using the same parameters as the brown curve but using $p = 2.9$ and artificially moving the spectrum up to see if the spectrum can fit the \textit{NICER} pulsed X-ray spectrum data. This emulates using a higher $B$-field as explained below. I use the blackbody spectra from \citetalias{Takata2019} to reduce any differences in results but I discuss how we previously produced these spectra in \citet{DuPlessis2019}. The \citetalias{Takata2017} best-fit spectrum is plotted in yellow (using $p=2.5$) and the \citetalias{Takata2017} best-fit spectrum is plotted in cyan.

From Figure~\ref{Spec_p3} one sees the $B_{\rm S}= 6.0\times 10^{8} \, \rm{G}$ and $B_{\rm S}= 4.0\times 10^{8} \, \rm{G}$ cases have far too high spectral fluxes to fit the observational SED peak at $\sim 1 \rm{eV}$ and the X-ray SED. The $B_{\rm S}= 2.0\times 10^{8} \, \rm{G}$ case plotted in purple has a spectral peak at a similar flux to the observations where the $B_{\rm S}= 4.0\times 10^{7} \, \rm{G}$ case is found to have a much too low flux. Interestingly all of the mentioned cases have a spectral peak at higher energies than the observed spectral peak in the SED. Furthermore, these spectral peaks do not move down and left when reducing the $B$-field as is expected for SR from Equation~(14) and the critical SR photon energy in Chapter~\ref{sec:Paper3}. The reason this does not occur is that in the higher $B$-field cases the particles radiate the bulk of their emission at higher altitudes in the magnetosphere vs. the lower $B$-field cases the particles radiate the bulk of their emission closer to the magnetic mirror. Thus at what radius the particles mirrors is a considerably more complicated answer depending on the particle $\gamma$, $\theta_{p}$, $B$-field, and $E$-field\footnote{$E_{\perp}$-field in this case.} since the RRF losses are higher for higher field and $\gamma$ values. This means the particles initialised in the higher-field cases will radiate the bulk of their emission higher up in the magnetosphere vs the low-field cases, since the particles in the latter case have to travel deeper into the magnetosphere before encountering higher field strengths. 

Noticeably, the derivation for $\gamma_{\rm min} = 50$ in \citetalias{Takata2017, Takata2019} is dependent on the WD surface $B$-field strength where they used $B_{\rm S}= 6.0\times 10^{8} \, \rm{G}$ to obtain this value. Thus, reasonably assuming that $\gamma_{\rm min}$ would be lower for lower WD surface $B$-field values, I investigate using $\gamma_{\rm min} = 10$. I use the same amount of divisions per decade as before simply lowering $\gamma_{\rm min}$ and taking this into account for the spectral normalisation. This yields the brown curve in Figure~\ref{Spec_p3} where one sees the spectral peak aligns well with the optical spectral peak of the observations at $1 \, \rm{eV}$. Using $B_{\rm S}= 2.5\times 10^{8} \, \rm{G}$ the overall flux is a bit low thus the best fit $B$-field would most likely be between $B_{\rm S}= 2.5\times 10^{8} \, \rm{G}$ and $B_{\rm S}= 3.0\times 10^{8} \, \rm{G}$ for our modelling of this scenario. If I model the spectrum for the $B_{\rm S}= 2.5\times 10^{8} \, \rm{G}$, use $p = 2.9$, and artificially move up the spectrum thereby mimicking a slightly higher $B$-field without rerunning the model, one sees this pink spectrum fits the observations quite well. It fits the optical peak and \textit{NICER} pulsed X-ray data very well, converging with the optical/UV observed photon spectral index varying between $0.8-1.4$ from \citet{Garnavich2019} (spectral index $= (p -1)/2 = 0.95$). Interestingly the photon spectral index for the pulsed X-rays constrained by \citet{Takata2021} was constrained to be $2.8^{+0.28}_{-0.25}$ . It is evident that the low-energy component of the model SED falls below the observed spectrum as well as there being a spectral bump at $0.05 \, \rm{eV}$ that is unaccounted for by our model spectrum. With the \textit{NICER} pulsed data indicating a similar slope to the later segment of the optical peak at $1 \, \rm{eV}$ it would be impossible to fit these X-ray data with a single particle population if the SR peak was at the $0.05 \, \rm{eV}$ spectral bump. The lower-energy spectrum can possibly be due to a different particle population producing this observed flux and spectral bump. A better fit can be found by varying $p$ slightly, varying $\zeta$, and using a slightly higher $B$-field but one would rather have a high-statistic run and obtain similar emission maps to the observations before trying to constrain the parameters too precisely. Interestingly, one sees all the $B_{\rm S}= 4\times 10^{8} \, \rm{G}$ cases mostly overlap in Figure~\ref{Spec_p3} namely the pale blue, green and magenta curves. Therefore, using $\alpha = 80^{\circ}$ instead of $\alpha = 60^{\circ}$ and excluding the $E_{\perp}$-field had no major impact on the spectrum but this does affect the emission maps significantly as we will see in the next subsection. Comparing our spectra to the plotted \citetalias{Takata2019}, spectrum one sees that using the same $p$ value their spectral slope is a lot harder than our own. Their spectral peak is also found to be at $\sim 0.1 \, \rm{eV}$ compared to our $\sim 1 \, \rm{eV}$. Their spectrum also fits the lower energy data of the SED quite well but they would have to make a lot of adjustments to shift the spectral peak to higher energies to fit the \textit{NICER} pulsed data and most likely lose their fit of the lower energy SED data. As mentioned in the previous section, the \citetalias{Takata2017} spectrum seems to be fitted to the thermal X-rays instead of the pulsed component. Notably, our modelling $B_{\rm S}= 4.0\times 10^{8} \, \rm{G}$ and $B_{\rm S}= 6.0\times 10^{8} \, \rm{G}$ cases' spectra start at higher photon energies due to the spectra not being calculated for the lower energies. This was done to save computational time with the higher field cases but the spectral peak is more than sufficiently covered in these calculations.

\begin{figure}[!h]
\centering
\includegraphics[width=\textwidth]{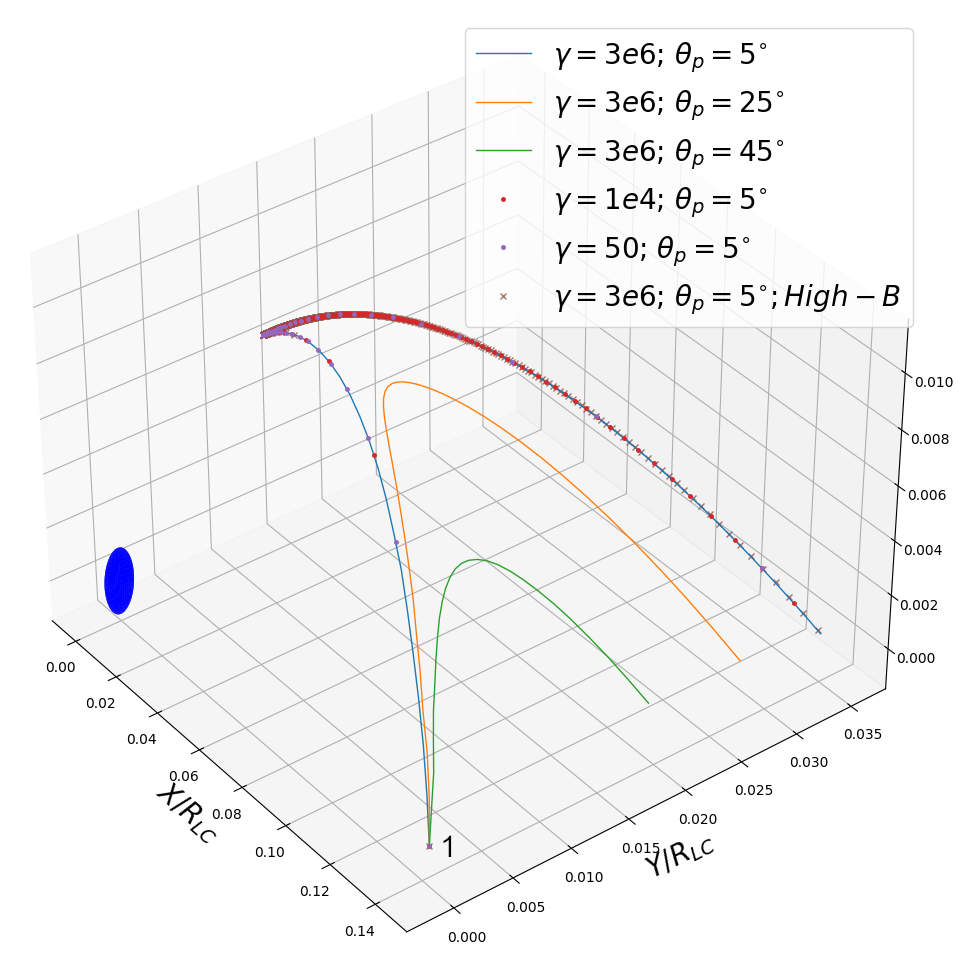} 
\caption[Mirrored Particle Trajectories]{The particle trajectories for various initial $\gamma$ and $\theta_{\rm p}$ values using $B_{\rm S}= 6.0\times 10^{7} \, \rm{G}$ and $B_{\rm S}= 3.0\times 10^{8} \, \rm{G}$ for the high $B$-field case. The initial particle parameters are as shown in the legend
%The initial particle parameters are $\gamma = 3\times 10^{6}$ and $\theta_{\rm p} = 5^{\circ}$ in blue, $\gamma = 3\times 10^{6}$ and $\theta_{\rm p} = 25^{\circ}$ in orange, $\gamma = 3\times 10^{6}$ and $\theta_{\rm p} = 45^{\circ}$ in green, $\gamma = 1\times 10^{4}$ and $\theta_{\rm p} = 5^{\circ}$ in red, $\gamma = 50$ and $\theta_{\rm p} = 5^{\circ}$ in purple, and the higher $B$-field case using $\gamma = 3\times 10^{6}$ and $\theta_{\rm p} = 5^{\circ}$ in brown. 
where the blue, red, purple and brown curves are overlapping. The axis is normalised to $R_{\rm LC}$ with the WD shown by the blue circle. The WD looks warped due to the scaling of the different axes. The `1' indicates the particles' origin.}
\label{Mirrored_traj}
\end{figure}

As a visual illustration and aid to help probe the micro-physics at play in the magnetic mirror scenario, I include a few particle parameter combinations in a trajectory plot and indicate how $\gamma$ evolves over the particle trajectory. In Figure~\ref{Mirrored_traj} I use the same magnetic mirror setup as described before, but use $B_{\rm S}= 6.0\times 10^{7} \, \rm{G}$ to reduce the computational time for each particle but mainly to reduce the file sizes of saving the particle parameters along their trajectories. This serves as an insight into their behaviour when varying $\theta_{\rm p}$, $\gamma$, and the $B$-field. For the particle labelled as `High-B' I used $B_{\rm S}= 3.0\times 10^{8} \, \rm{G}$. In Figure~\ref{Mirrored_traj} I plot the particle trajectory when using the initial parameters $\gamma = 3\times 10^{6}$ and $\theta_{\rm p} = 5^{\circ}$ in blue, $\gamma = 3\times 10^{6}$ and $\theta_{\rm p} = 25^{\circ}$ in orange, $\gamma = 3\times 10^{6}$ and $\theta_{\rm p} = 45^{\circ}$ in green, $\gamma = 1\times 10^{4}$ and $\theta_{\rm p} = 5^{\circ}$ in red dots, $\gamma = 50$ and $\theta_{\rm p} = 5^{\circ}$ in purple dots, and the higher $B$-field case using $\gamma = 3\times 10^{6}$ and $\theta_{\rm p} = 5^{\circ}$ in brown crosses. The WD is represented to scale as the blue dot in the figure but is a little warped by the axis scaling since the movement is not equal in each direction. I have also normalised $X$, $Y$, and $Z$ to $R_{\rm LC}$. The first notable trait in the figure is that the particles with the same $\gamma$ and lower $\theta_{\rm p}$ mirror closer to the WD surface due to the $\nabla B$-drift scaling as $v_{\perp}^{2}$. In Equation~(\ref{v_grad}) one sees the direction of this drift velocity is $\nabla\times\mathbf{B}$, thus perpendicular to $\mathbf{B}$ and increasing the particle $\theta_{\rm p}$. This means the larger initial $v_{\perp}$ (due to the larger $\theta_{\rm p}$) leads to a larger $\nabla \mathbf{B}$-drift, causing the particle $\theta_{p}$ to increase until the particle is mirrored at $\theta_{p}=90^{\circ}$. The next noticeable feature is that the blue, red, purple, and brown curves overlap meaning the particles with the same initial $\theta_{\rm p}$ have very similar trajectories. If one were to zoom into the mirror point one would find that their mirror points do not perfectly align but they have very similar trajectories. In all of the curves, one sees the particle's trajectories follow the open field line towards the WD, slightly curving on their inward path but having a much more arched path on their outward trajectory. This is due to their inward motion `chasing' the rotating fields vs. their outward motion following the rotating fields.

\begin{figure}[!h]
\centering
\includegraphics[width=\textwidth]{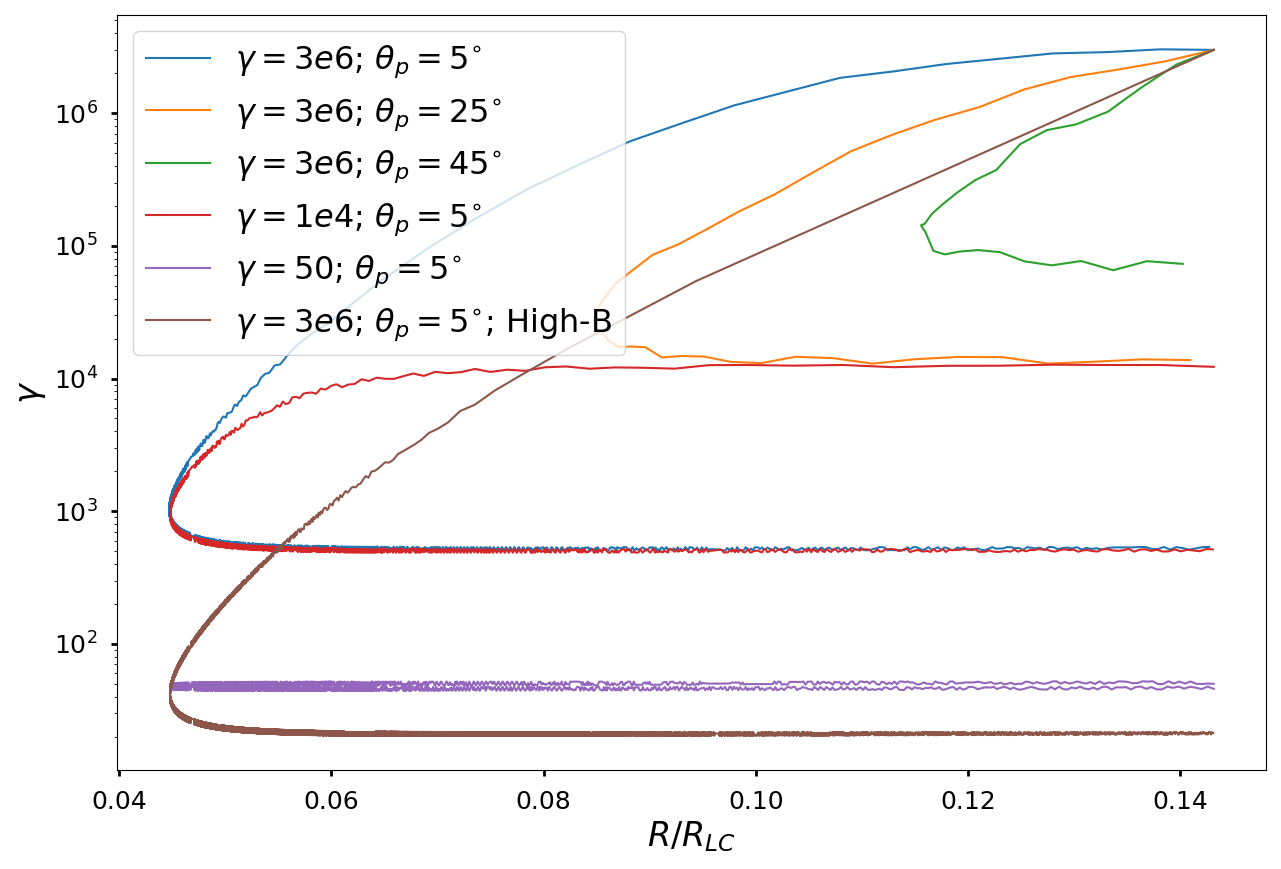} 
\caption[Mirrored Particle $\gamma$]{The same initial particle parameters used in Figure~\ref{Mirrored_traj} showing the particle $\gamma$ value vs its radial distance normalised to $R_{\rm LC}$.}
\label{Mirrored_gam}
\end{figure}

In Figure~\ref{Mirrored_gam} I probe some of the micro-physics involved with the $\gamma$, $\theta_{\rm p}$, and $B$-field dependence with the RRF. In this figure, each coloured curve corresponds to the scenario and initial parameters used for the particles in Figure~\ref{Mirrored_traj} where I have plotted these particles' $\gamma$ vs their radial position normalised to $R_{\rm LC}$. Starting by assessing the different $\theta_{\rm p}$ values with the same $\gamma$, one sees the particles with lower values experience more total losses, since they travel deeper into the magnetosphere where the $B$-field is higher. However, by looking at the slope of their initial losses inward, one sees the green curve with the highest $\theta_{\rm p}$ has the most aggressive losses but is turned around early and thus has the lowest total losses. Comparing the particles with the same $\theta_{\rm p}$ and different $\gamma$ values, one sees the higher $\gamma$ value particles experience significantly more total losses and have larger loss rates (the slope of their $\gamma$ values) due to the $\gamma^{2}$ dependence in the dominant term of the RRF in Equation~(12) as in Chapter~\ref{sec:Paper2} and further discussed in Chapter~\ref{sec:Paper3} for non-uniform fields. They also have higher loss rates starting higher in the magnetosphere, contrary to very close to the magnetic mirror as is the case for the low-energy particles. Due to the log scale, it is difficult to see the change in the purple curve, as a description the $\gamma$ drops from $50$ to $\sim 35$, very close to the mirror point and does not have losses elsewhere\footnote{Using higher $B$-fields the $\gamma=50$ loses much more energy than this case.}. This is displayed much clearer in Chapter~\ref{sec:Paper2} Figure~10 when comparing the $\gamma=50$ changes to the \citetalias{Takata2017} results. As expected when using the same $\gamma$ and $\theta_{\rm p}$ but changing the $B$-field, one sees that the loss rate and total losses are much higher for the brown curve than the blue curve, especially in the higher magnetosphere. All of the particles' $\gamma$ values oscillate due to the $\mathbf{E}\times\mathbf{B}$ effects discussed in Chapter~\ref{sec:Paper2} thus the jagged nature of the curve are due to saving resolution, not numerical inaccuracies. Therefore, Figures~\ref{Mirrored_traj} and~\ref{Mirrored_gam} show there is a complex dependence on initial particle $\gamma$, $\theta_{\rm p}$, and field strengths affecting their trajectory, emission position, and emission intensity.       

\begin{figure}[!h]
\centering
\includegraphics[width=\textwidth]{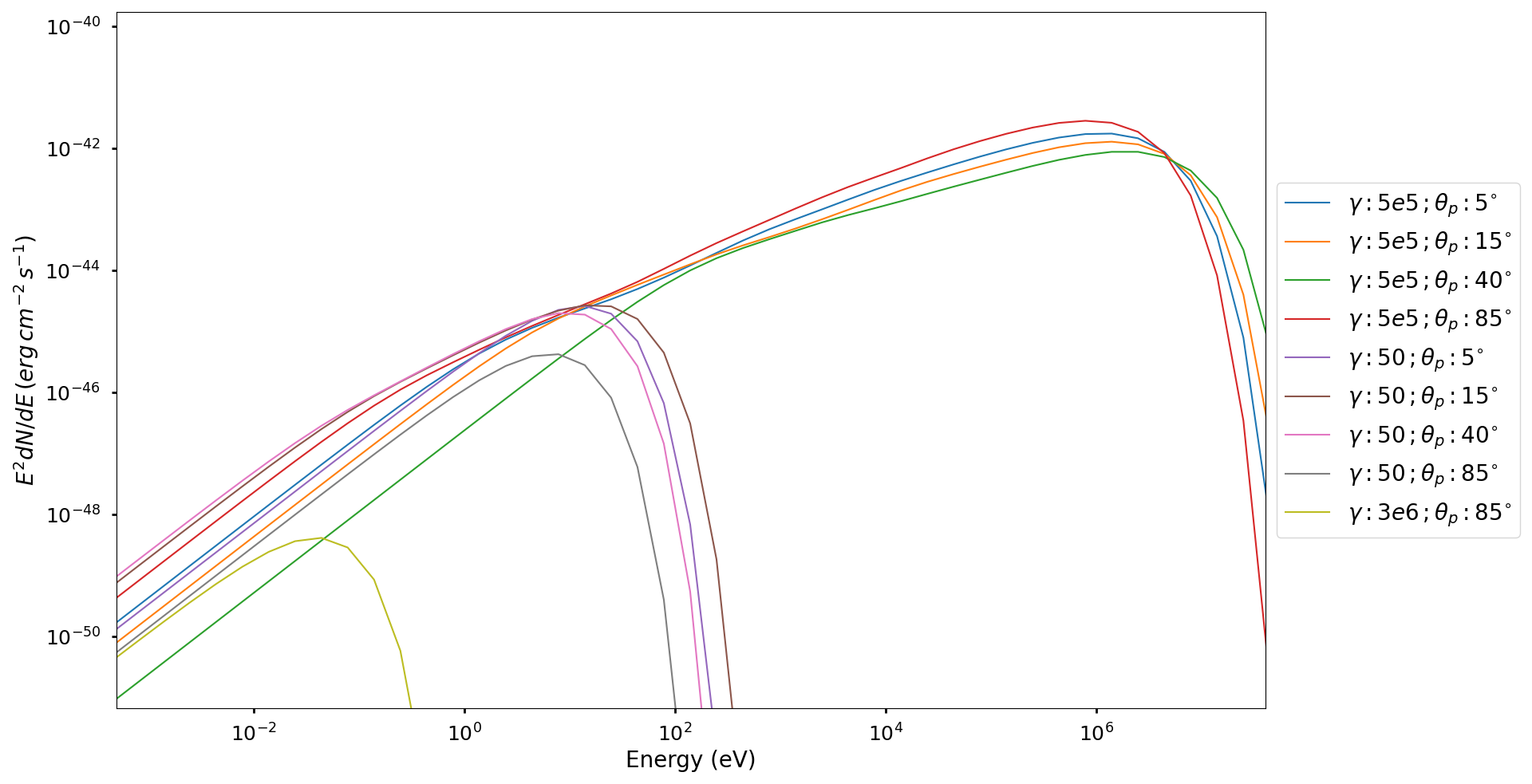}  
\caption[AR Sco Particle Case Spectra]{I have plotted the non-normalised single-particle spectra using $B_{\rm S}= 4.0\times 10^{8} \, \rm{G}$, $\alpha=60^{\circ}$, and $\zeta = 60^{\circ}$. For the first set of spectra, I show the spectrum for 4 particle pitch angles with $\gamma \sim 5\times 10^{5}$ namely, $\theta_{\rm p} = 5^{\circ}$ plotted in blue, $\theta_{\rm p} = 15^{\circ}$ plotted in orange, $\theta_{\rm p} = 40^{\circ}$ plotted in green, and $\theta_{\rm p} = 85^{\circ}$ plotted in red. For the second set of spectra, I show the spectrum for 4 particle pitch angles with $\gamma = 50$ namely, $\theta_{\rm p} = 5^{\circ}$ plotted in purple, $\theta_{\rm p} = 15^{\circ}$ plotted in brown, $\theta_{\rm p} = 40^{\circ}$ plotted in pink, and $\theta_{\rm p} = 85^{\circ}$ plotted in grey. Lastly, I include the spectrum for a particle with $\gamma = 3\times 10^{6}$ and $\theta_{\rm p} = 85^{\circ}$ where the particle was unconstrained by the fields.} 
\label{Cases_spec}
\end{figure}

To investigate how the spectra of Figure~\ref{Spec_p3} are built up by the various macro particles with different initial $\gamma$ and $\theta_{\rm p}$ values, I plot their single-particle spectra in Figure~\ref{Cases_spec}. For this scenario, I used the \citetalias{Takata2019} `best-fit' parameters, namely the $B_{\rm S}= 4.0\times 10^{8} \, \rm{G}$ case using $\alpha=60^{\circ}$, and $p=3.0$ but using a uniform $\theta_{\rm p}$ as they used in \citetalias{Takata2017}, and one of the cases in \citetalias{Takata2019}. In Figure~\ref{Cases_spec} I plot the high-energy particles with initial $\gamma \sim 5\times 10^{5}$ initialising with $\theta_{\rm p} = 5^{\circ}$ plotted in blue, $\theta_{\rm p} = 15^{\circ}$ plotted in orange, $\theta_{\rm p} = 40^{\circ}$ plotted in green, and $\theta_{\rm p} = 85^{\circ}$ plotted in red. For the low-energy particles using initial $\gamma = 50$ I show initialising with $\theta_{\rm p} = 5^{\circ}$ plotted in purple, $\theta_{\rm p} = 15^{\circ}$ plotted in brown, $\theta_{\rm p} = 40^{\circ}$ plotted in pink, and $\theta_{\rm p} = 85^{\circ}$ plotted in grey. The reason I have used $\gamma \sim 5\times 10^{5}$ for our high-energy contribution instead of our $\gamma_{\rm max} = 3.4\times 10^{6}$ is due to when these particles have a high initial $\theta_{\rm p}$ close to $90^{\circ}$, they are unconstrained by the field. This is due to the particle motion being almost perpendicular to the local $B$-field as well as having large $\gamma$ values, meaning they simply fly off in the initialised direction. Thus $\gamma \sim 5\times 10^{5}$ is a case where the larger $\theta_{\rm p}$ value particles were also constrained by the fields. In yellow, I have plotted the single-particle spectra for the unconstrained particle with initial $\gamma = 3\times 10^{6}$ and $\theta_{\rm p} = 85^{\circ}$. Looking at Figure~\ref{Cases_spec} one sees the $\gamma=50$ particle's spectral peaks correspond to the cumulative spectral peak for the same case in Figure~\ref{Spec_p3}. This makes sense due to the majority of the particles having low energy due to the steep $p = 3$ used for the particle energy distribution. Assessing the initial $\theta_{\rm p}$ values of the particle is more difficult. This is from the fact that a higher $\theta_{\rm p}$ means higher SR flux and higher energy cutoff from Equation~(14) and the critical SR photon energy in Chapter~\ref{sec:Paper3}. Conversely, a higher $\theta_{\rm p}$ means the particle is mirrored much higher in the magnetosphere, meaning the particle experiences lower local $B$-fields leading to lower losses and energy cutoff via the same equations. There is thus a balance to these effects for each parameter combination as shown in Figures~\ref{Mirrored_traj} and~\ref{Mirrored_gam}. This is also illustrated in the $\gamma = 50$ spectra cases: the $\theta_{\rm p } = 5^{\circ}$ particle spectrum is higher than that of $40^{\circ}$ but lower than that of $15^{\circ}$, with $85^{\circ}$ being the lowest. The $\gamma \sim 5 \times 10^{5}$ spectra cases show a different order, with $85^{\circ}$ being the highest but lower energy cutoff than the rest followed by $5^{\circ}$, then $15^{\circ}$, and $40^{\circ}$ the lowest. To add to the complicated spectra, one sees the unconstrained particle shown by the yellow curve yielding the lowest spectrum contribution due to such particles experiencing little losses since they are unconstrained by the field. Therefore, these particles will contribute little to the cumulative spectrum. It is evident that realistic parameters are needed when solving the general equations of motion since one does not expect the particles to have initial $\theta_{\rm p}$ close to $90^{\circ}$. It is also crucial to use a more realistic $\theta_{\rm p}$ distribution since if one has less contribution from the high-$\theta_{\rm p}$ particles this could shift the spectral peak slightly and lower the overall flux.

\begin{figure}[!h]
\centering
\includegraphics[width=\textwidth]{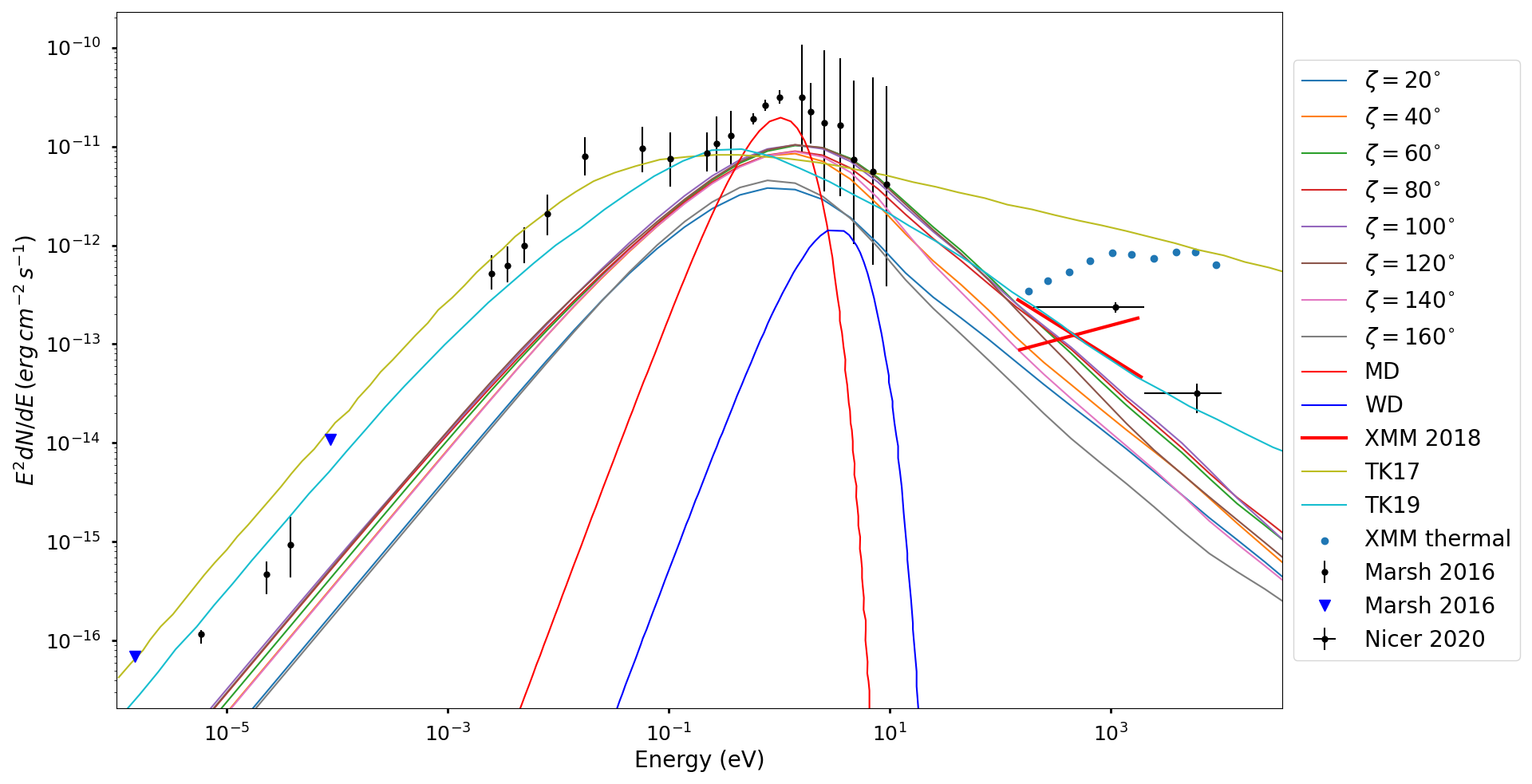} 
\caption[AR Sco Spectrum $p=3.0$ for different $\zeta$ values]{The same as in Figure~\ref{Spec_p3}, but for $B_{\rm S}= 2.5\times 10^{8} \, \rm{G}$, $\alpha=60^{\circ}$, $\gamma_{\rm min}=10$, $p=3.0$, and varying $\zeta$. Here using $\zeta = 20^{\circ}$ is plotted in pale blue, $\zeta = 40^{\circ}$ is plotted in orange, $\zeta = 60^{\circ}$ is plotted in green, $\zeta = 80^{\circ}$ is plotted in magenta, $\zeta = 100^{\circ}$ is plotted in purple, $\zeta = 120^{\circ}$ is plotted in brown, $\zeta = 140^{\circ}$ is plotted in pink, and $\zeta = 160^{\circ}$ is plotted in grey.}
\label{Spec_zeta}
\end{figure}

In Figure~\ref{Spec_zeta} I plot the magnetic mirror scenario spectrum using $B_{\rm S}= 2.5\times 10^{8} \, \rm{G}$, $\alpha=60^{\circ}$, $p=3.0$, setting $\gamma_{\rm min} = 10$, $\gamma_{\rm{max}}= 3.4\times10^{6}$, and varying $\zeta$. Here I have plotted $\zeta=20^{\circ}$ in pale blue, $\zeta=40^{\circ}$ in orange, $\zeta=60^{\circ}$ in green, $\zeta=80^{\circ}$ in magenta, $\zeta=100^{\circ}$ in purple, $\zeta=120^{\circ}$ in brown, $\zeta=140^{\circ}$ in pink, and $\zeta=160^{\circ}$ in grey. The WD and M-dwarf blackbody spectra are plotted in blue and red, respectively, with the \citetalias{Takata2017} `best-fit' spectrum plotted in yellow and the \citetalias{Takata2019} `best-fit' spectrum plotted in cyan. Here one sees the spectra using $\zeta$ from $40^{\circ}$ to $140^{\circ}$ are reasonably similar, with their spectral peaks at similar energies, but their high-energy slopes seem to deviate from one another. The $\zeta=20^{\circ}$ and $\zeta=160^{\circ}$ spectra are found to deviate the most, having lower flux and lower energy cutoffs.  

\subsection{Emission Maps} \label{sec:ARSCO_s3.2}   
In this subsection, I will show the emission maps from the different cases of the magnetic mirror model for AR Sco discussed above and their corresponding spectra shown in Figure~\ref{Spec_p3}. As mentioned, these emission maps were produced by sampling only one field line as an exploration into the magnetic mirror proposed by \citetalias{Takata2017}. Due to this model approach being computationally heavy, using one field line also allows one to save computational time while I first asses if the spectra and emission maps are feasible before we use multiple field lines to increase the statistics of the emission map and spectra. Using multiple field lines will drastically improve the statistics, smoothing out the emission maps and light curves, yielding enough resolution to better identify caustics and more specific spatial origins of the emission. One can see this effect in the emission maps of Chapter \ref{sec:Paper3} when using one polar cap rim in the active region vs several for both our model and the \citetalias{Harding2015, Harding2021, Barnard2022} models. Another factor that needs to be accounted for is the $E_{\perp}$-field oscillating the parameters as discussed in Chapters \ref{sec:Paper2} and \ref{sec:Paper3}, where better statistics is required to smoothen out this oscillatory emission in the emission maps by using multiple field lines. These energy-dependent emission maps presented in this section are taken over the whole AR Sco photon energy spectrum displayed in our SED results, thus from $10^{-5} \, \rm{eV}$ to $10^{5} \, \rm{eV}$.   

\begin{figure}[!h]
\centering
\includegraphics[width=\textwidth]{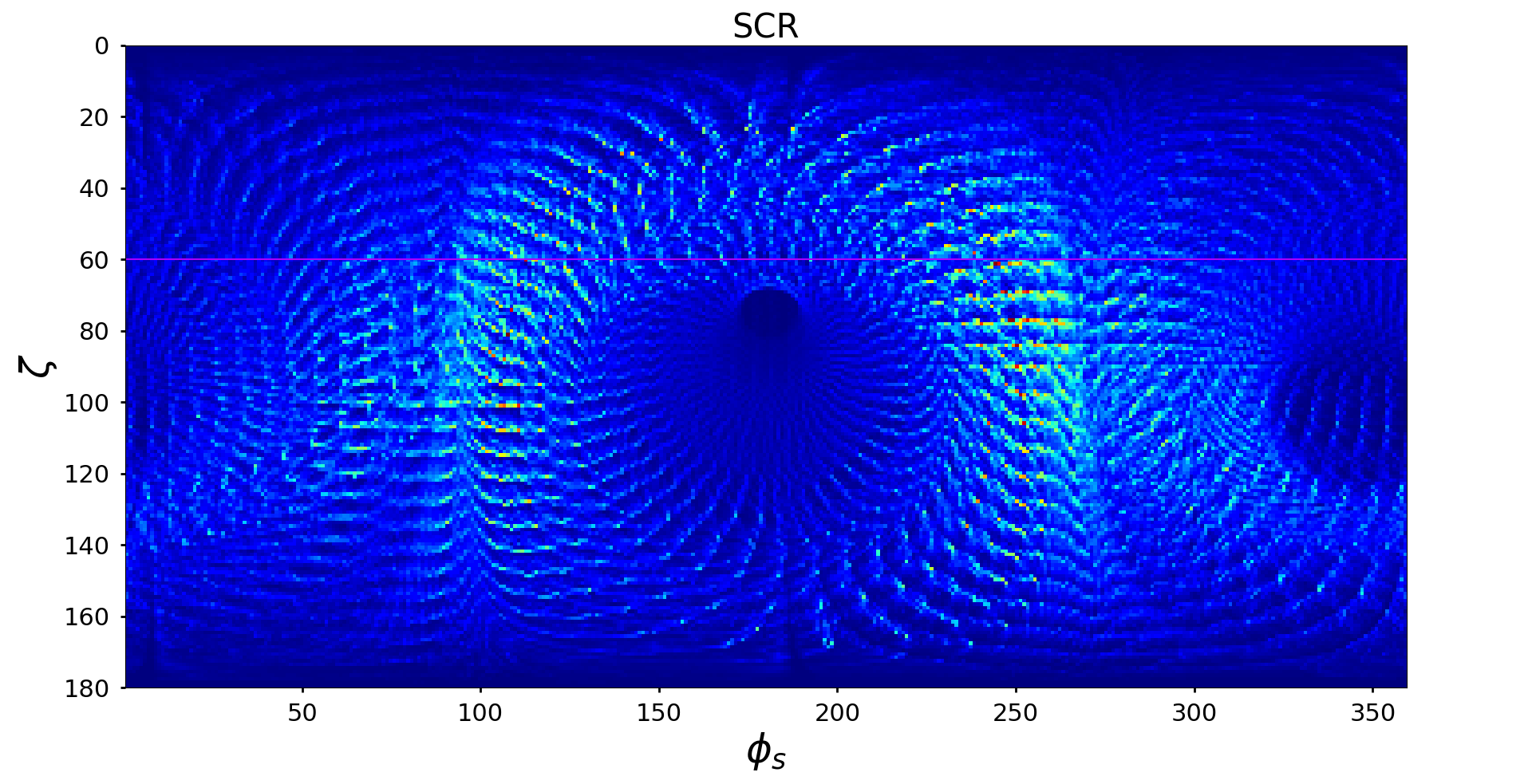}  
\caption[AR Sco Emission Map for $B_{\rm S}= 6\times 10^{8} \, \rm{G}$ and $\alpha=60^{\circ}$ Case]{AR Sco emission map produced using $B_{\rm S}= 6\times 10^{8} \, \rm{G}$, $\alpha = 60^{\circ}$, and $p=3.0$. I show the $\zeta=60^{\circ}$ cut in purple used to concurrently produce the spectra in Figure~\ref{Spec_p3} and this emission map.}
\label{E_map_EB_6e8}
\end{figure}

\begin{figure}[!h]
\centering
\includegraphics[width=\textwidth]{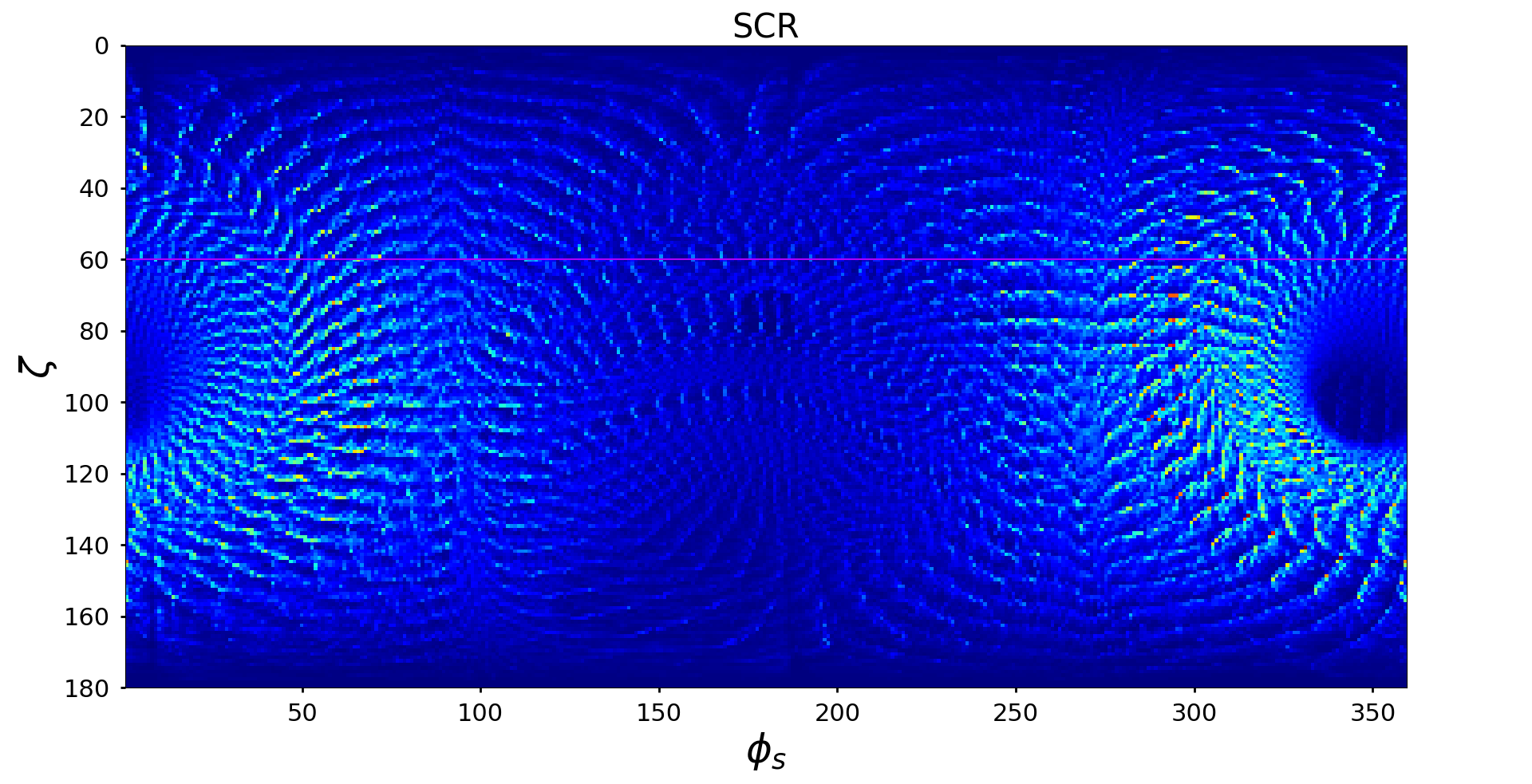}  
\caption[AR Sco Emission Map for $B_{\rm S}= 4\times 10^{8} \, \rm{G}$ and $\alpha=60^{\circ}$ Case]{AR Sco emission map produced using $B_{\rm S}= 4\times 10^{8} \, \rm{G}$, $\alpha = 60^{\circ}$, and $p=3.0$. I show the $\zeta=60^{\circ}$ cut in purple used to concurrently produce the spectra in Figure~\ref{Spec_p3} and this emission map.}
\label{E_map_EB_4e8}
\end{figure}

\begin{figure}[!h]
\centering
\includegraphics[width=\textwidth]{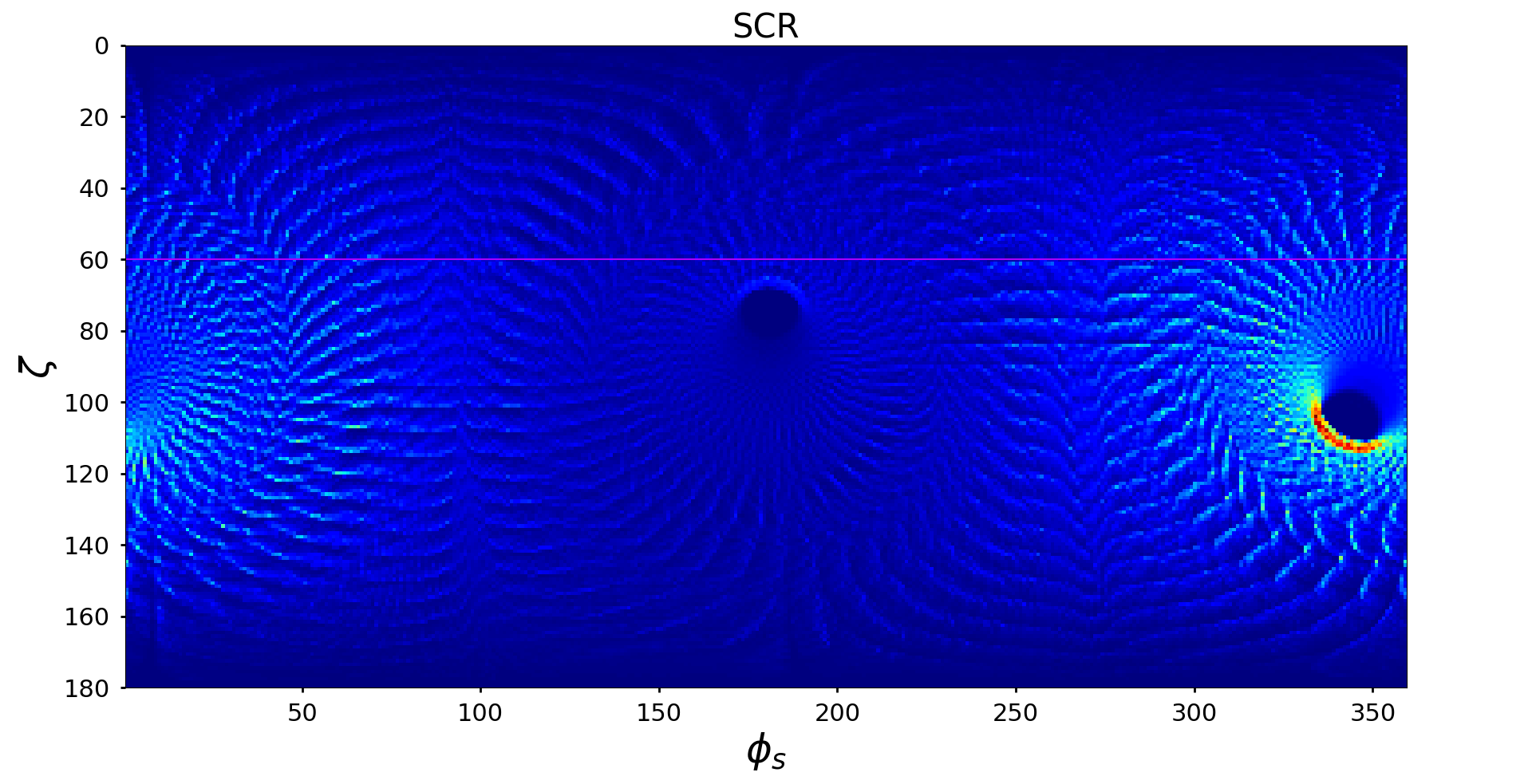}  
\caption[AR Sco Emission Map for $B_{\rm S}= 2.5\times 10^{8} \, \rm{G}$ and $\alpha=60^{\circ}$ Case]{AR Sco emission map produced using $B_{\rm S}= 2.5\times 10^{8} \, \rm{G}$, $\alpha = 60^{\circ}$, $p=3.0$, and $\gamma_{\rm min}=10$. I show the $\zeta=60^{\circ}$ cut in purple used to concurrently produce the spectra in Figure~\ref{Spec_p3} and this emission map.}
\label{E_map_EB_2_5e8}
\end{figure}

In Figure~\ref{E_map_EB_6e8} I show the emission map produced by our model for the AR Sco magnetic mirror scenario using $B_{\rm S}= 6\times 10^{8} \, \rm{G}$, $\alpha = 60^{\circ}$ and $p=3.0$. I show the $\zeta = 60^{\circ}$ cut in the emission map for particles injected in one hemisphere used to produce the spectrum in Figure~\ref{Spec_p3} in purple. In Figure~\ref{E_map_EB_6e8} one can see the two poles at $\phi_{s} \sim 350^{\circ}$ and $\phi_{s} \sim 180^{\circ}$ where these are formed where the particles are turned around by each of their respective magnetic mirror points. Two poles are visible due to the particle heading towards the WD surface, emitting in that direction forming one pole and as the particle is turned around by the magnetic mirror it travels outwards emitting in the opposite direction and forming a secondary pole in the emission map. These poles are separated by $\sim180^{\circ}$ due to the opposite direction of the emission, but when the particle moves outward it is dragged along by the $\mathbf{E}\times \mathbf{B}$-drift, causing the two poles to be shifted slightly closer than $180^{\circ}$ to one another as seen in the figure. The effect of the direction the particle is travelling on its trajectory is shown in Figure~\ref{Mirrored_traj}, showing the effect beyond the $\mathbf{E}\times \mathbf{B}$-drift. One would only expect the poles to be exactly separated by $180^{\circ}$ if the emission is taken from out-flowing particles from each of the WD's poles. Interestingly, in this high-field case the emission seems to be located further up in the magnetosphere as it is seen to be located at a distance from the magnetic mirror points. This suggests the low energy particles that make up the majority of the particle energy spectrum radiate further away from the magnetic mirror point due to the already high $B$-field experienced by the particles as supported by my finding in Figure \ref{Mirrored_gam}. The sweeping features in the emission maps are due to the large $\theta_{\rm p}$ particles that only travel a short distance towards the WD surface before being mirrored. This single-particle emission map for large $\theta_{\rm p}$ is shown in Figure~\ref{E_map_large_theta} and this emission seems to be more prominent in the high $B$-field cases. Next I look at the the emission map using $B_{\rm S}= 4\times 10^{8} \, \rm{G}$, $\alpha = 60^{\circ}$, and $p=3.0$ shown in Figure~\ref{E_map_EB_4e8}. Here one sees the emission is closer to the one pole at $\phi_{s} \sim 350^{\circ}$ compared to Figure~\ref{E_map_EB_6e8}. To see if the emission gets closer to the magnetic mirror point I investigate Figure~\ref{E_map_EB_2_5e8} that was generated using $B_{\rm S}= 2.5\times 10^{8} \, \rm{G}$, $\alpha=60^{\circ}$, $\gamma_{\rm min}=10$, and $p=3.0$. Here one sees that most of the emission is located at the mirror point of the $\phi_{s} \sim 350^{\circ}$ pole. This result of the bulk emission forming the spectral peak moving from the higher magnetosphere to the mirror point motivated our discussion about the spectral peak not shifting as expected in the previous section. Upon further analysis, one finds an intense caustic visible at the lower left edge of the pole as well, where this caustic could extend upward or have an extended structure if I include more field lines. I believe the caustic is more prominent at the leading edge of the pole due to the particle being co-rotated by the fields (dragged) as it gyrates around the field. Hence at the leading edge, the particle will radiate and as it co-rotated in the direction of the leading edge, other points on the particles' gyration path now align with this initial photon direction causing the photon bunching (caustic). A similar but more dispersed caustic is observed in Figure~\ref{E_map_EB_4e8} but located further in the magnetosphere. Interestingly, the pole at $\phi_{s} \sim 350^{\circ}$ seems to become smaller for the lower $B$-field cases as the particles radiate their bulk emission closer and closer to the magnetic mirror. For pulsar cases \cite{Morini1983} proposed caustics to form from photon bunching occurring at the trailing edge due to aberration effects and time-of-flight delays for photons emitted outward over a range of magnetic field lines near the LC. Therefore, the caustics in our emission maps are different to the \cite{Morini1983} caustics due to the low altitude of our caustics since the time-of-flight delays are not as prominent close to the stellar surface vs. at the light cylinder.

\begin{figure}[!h]
\centering
\includegraphics[width=\textwidth]{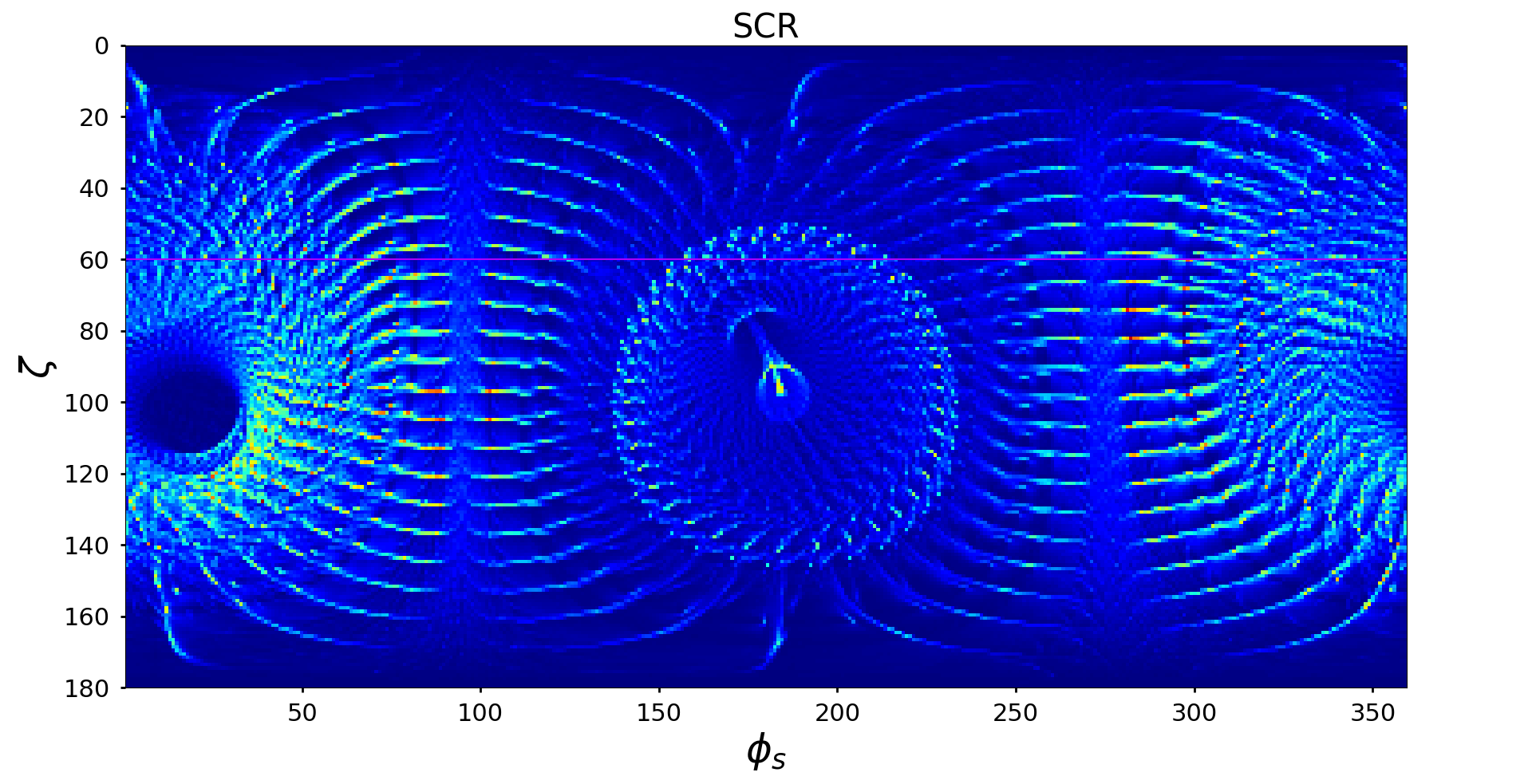}  
\caption[AR Sco Emission Map for $B_{\rm S}= 4\times 10^{8} \, \rm{G}$ and $\alpha=60^{\circ}$ Case with no $E_{\perp}$]{AR Sco emission map produced using $B_{\rm S}= 4\times 10^{8} \, \rm{G}$, $\alpha = 60^{\circ}$, and $p=3.0$ with the $E_{\perp}$-field excluded. I show the $\zeta=60^{\circ}$ cut in purple used to concurrently produce the spectra in Figure~\ref{Spec_p3} and this emission map.}
\label{E_map_B}
\end{figure}

\begin{figure}[!h]
\centering
\includegraphics[width=\textwidth]{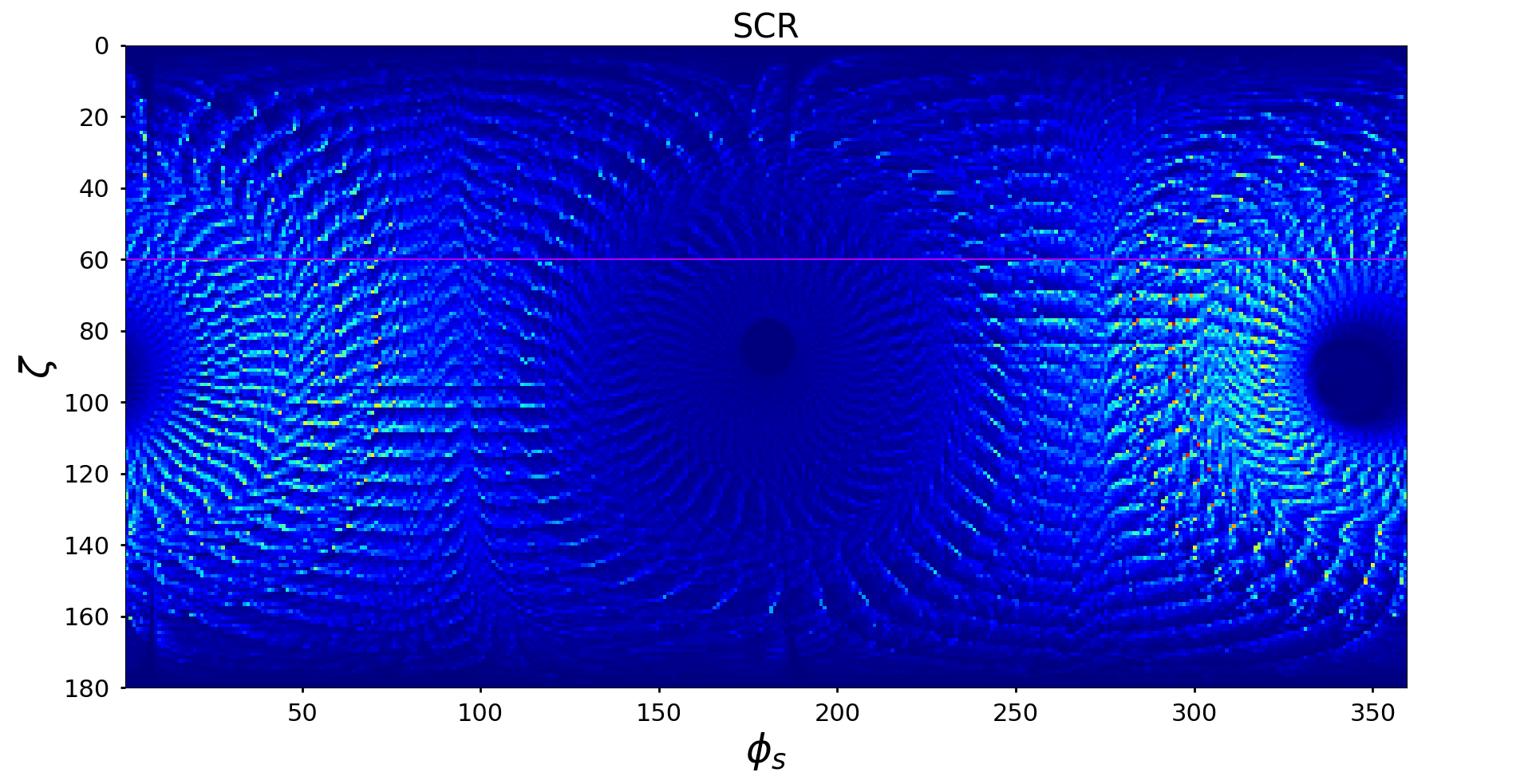}  
\caption[AR Sco Emission Map for $\alpha=80^{\circ}$ Case]{AR Sco emission map produced using $B_{\rm S}= 4\times 10^{8} \, \rm{G}$, $\alpha = 80^{\circ}$ and $p=3.0$. I show the $\zeta=60^{\circ}$ cut in purple used to concurrently produce the spectra in Figure~\ref{Spec_p3} and this emission map.}
\label{E_map_alpha}
\end{figure}

\begin{figure}[!h]
\centering
\includegraphics[width=\textwidth]{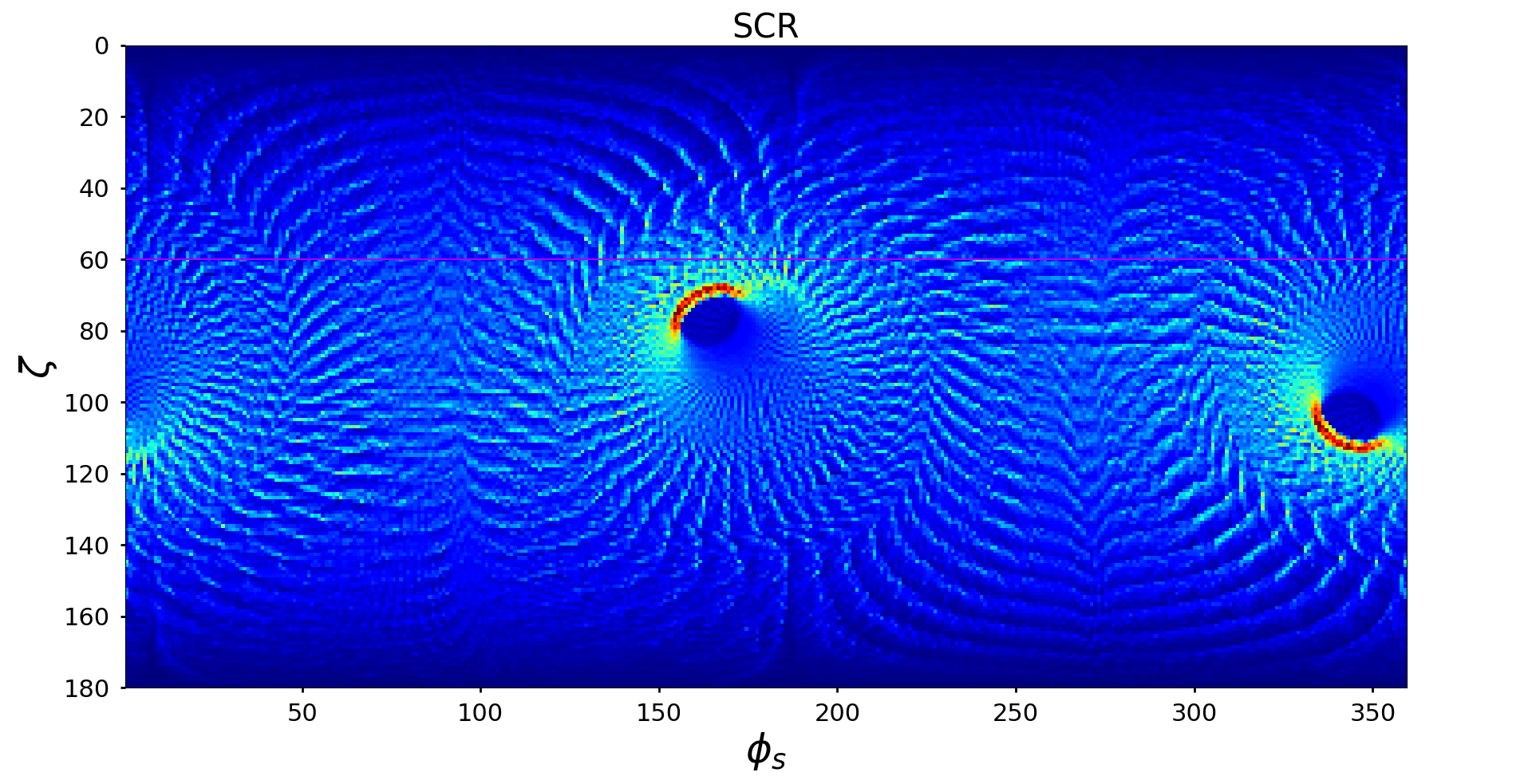}  
\caption[AR Sco Emission Map for 2 Pole Injection]{AR Sco emission map produced using the same parameters as Figure~\ref{E_map_EB_2_5e8}. Here I show the flipped and shifted emission for the southern pole included as well. I show the $\zeta=60^{\circ}$ cut in purple.}
\label{E_map_EB_2_5e8_2pole}
\end{figure}

As a further assessment of the emission maps for the magnetic mirror scenario, I show some of the extra cases presented in Figure~\ref{Spec_p3} as well. In Figure~\ref{E_map_B} I show the case using $B_{\rm S}= 4\times 10^{8} \, \rm{G}$, $\alpha = 60^{\circ}$, and $p=3.0$ when excluding the $E_{\perp}$ field. In the figure, one sees the emission map still has low statistics requiring more $B$-fields lines for a smoother emission map but the emission is more `stringy' and less `blobby' due to the lack of the $E_{\perp}$-field oscillating the parameters. I found that since the particles are not drifted by the $E_{\perp}$-field they mirror much closer to the WD surface as seen in Figure~(10) from Chapter~\ref{sec:Paper2}. I also found that some of the particles with small $\theta_{\rm p}$ values do not mirror and simply head towards the surface as suggested by \citetalias{Takata2017} but in all the cases with included $E_{\perp}$-fields all the particles were mirrored due to the added $\mathbf{E}\times\mathbf{B}$-drift effect on the particle, causing an additional increase in $p_{\perp}$. These non-mirrored particles' effects can be seen at the $\phi_{s} \sim 180^{\circ}$ pole in the emission map. In this figure, one also sees the second pole at $\phi_{s} \sim 10^{\circ}$ instead of $\phi_{s} \sim 350^{\circ}$, showing the large effect excluding the $E_{\perp}$-field in the emission maps. Another effect is that the caustic appears on the lower right of the pole instead of the lower left as was seen when including the $E_{\perp}$-field. Hence, there are quite a lot of changes in the emission maps and by extension the light curves if the $E_{\perp}$-field is not included. For Figure~\ref{E_map_alpha} I show the emission map for the magnetic mirror scenario using $B_{\rm S}= 4\times 10^{8} \, \rm{G}$, $\alpha = 80^{\circ}$, and $p=3.0$. In the figure one can see the 2 poles are closer to alignment along $\zeta=90^{\circ}$, as expected from using a $\alpha$ value close to $90^{\circ}$. This also seems to have shifted the caustic to the more central left of the pole at $\phi_{s} \sim 350^{\circ}$ than what is observed in Figure~\ref{E_map_EB_4e8} but as mentioned, more field lines need to be included to obtain a more robust assessment of these features. In Figure~\ref{E_map_EB_2_5e8_2pole}, I plot the emission map using the same parameters as Figure~\ref{E_map_EB_2_5e8} but flip and shift the emission from the northern pole, as discussed and done in Chapter~\ref{sec:Paper3}, to emulate particles injected towards the southern pole as it points towards the companion. This is done since all the previous emission maps only have particles injected when the open field lines of the northern pole points towards the companion. This is done since time-dependent particle injection would be a large increase in difficulty to simulate and is why it has only been done with orbital dependence in the geometric model by \citet{Potter2018b}. In the \citetalias{Takata2017, Takata2019} it appears they suggest particles are injected towards the southern pole of the WD from the companion surface pointing away from the WD. They seem to have simply used this set-up to invoke a second pole in the emission to get the double-peaked structure in the light curves, there does not seem to be modelling of these particles. The observations by \citet{Garnavich2019} indicate some thermal emission at the back face of the companion from their velocity plots, suggesting some heating due to the magnetic field interaction. The problem is these particles injected at the companion's face pointing away from the WD will only be captured and guided towards the southern pole by closed magnetic field lines, since the open magnetic field lines will simply guide these particles beyond the light cylinder. In Figure~\ref{E_map_EB_2_5e8_2pole} one sees the two prominent poles well separated by $180^{\circ}$ but this could possibly shift if the southern pole emission is properly simulated. Thus, one can see this emission map produces a double-peaked light curve due to the two prominent poles. The dominance of each peak in the light curve will change depending on where the observer cuts along $\zeta$ as well as the magnetic inclination angle, since using lower $\alpha$ will move the two poles up and down, respectively. This could lead to the observer missing one of the poles completely as is also seen for pulsar emission maps.     

\subsection{Particle Cases} \label{sec:ARSCO_s3.3}
\begin{figure}[!h]
\centering
\includegraphics[width=\textwidth]{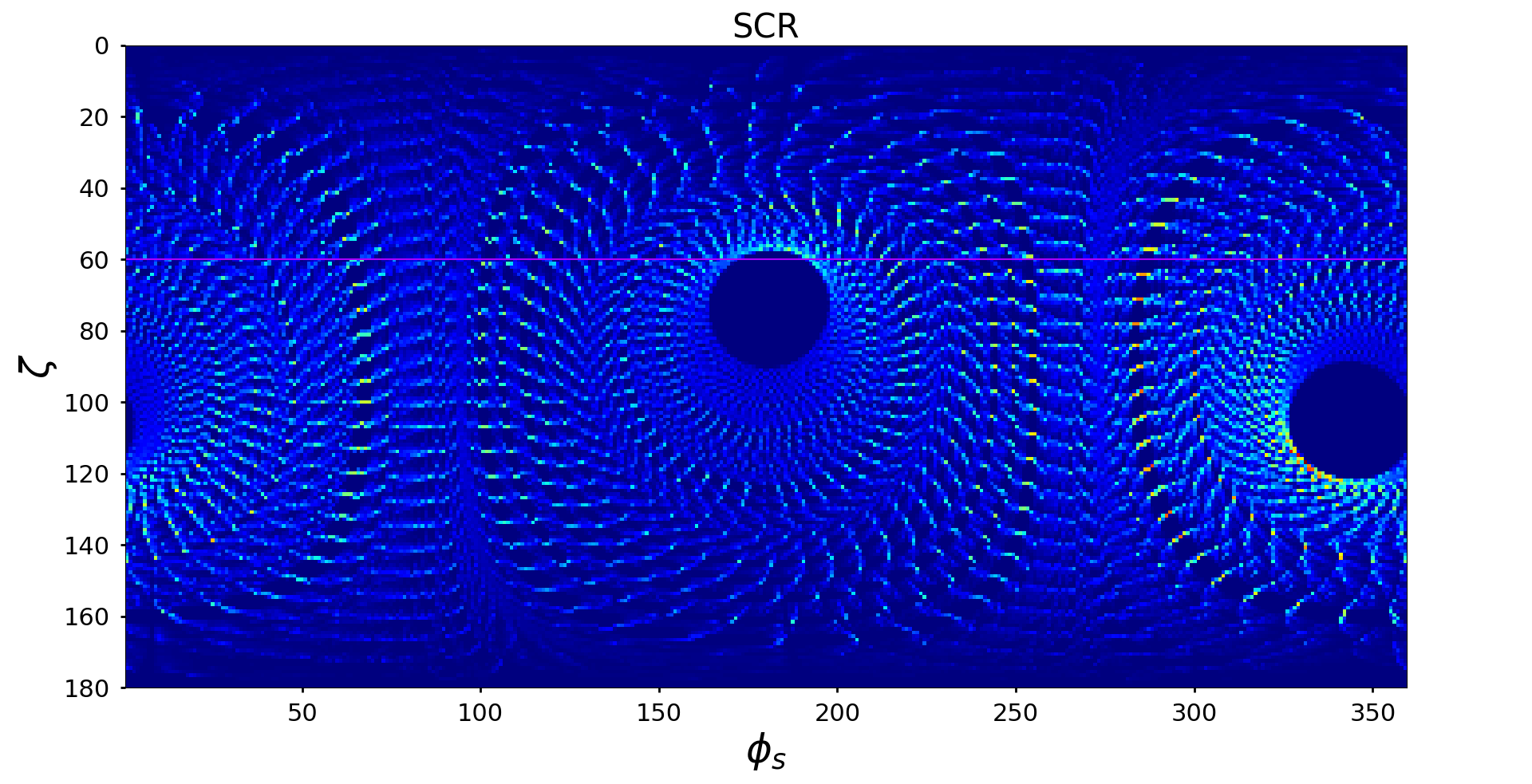}  
\caption[AR Sco Emission Map for a Standard Particle]{AR Sco emission map for the same scenario as Figure \ref{E_map_EB_4e8} but showing the results for a single mirrored particle with an initial $\theta_{p} = 15^{\circ}$ and $\gamma = 5\times 10^{5}$.}
\label{E_map_normal}
\end{figure}

\begin{figure}[!h]
\centering
\includegraphics[width=\textwidth]{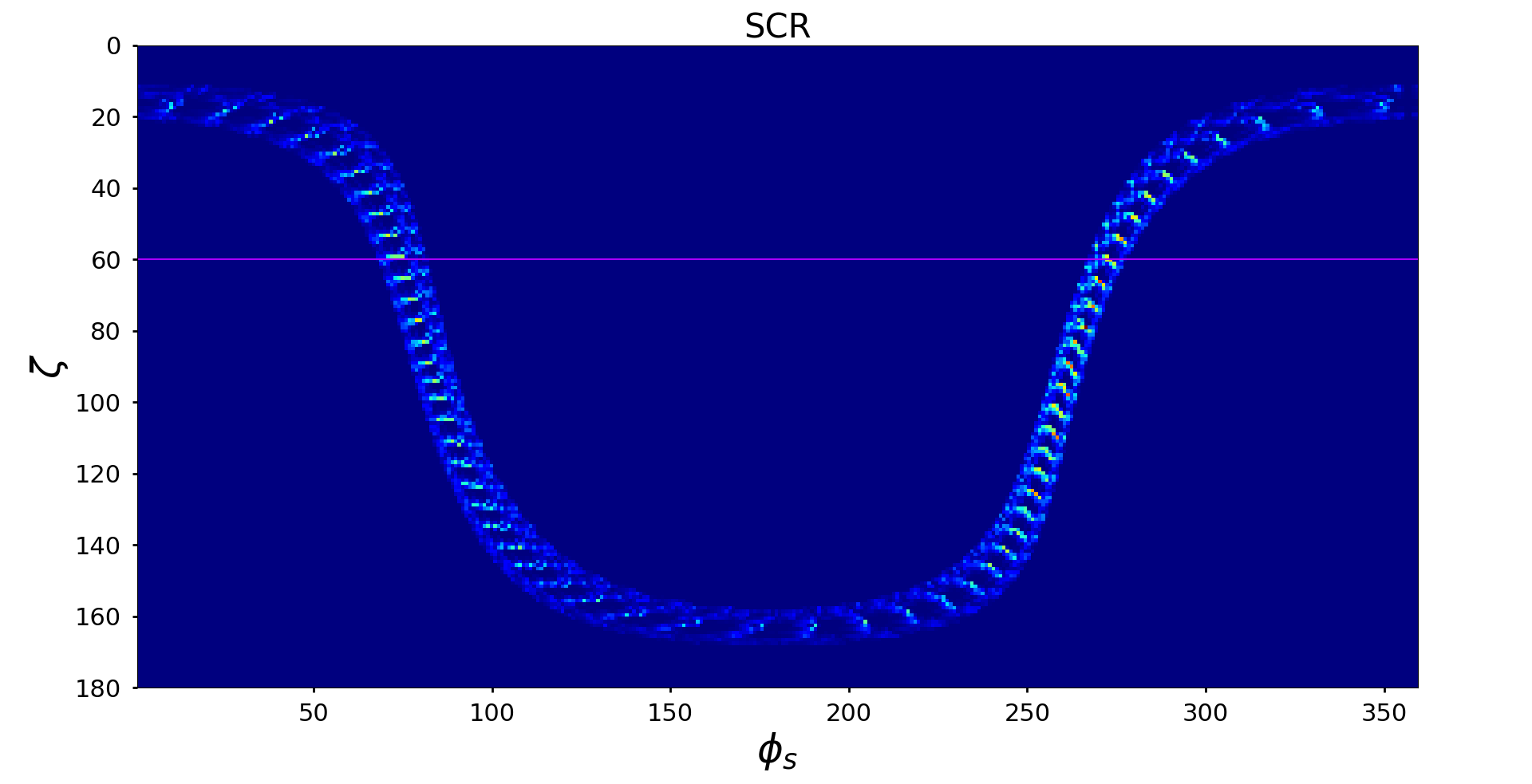}  
\caption[AR Sco Emission Map for Large $\theta_{p}$ Particle]{AR Sco emission map for the same scenario as Figure \ref{E_map_EB_4e8} but showing the results for a single particle with a large initial $\theta_{p} = 85^{\circ}$ and $\gamma = 5\times 10^{5}$ which mirrors almost immediately.}
\label{E_map_large_theta}
\end{figure}

\begin{figure}[!h]
\centering
\includegraphics[width=\textwidth]{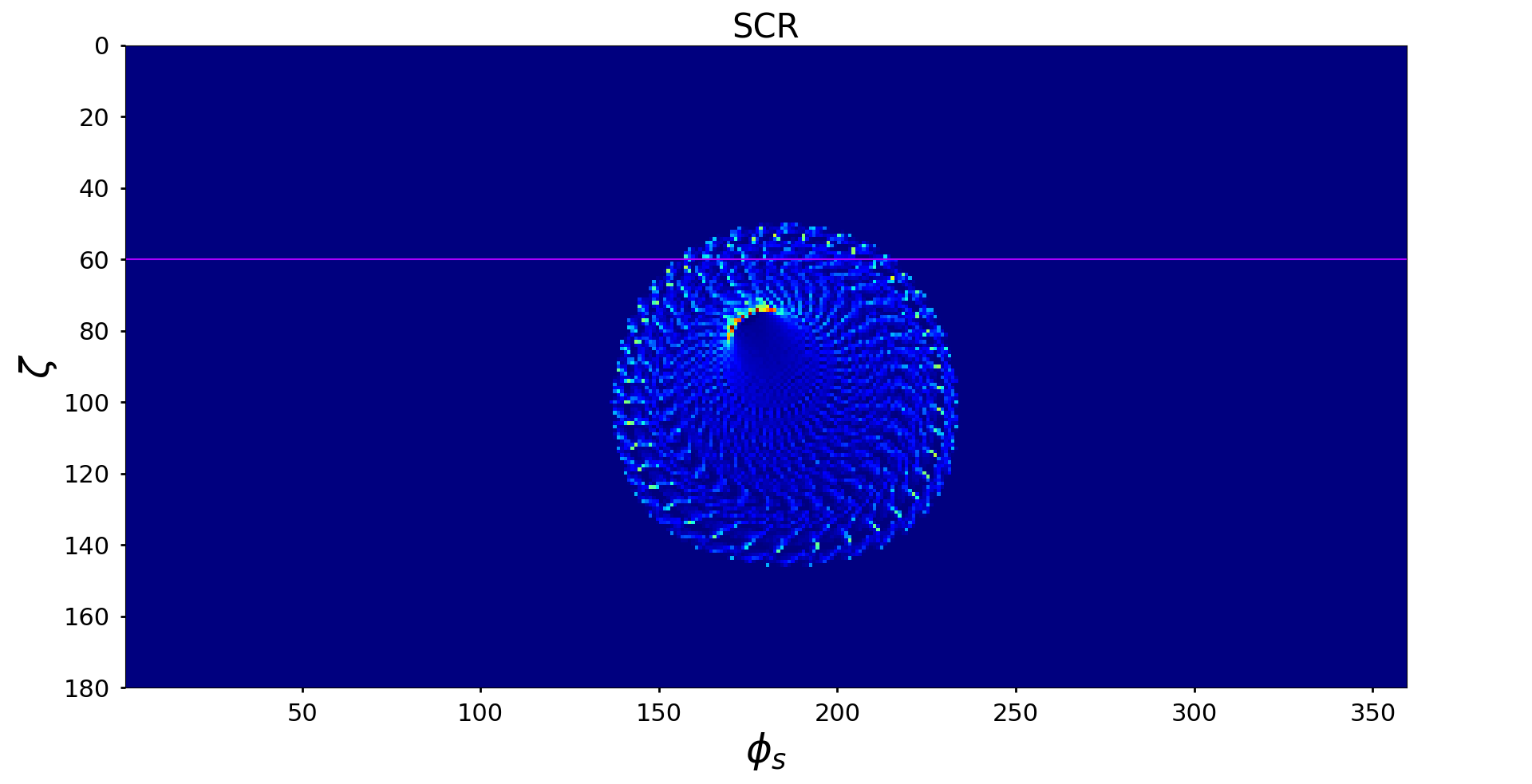}  
\caption[AR Sco Emission Map for Non-Mirrored Particle]{AR Sco emission map for the same scenario as Figure \ref{E_map_B} but showing the results for a single particle with a small initial $\theta_{p} = 5^{\circ}$ and $\gamma = 5\times 10^{5}$ that is not mirrored.}
\label{E_map_inward}
\end{figure}

In this subsection, I assess some of the individual particle contributions to the emission maps namely a standard mirrored particle, a particle with large $\theta_{\rm p}$ being mirrored very quickly, and a non-mirrored particle in the case without the $E_{\perp}$-field included. For Figures~\ref{E_map_normal} and~\ref{E_map_large_theta}, I use the same parameters as Figure~\ref{E_map_EB_4e8}. In Figure~\ref{E_map_normal}, I show the emission map contribution of how the majority of the particles would appear as they head towards the WD surface are mirrored and head outwards towards the companion. The pole size does change for each parameter combination, since each particle is mirrored at a different height, depending on the initial $\theta_{\rm p}$ and $\gamma$ thus being affected by the RRF losses as shown in Figures~\ref{Mirrored_traj} and~\ref{Mirrored_gam}. The spread of the emission in the emission map is also different for the different particle $\theta_{\rm p}$ values. The specific initial parameters for this particle are $\theta_{\rm p} = 15^{\circ}$ and $\gamma = 5\times 10^{5}$. In Figure~\ref{E_map_large_theta} I show the emission map contribution for a particle with initial parameters $\theta_{\rm p} = 85^{\circ}$ and $\gamma = 5\times 10^{5}$. In this case, the particle barely travels toward the WD surface before it is mirrored. Therefore these are the particles that cause the sweeping patterns in the high $B$-field emission maps especially seen in Figure~\ref{E_map_EB_6e8}. Therefore, as mentioned, it is important to assess the realistic contribution of the high-$\theta_{\rm p}$ particles in one's particle distribution. In Figure~\ref{E_map_inward}, I show the emission map contribution for a non-mirrored particle using the same parameters as in Figure~\ref{E_map_B} and neglecting the $E_{\perp}$-field. The initial parameters are $\theta_{p} = 5^{\circ}$ and $\gamma = 5\times 10^{5}$. One sees the particle moving inward toward the surface and stopped once the particle $\gamma$ was below the specified threshold. This indicates that the pole at $\phi_{s} \sim 180^{\circ}$ is due to the inward motion of the particle and the pole at $\phi_{s} \sim 350^{\circ}$ is due to the outward motion of the particle that was injected at the open northern pole field lines. This explains the strange features due to the non-mirrored particles at $\phi_{s} \sim 180^{\circ}$ in Figure~\ref{E_map_B}. Hence, from Figure~\ref{E_map_EB_2_5e8} one sees that the emission is much more prominent for the particles just after they have been mirrored and start travelling outward. I believe the outward pole emission is more prominent due to the emission being directed inward (away from the observer) vs being directed outward and beamed towards the observer. This effect will be overshadowed if particles are injected towards both poles as each pole points toward the companion as the WD completes its rotation as is illustrated in Figure~\ref{E_map_EB_2_5e8_2pole}.    

\subsection{Light Curves} \label{sec:ARSCO_s3.4}
\begin{figure}[!h]
\begin{minipage}{20pc}
\centering
\includegraphics[width=\textwidth]{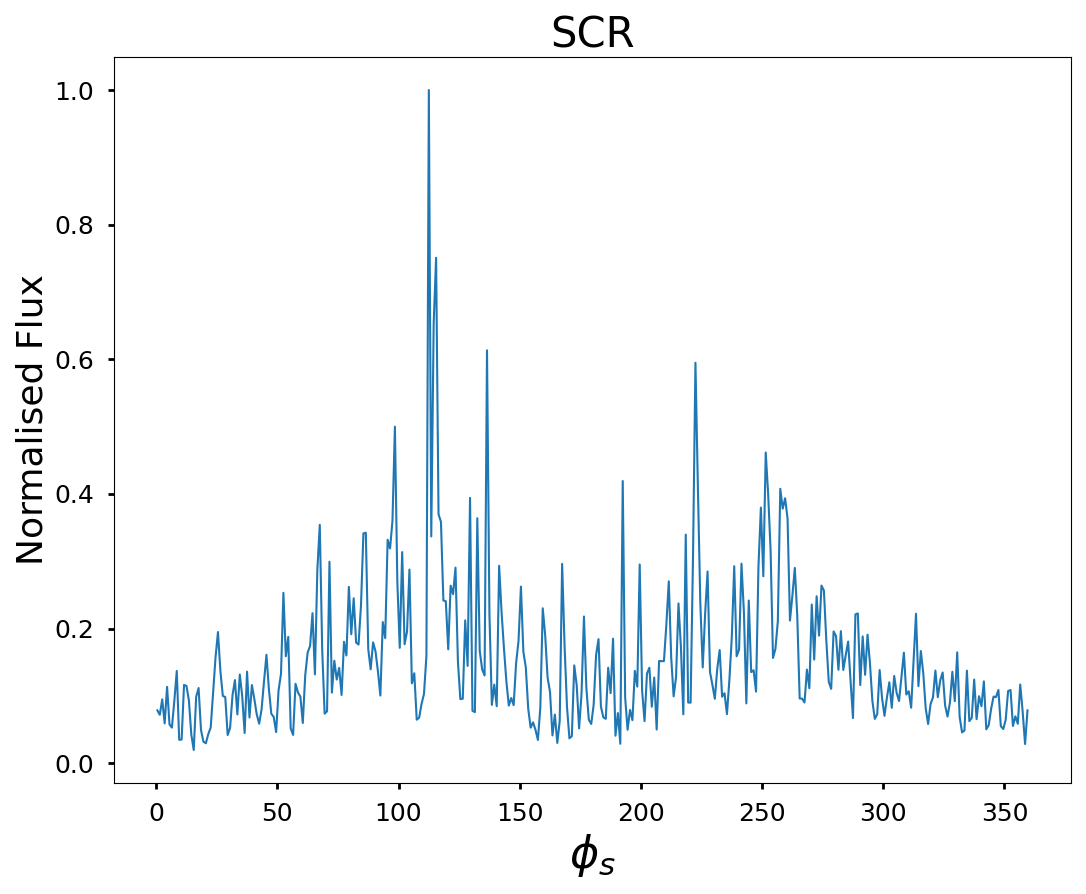}  
\caption[AR Sco Light Curve for Higher Fields and $\alpha=60^{\circ}$ Case]{AR Sco light curve produced from the emission map in Figure \ref{E_map_EB_6e8} using $\zeta = 60^{\circ}$.}
\label{LC_EB_6e8}
\end{minipage}%\hspace{0pc}
\begin{minipage}{20pc}
\centering
\includegraphics[width=\textwidth]{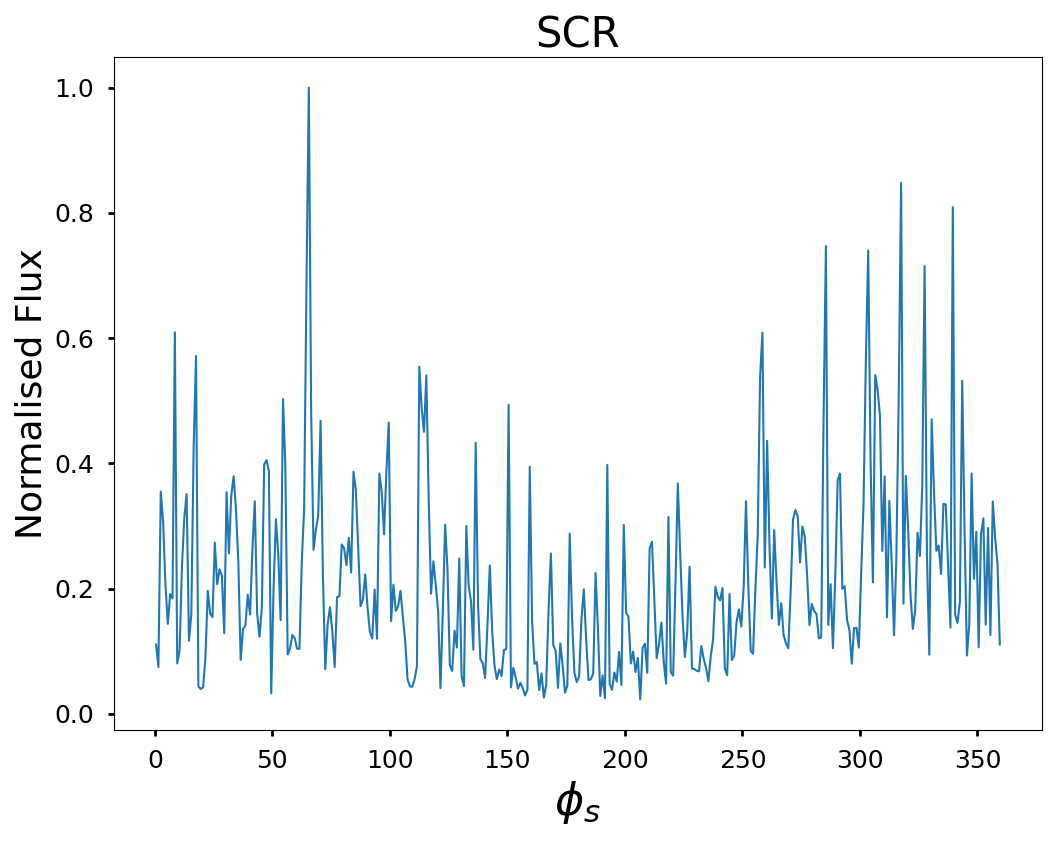}  
\caption[AR Sco Light Curve for Low $E$ and $B$-field Case]{AR Sco light curve produced from the emission map in Figure \ref{E_map_EB_4e8} using $\zeta = 60^{\circ}$.}
\label{LC_EB_4e8}
\end{minipage}
\end{figure}

\begin{figure}[!h]
\begin{minipage}{20pc}
\centering
\includegraphics[width=\textwidth]{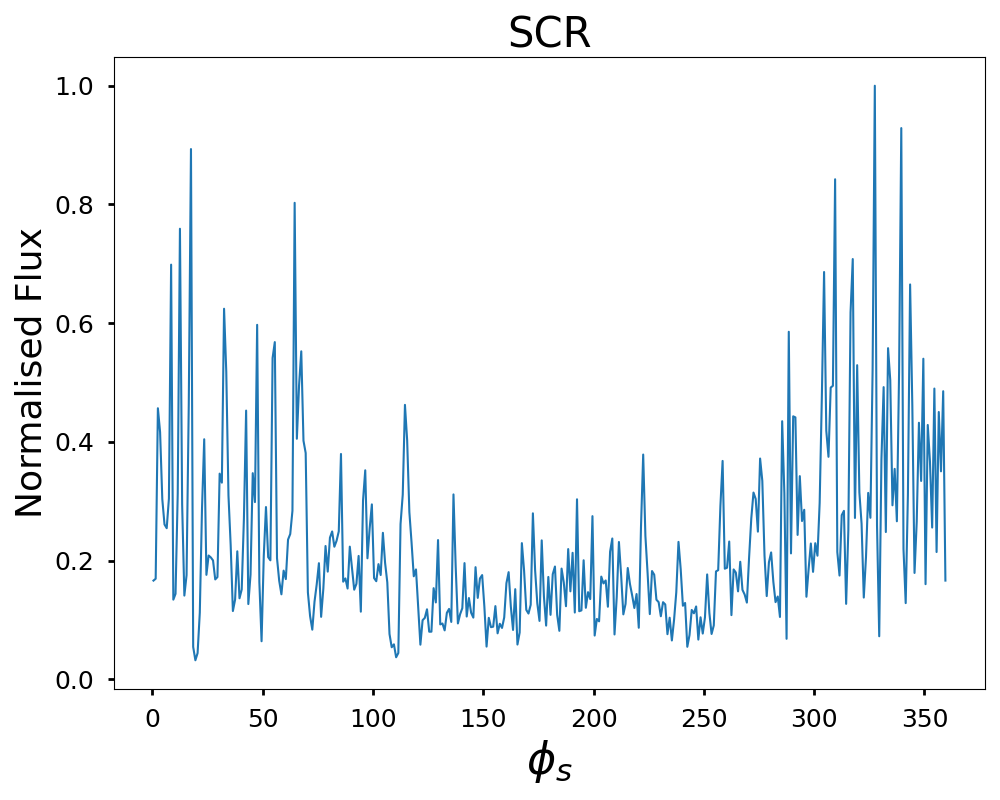}  
\caption[AR Sco Light Curve for $E$ and $B$-field Case with $B_{\rm S}= 2.5\times 10^{8} \, \rm{G}$]{AR Sco light curve produced from the emission map in Figure \ref{E_map_EB_2_5e8} using $\zeta = 60^{\circ}$.}
\label{LC_EB_2e8}
\end{minipage}%\hspace{0pc}
\begin{minipage}{20pc}
\centering
\includegraphics[width=\textwidth]{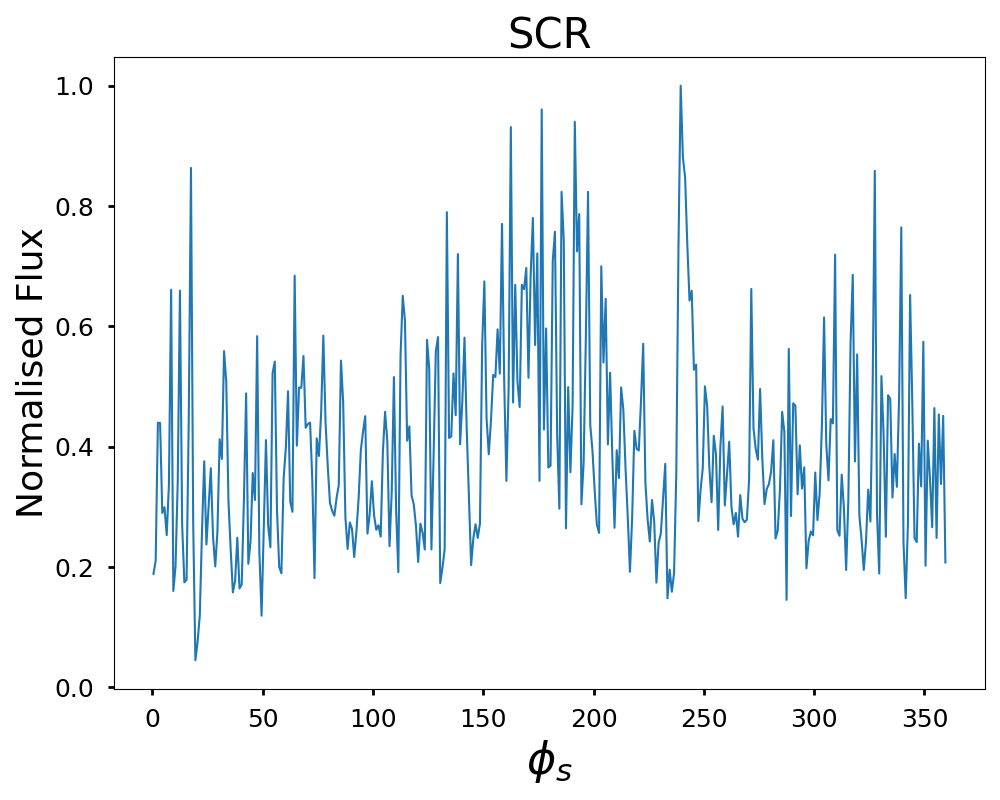}  
\caption[AR Sco Light Curve for $E$ and $B$-field Case with 2 Poles]{AR Sco light curve produced from the emission map in Figure \ref{E_map_EB_2_5e8_2pole} using $\zeta = 60^{\circ}$.}
\label{LC_EB_2e8_2pole}
\end{minipage}
\end{figure}

In this section, I show the normalised light curves produced by our model from some of the emission maps in the previous section. As mentioned due to the low statistics of only using one field line, our emission maps and light curves are blobby/oscillatory and require more field lines for smooth/high-statistic results. Thus I will only show these results to illustrate that I can produce light curves and look at the general trend of the light curves, not the finer structure. In Figure~\ref{LC_EB_6e8}, I plot the light curves for the emission maps in Figure~\ref{E_map_EB_6e8} of the illustrated $\zeta=60^{\circ}$ cut. In this figure, one sees the light curve is quite oscillatory as expected due to the low statistics in the emission maps. One does see two clear peaks namely at $\phi_{\rm s} \sim 120^{\circ}$ and $\phi_{\rm s} \sim 250^{\circ}$ with a fainter secondary peak. Due to the emission originating in the extended magnetosphere, the peaks are not separated by $180^{\circ}$. In Figure~\ref{LC_EB_4e8}, I plot the light curve for Figure~\ref{E_map_EB_4e8} were one sees some form of peak at $\phi_{\rm s} \sim 60^{\circ}$ and a fainter peak at $\phi_{\rm s} \sim 330^{\circ}$. These peaks are therefore found to be separated by more than $180^{\circ}$. In Figure~\ref{LC_EB_2e8} I plot the light curve of Figure~\ref{E_map_EB_2_5e8} where one finds a broad single peak centred around $\phi_{\rm s} \sim 350^{\circ}$. This single peak structure is due to the emission being more localised closer to the magnetic mirror of the out-flowing particles from the WD surface. Finally, I plot the light curve for Figure~\ref{E_map_EB_2_5e8_2pole} in Figure~\ref{LC_EB_2e8_2pole}. Here one sees somewhat of a double-peaked structure with the fist peak around $\phi_{\rm s} \sim 170^{\circ}$ and the second peak around $\phi_{\rm s} \sim 360^{\circ}$ with the second peak's centre being unclear. It is thus apparent that higher statistics are required in each of these cases to extract information beyond the general trend of the light curves.  

\section{Conclusions} \label{sec:ARSCO_s4}
In this Chapter, I showed that I could reproduce the magnetic mirror model for AR Sco assuming uniform particle pitch angles as proposed by \citetalias{Takata2017}. I demonstrate I could produce an SCR spectrum that fits the higher-energy (optical and X-ray) observational AR Sco SED including the latest \textit{NICER} pulsed, non-thermal X-rays with my exploratory simulations. Additionally, I showed the emission maps and light curves produced by these exploratory low-statistics runs. I found quite a few deviations from the results of the \citetalias{Takata2017, Takata2019} models but explain these differences with regard to the particle dynamics involved in this scenario. These results highlighted the importance of modelling the full particle dynamics for this scenario.

In Figure~\ref{Spec_p3} I found that using the same parameters as both \citetalias{Takata2017, Takata2019} the model spectrum exceeded the observational SED and had a spectral peak that was at a higher photon energy than the observational peak at $\sim 1 \, \rm{eV}$. The spectra produced by our model were found to have a softer spectral slope than that of the \citetalias{Takata2019} spectrum using the same $p=3$. As mentioned in the text we do not know why the \citetalias{Takata2019} spectral slope is harder than expected when using an index of $p=3.0$. I found that I could better fit the observational SED with our model using a lower $B$-field for the WD and using lower minimum particle energies namely $\gamma_{\rm min} = 10$. This is a reasonable change since the $\gamma_{\rm min} = 50$ used in \citetalias{Takata2017, Takata2019} is derived from the magnetic dissipation assuming $B_{\rm S} \sim 6 \times 10^{8} \, \rm{G}$. Hence, a lower $B$-field would mean $\gamma_{\rm min}$ would be lower. Using $B_{\rm S} \sim 2.5 \times 10^{8} \, \rm{G}$ and $\gamma_{\rm min} = 10$ I found our model to fit the spectral peak at $\sim 1 \, \rm{eV}$ but having a slightly lower overall flux in Figure~\ref{Spec_p3}. I found artificially increasing the flux via normalisation and using $p=2.9$ to fit the \textit{NICER} pulsed data points, that our spectrum fit the SED very well. In the future, we will do more robust simulations, but with these exploratory results the $B$-field seems to fall somewhere between $B_{\rm S} \sim 2.5-3.0 \times 10^{8} \, \rm{G}$. Our result of $p=2.9$ also conforms very well with the observationally constrained photon index from \citet{Garnavich2019}. I do find that our spectrum does not fit the lower photon energy segment of the SED. If our spectrum modelled the lower energy SED bump at $\sim 0.05 \, \rm{eV}$ our model would fit the lower energy SED but it would be impossible to fit the NICER pulsed X-ray data points. Therefore, it is possible that there are two particle populations present, one to fit the optical SED peak and pulsed X-rays and one to fit the lower-energy SED and bump. This could also possibly be explained by using a double Gaussian pitch angle distribution but this assumption seems somewhat unphysical.

Upon investigating the trajectories, $\gamma$ losses, and single-particle spectra for the different particle $\gamma$ and $\theta_{\rm p}$ combinations in Figures~\ref{Mirrored_traj}, \ref{Mirrored_gam}, and \ref{Cases_spec}, respectively, I found that the magnetic mirroring was more complex than initially anticipated. In my analysis of Figure~\ref{Cases_spec}, I found that the particle SED depends on the initial $\gamma$ and $\theta_{\rm p}$ due to the SR calculations in Equation~(14) and the critical SR photon energy in Chapter~\ref{sec:Paper3}\footnote{I refer to the SR equations since the SCR is in the SR regime and more intuitive to follow but the same applies for $\tilde{B}_{\perp}$ where the pitch angle is encapsulated in the $\mathbf{p}\times\mathbf{B}$ terms.}. However, these spectra and Equations also depend on at which altitude the particle mirrors in the magnetosphere since a higher altitude would mean a lower $B$-field experienced by the particle. Where the particle mirrors are also dependent on the initial $\theta_{\rm p}$ as seen in Figure\ref{Mirrored_traj}. In Figure~\ref{Mirrored_gam}, I found that particles with higher $\gamma$ values have higher loss rates and radiate their energy higher up in the magnetosphere. This higher altitude magnetosphere emission is also seen for particles with the same $\gamma$ values but using higher $B$-fields. Thus, the lower-energy particles that make up the bulk of the emission will radiate higher up in the magnetosphere for larger WD $B$-field strengths. Investigating the emission maps in Figures~\ref{E_map_EB_6e8}, \ref{E_map_EB_4e8}, and~\ref{E_map_EB_2_5e8} showed that the bulk emission (low-energy particles) is located higher up in the magnetosphere in the higher $B$-field cases than the lower $B$-field cases. This conformed with the losses seen in Figure~\ref{Mirrored_gam}. This is also visible in the high $B$-field case of Figure~\ref{E_map_EB_6e8} showing the bulk emission further away from the WD poles. This explains the spectral peaks not shifting with the $B$-field as expected in Figure~\ref{Spec_p3} since the $B$-field the particle experiences along its trajectory and its $\gamma$ in each $B$-field case is not the same due to the different mirror points and RRF dominance. Another important factor to take into account is the effect of an $E_{\perp}$-field since there is no reasonable justification to neglect it from the calculations, unlike the screened $E_{\parallel}$-field. In Chapter~\ref{sec:Paper2} Figure~10 I show the effect on the particle $\gamma$ as well as causing the particle to mirror higher in the magnetosphere. I also found that the inclusion of the $E_{\perp}$-field caused all the particles to mirror even for $\theta_{\rm p} = 0^{\circ}$ due to the standard gradient drift effects and the $\mathbf{E}\times\mathbf{B}$-drift effect on the particle. The positions of the poles and emission were also affected in the emission maps of Figure~\ref{E_map_B} due to the different particle trajectories with the inclusion of $\mathbf{E}\times\mathbf{B}$ drifts. All this shows how important it is to model the general particle dynamics for magnetic mirroring scenarios, since the particle trajectories and dynamics are greatly impacted by the initial $\gamma$, $\theta_{p}$, and $B$-field and $E_{\perp}$-field experienced by the particle. The \citetalias{Takata2017,Takata2019} models do not include any of the drift effects on the particles and we are unsure how they mirror their particles. Therefore, I do not know if they use their theoretical mirror radius equation to account for the different mirror radii due to the initial $\gamma$ and $\theta_{p}$. Due to the many differences and approximations used in the \citetalias{Takata2017,Takata2019} models, I believe this is why there are differences in our results using the same parameters.

Due to these being exploratory simulations, it is evident from the emission maps and light curves results that one needs to sample more field lines for higher statistics and smoother results. For an upcoming publication, we will include these results, identify a better fitting $B$-field, probe a Gaussian particle pitch angle distribution centred on a chosen value, and simulate more field lines. It is evident from these results that there is much-needed exploration to be done using our model into the injection scenarios and particle dynamics present in AR Sco. This is due to the possible second particle population needed for the lower-energy SED fit, the fact that I do find double-peaked light curves but still need to fit the observational light curves, and there is a massive computational undertaking to reproduce the orbital phase-resolved observations for AR Sco shown in the optical band in \citet{Potter2018b} and for the X-rays in \citet{Takata2021}. As a start to these future undertakings, the spectrum and emission maps should be simulated with different initial spin phases of the WD to compare the contribution a time-dependent injection has as the WD completes one rotation. This could possibly address the lower-energy SED spectrum, since the most prominent and higher-energy emission would occur when the particles are injected into the open field lines $(\phi_{s} = 0.0$ or $0.5)$ close to the WD instead of the closed field lines $(\phi_{s} = 0.25$ or $0.75)$ where they have to travel much further to the magnetic poles. This would also drastically change the emission maps leading to two poles formed from the particles injected as both the northern pole and southern pole point towards the companion, amplified by the particle being turned around by the magnetic mirror. It would thus lead to interesting caustics in the emission maps. The inclusion of the closed field line injection could also lead to the extra contribution of higher-magnetosphere emission. Next, one should asses using a delta approximation for the particle $\theta_{\rm p}$ using the most contributing value for all the particles and determine what would be lost from the emission maps and spectra. This is due to how computationally demanding it is to simulate a $\theta_{\rm p}$ distribution and a particle $\gamma$ distribution. Running the code on a larger cluster with more resources would drastically improve the ability to probe multiple field lines and pitch angle distributions beyond the 30 to 100 CPUs I was able to use for our current results. The inclusion of some form of accelerating $E_{\parallel}$ can be probed as well, since currently we assume no $E_{\parallel}$. After assessing variations in $\alpha$ and the spectra and light curves agree with the observations, we can use higher-statistic runs to probe more of the emission regions and caustics. From this point, we can then build up the orbital-resolved emission maps by simulating different orbital regions. One can then take the light curve from these emission maps for different orbital phases and populate the orbital phase vs. spin phase emission grid using the light curve for a specified $\zeta$ and orbital phase reproducing the orbital phase vs spin phase plot in \citet{Potter2018}. This is the closest we can get to modelling the emission on the orbital scale of the system, since it is highly impractical to simulate the whole orbital time scale when resolving the full particle gyrations. Originally, we also planned to model the polarisation for the source but due to time constraints and unpredicted delays were unable to include them in this thesis. Modelling the polarisation would allow us to constrain the models and scenarios even further by adding the requirement to reproduce the linear flux, degree of polarisation, and polarisation position angle from the observations in \citet{Potter2018b}. This will be pursued in future work.

%%%%%%%%%%%%%%%%%%%%%%%%%

\chapter{Conclusions} \label{sec:Conclusion}
From the extensive amount of observational data, AR Sco is found to be a unique and complicated source serving as the prototype and archetype for the now-established class `white dwarf pulsar'. It is clear that more robust modelling is required to probe and elucidate the emission and injection mechanisms operating in the system, especially with the discovery of the sibling source by \citet{Pelisoli2023} and the active searches for similar sources. It is crucial for our own model and future models to concurrently reproduce the observational orbital phase vs spin phase plots, light curves, spectra, and polarisation signatures to incorporate all the available model constraints. Thus, in this thesis, the main aim was to develop a general emission model to predict the emission maps, light curves, and spectra of AR Sco at various orbital phases. Additionally, I modelled a pulsar scenario, calibrating our results with an existing pulsar emission model of \citetalias{Harding2015, Harding2021}to establish confidence in our modelling results for AR Sco.

Due to the article format of this thesis, each article and result chapter has its own extensive conclusion. To avoid simply repeating every conclusion segment, I will only briefly summarise each conclusion, highlight the scientific contribution the results yielded, and end with future avenues to pursue with this work. 
  
\section{Results and Contributions}

\subsection{AR Sco Geometric Emission Features}
The published work in Chapter \ref{sec:Paper1} showed that I could constrain $\alpha$ and $\zeta$ for AR Sco over the whole binary orbit by fitting our geometric RVM from \citet{DuPlessis2019} to the orbital phase-resolved observations from \cite{Potter2018}. This further emphasised that the emission appears to be coming from close to the WD surface and that the $B$-field has a dipole-like structure as found in our previous work. I also found variation in $\alpha$ and $\zeta$ on the orbital scale of $\sim 10^{\circ}$ and $\sim 30^{\circ}$, respectively. We speculated that the variation could be due to precession\footnote{Precession on this time scale is highly unlikely.}, inhomogeneous sampling of the orbit, a wobbling emission source, different particle populations contributing to the emission, depolarisation effects from the WD and companion, and changes in the WD $B$-field structure due to the interacting fields. Recently, \citet{Garnavich2023} showed with their latest long-term observations, that the WD is indeed precessing (estimated between $40 - 100$ years), supporting the idea first proposed by \cite{Katz2017}. Doing a deeper period analysis of the spin and beat-coupled linear polarisation showed that between orbital phases $0.1- 0.6$, the spin frequency and its harmonics were equivalent or higher than the beat frequency and its accompanying harmonics. Interestingly, between orbital phases $0.6 - 1.1$ the spin frequency and its harmonics were found to be absent. This could thus support the idea of multiple particle pitch angle populations suggested by \citetalias{Takata2019}. What is clear is the need to model the general particle dynamics to elucidate the spin-beat emission coupling, the magnetic mirroring of the particles, the time-dependent particle injection scenarios, as well as what happens to the particles as they cool and are finished cooling.       

\subsection{Particle Dynamics}
The work presented in Chapter \ref{sec:Paper2} is the published methods for the general particle dynamics of the emission modelling code I developed. I showed that using the higher-order Dormand-Prince 8(7) numerical integrator coupled with the implemented adaptive time-step methods, I could achieve improved accuracy as well as computational time. In the high field and high-RRF test scenarios required for pulsars and pulsar-like sources, I found that when balancing accuracy, stability, and runtime the Dormand-Prince scheme was the best for our use case. In my assessment, I found the commonly-used Vay symplectic integrator to be insufficiently accurate to deal with the large $E_{\perp}$-field required for pulsar or pulsar-like magnetosphere modelling. This would require impractically small timesteps for adequate pulsar modelling with the Vay scheme, thus leading to our adaptive Dormand-Prince scheme outperforming the Vay scheme in accuracy and computational time. This is separate from the added problem of including the RRF to the symplectic integrators without re-deriving the Hamiltonian. In additional investigation of the RRF's effect on the particle $\theta_{\rm p}$, I found that the RRF does not significantly decrease the $\theta_{\rm p}$ for super-relativistic particles but only once the $\gamma$ value becomes small. This is due to the $\gamma^{2}$ dependence in the dominant term of the RRF that is directly opposite to the particle's velocity. This highlights the problem in models assuming the $p_{\perp}$ is immediately depleted or negligently small due to the RRF. This is also contrary to the fact that the $\mathbf{E}\times\mathbf{B}$-drift will sustain large $\theta_{\rm p}$ values even with included RRF as is also confirmed by our result. Finally, I was able to show how our particle converges to the radiation-reaction-limit regime of AE for uniform fields. This gives great confidence in our RRF calculations as well as being a novel result, since we have seen no other author simulate a particle entering the AE equilibrium state with one set of equations of motion. The closest I could find was that of \citet{Yangyang2022} who use three different regions with different approximated equations to simulate the particle entering equilibrium. These results further validate the AE approach, since it is a solution to the equilibrium state. All of these results give us high confidence in our particle dynamics for modelling AR Sco.                

\subsection{Pulsar Scenario Calibration}
In Chapter \ref{sec:Paper3} I present the draft of our paper to be submitted to MNRAS shortly after the submission of my thesis. In this work, I showed that using my particle dynamics calculated from first principles, I could reproduce similar emission maps and spectra to the \citetalias{Barnard2022} model for a pulsar with the same parameters as Vela but $10\%$ the $B$-field strength. For this comparison, I initialised my equations of motion higher in the magnetosphere when using their electromagnetic fields due to encountering divergence in the fields at the ramp function regions. Additionally, I also had to use $10\%$ the $B$-field of Vela due to the particles entering the non-classical RRF regime. Futhermore, I showed that the particles converged to the AE results when using FF-fields and RVD fields. These are novel results since I illustrated how the particles enter an equilibrium of the Lorentz force and the RRF using one set of equations of motion. I confirmed that following the $\mathbf{E}\times\mathbf{B}$ trajectory and using the AE parameters for the \citet{Vigano2015} SCR calculations cause these results to converge with those of \citetalias{Cerutti2016} in most cases. However I identified the latter to be more accurate, computationally inexpensive, and applicable to my use cases. I, therefore, showed that my results validated the approach of \citetalias{Harding2021} following the AE trajectory and using the modelled AE parameters for the \citet{Vigano2015} SCR calculations in the limit of a small general pitch angle. In this work, I emphasised the importance of including the $\mathbf{E}\times\mathbf{B}$-drift effects in the particle trajectories and radiation calculations when a large $E_{\perp}$-field is present. Additionally, I showed the CR, SR, and SCR emission maps produced for a Vela-like pulsar scenario using my model, which yielded the expected caustics and emission features. Lastly, I demonstrated that my model and implemented numerical methods can better deal with the high RRF regime and high fields required for realistic pulsar magnetospheres compared to current PIC models due to using a higher-order numerical scheme and adaptive time-step methods. In Chapter~\ref{sec:Paper2}, I discussed that this was due to our implementation of the adaptive Prince-Dormand scheme outperforming the Vay scheme (used in most PIC models) in accuracy and computational time (with sufficient accuracy and stability as a requirement). In Chapter~\ref{sec:Paper3}, I further showed and discussed that my model can more accurately deal with higher surface $B$-fields and higher RRF regimes without experiencing numerical runaway \citep{Timokhin2024, Soudais2024, Mottez2024}. All of these mentioned results give us high confidence in our radiation calculations, emission maps, light curves, and spectral modelling AR Sco.     

\subsection{AR Sco Magnetic Mirror Modelling}
In Chapter \ref{sec:ARSCO}, I showed our results for modelling the AR Sco magnetic mirror scenario. I show in this initial low-statistic exploratory modelling that I can simulate the magnetic mirror scenario and capture the required micro-physics which has not previously been done for AR Sco. In this initial modelling, I showed that I can fit the observational SED for AR Sco including the recent \textit{NICER} pulsed X-ray data. Our fitted model required a power-law index of $p=2.9$, well in agreement with the observational constraints from the \citet{Garnavich2019} Optical/UV photon index of between 0.8-1.4. Using $p =2.9$ would yield a photon index of $0.95$. Using our exploratory modelling I thus constrained the WD $B$-field to be $B_{\rm S} = (2.5 - 3.0) \times 10^{8} \rm{G}$. Probing deeper into the operating micro-physics I assessed the effect the initial particle $\theta_{\rm p}$ and $\gamma$ had on the mirror point of the particles. As expected, the large $\theta_{\rm p}$ particles mirrored much further in the magnetosphere due to the $v_{\perp}^{2}$ dependence of the $\nabla \mathbf{B}$-drift in Equation~(\ref{v_grad}). This is because of the larger initial $v_{\perp}$ leading to a larger $\nabla \mathbf{B}$-drift and causing the particle $\theta_{p}$ to increase until the particle is mirrored at $\theta_{p}=90^{\circ}$. Additionally, I found that when including the $E_{\perp}$-field the particles mirrored higher up in the magnetosphere due to the $\mathbf{E}\times\mathbf{B}$-drift effects on the particle $\theta_{\rm p}$. Investigating the RRF I found the higher $\gamma$ particles had a higher loss rate and total losses due to the $\gamma^{2}$ dependence of the dominant term in the RRF. This caused these particles to yield significant radiation contributions higher up in the magnetosphere than the lower-energy particles which predominantly radiate close to the mirror points. Investigating the effect of higher fields in this model I found that particles with the same energy have much higher loss rates and total losses in the larger $B$-field cases as one expects. This means the low-energy particles that contribute to the bulk of the emission (due to the soft power-law index of the particle energies) radiate most of their energy higher up in the magnetosphere instead of at the mirror point as found in the low field cases. This was confirmed in our simulated emission maps for the different $B$-field cases. Due to the effect the initial $\theta_{\rm p}$ has on the mirror point and particle loss rate, and the effect $\gamma$ and the $B$-field had on the loss rate, changing the $B$-field did not shift the modelled spectral peak as expected for standard SED fitting (shifting the peak down and left for lower B-fields). Thus, I had to use a lower $B$-field and particle energies (justified by the lower $B$-field) to fit the observational SED peak. It is thus important to use realistic particle parameters, namely $\theta_{\rm p}$ distributions, due to the effect this has on both the emission maps and spectra. I found with our modelling that I could fit the optical peak at $1\, \rm{eV}$ and pulsed X-ray of the SED but was unable to fit the lower-energy peak at $0.05\, \rm{eV}$ as well as the lower-energy component of the SED. One would need multiple particle populations or double Gaussian peaks in the pitch angle distribution to fit both peaks as well as the pulsed X-ray component due to the soft index of the pulsed X-ray spectrum. I found the inclusion of the $E_{\perp}$-field to be important due to its large effect on the particle $\theta_{\rm p}$, mirror points, and subsequently the losses. Due to the large induced $E_{\perp}$-field present in the source, similar to pulsar sources, it therefore required our SCR calculations following the AE trajectory due to the standard SR and SCR being derived in the absence of an $E$-field. The large effects of excluding the $E_{\perp}$-field were also seen in the emission maps namely the highly shifted poles, altered caustics, and features due to the non-mirrored particles. This work highlights the importance of including the micro-physics (relevant drift effects and RRF) in magnetic mirror modelling due to the effects on the emission heights and caustics in the emission maps and the effects on the radiation calculations for the SEDs. Thus, there is much needed future modelling of AR Sco required with this code beyond our exploratory first modelling.          

\section{Suggestions for Future Work}
There are many future development avenues available for the work presented in this thesis. Firstly, simulating additional scenarios for AR Sco and further developing the code for AR Sco modelling, and secondly developing the code further to better model pulsars or use the code to model other sources. Therefore, I will break down the future developments into these two avenues. The code is parallelised therefore running the code on a cluster with more CPUs and resources would drastically improve the runtime when probing multiple field lines and using particle energy and pitch angle distributions for each field line. This will however not solve the problem of having to resolve the gyro-radius for large $B$-fields and high RRF cases. Thus, single-thread optimisation is required to address this problem as discussed below.

As mentioned in Chapter~\ref{sec:ARSCO_s4} there is still much potential for modelling AR Sco with our model/code. Firstly, we will identify better-fitting parameters and do higher-statistics runs by simulating multiple field lines. This will be published in a future work along with some of the results in Chapter~\ref{sec:ARSCO}. Next, we will assess the contribution to the emission maps and spectra if particles are injected at different WD spin phases instead of only injecting particles when the WD pole is pointing towards the companion. This yields time-dependent particle injection and leads to particles being injected at both poles as the WD rotates and could consolidate the low energy SED contribution. In my modelling presented in Chapter~\ref{sec:ARSCO}, I found the lower pitch angles to contribute more to the emission due to them mirroring closer to the WD surface. Hence, we will start by investigating a Gaussian particle $\theta_{\rm p}$ distribution yielding more physical pitch angles expected for the magnetic mirror scenario. This will also allow us to assess if it is appropriate to approximate by only using one $\theta_{\rm p}$ to save computational time by not simulating the other $\theta_{\rm p}$ combinations. Initially, I intended to include polarisation calculations into the model but due to time constraints and unforeseen delays, it could not be included. Including the polarisation calculations in the modelling would help constrain the possible injection scenarios, since one would have to reproduce the linear polarisation, degree of polarisation, and polarisation position angle of the observations as well. The biggest hurdle would be to reproduce the orbital phase-resolved emission maps, since I am currently only modelling one point in the orbital phase. It is impractical to simulate the whole binary orbit time scale with these simulations, thus as suggested it is better to model these various points in the orbital phase and use these results to build up an orbital phase vs. spin phase emission map to reproduce the observational maps from \citet{Potter2018b} and \citet{Takata2021}. There is also room to do magnetohydrodynamics (MHD) simulations to model the interacting $B$-fields of the WD and its companion for more realistic field structures. This same code can also be used to model the new AR Sco sibling discovered by \citet{Pelisoli2023} as well as similar future sources.         

Regardless of the source modelled, the inclusion of polarisation calculations would be a great improvement as extra constraints to the modelling. This can be accomplished reasonably easily by following either the implementation by \citet{Harding2017} or using the already calculated $\tilde{B}_{\perp}$ and following the approach of \citet{Cerutti2016pol}. The latter would be ideal since it is proposed to account for the SCR polarisation, since $\tilde{B}_{\perp}$ accounts for the SCR and the $E_{\perp}$ inclusion. As shown I can simulate high-field pulsars by starting higher in the magnetosphere similar to PIC models. For the application to realistic high $B$-field pulsars, we would have to calculate our own self-consistent fields along the particle trajectory. This would avoid oversampling coarse field grids or generating the field grids on the unrealistically small scale of the particle gyro-radius as discussed in Chapter~\ref{sec:Paper3}. These computations would not dramatically increase the computational cost of the code since the majority of the cost is in solving the equations of motion and identifying a sufficiently stable time step. We can use similar approaches to \citet{Cerutti2016, Kalapotharakos2018} to calculate the self-consistent fields, account for the nuance in charge conservation for macro particles as highlighted by \citet{Vranic2016}, and include the dissipative $E_{\parallel}$-fields to accelerate the particles as done by \citet{Kalapotharakos2018}. Another addition to these sources would be to deal with the RRF in the QED regime. This would be highly relevant if one is interested in simulating the particle dynamics and losses close to the pulsar surface or high-field regimes. It would also avoid numerical runaway solutions in PIC simulations or forcing the particle to stay in the classical regime which I discussed in Chapter~\ref{sec:Paper3}. Therefore, one would have to implement RRF equations that hold for the quantum regime or take a stochastic approach where the particle probabilistically loses energy. Unfortunately, there is no way to test the accuracy of the equations or approaches in these regimes. This is due to the classic RRF being tested with laser ablation plasma experiments and they have only started to approach laser energies required to test the high-radiation-reaction regime. If these implementations are successful, one would also be able to probe the quantum SR effect on the particle's perpendicular momentum in the large $B$-fields close to the stellar surface as proposed by \citet{Harding1987}. It would also be highly relevant for the PIC models probing the origin of the coherent radio emission and pair production from pulsars as investigated by \citet{Timokhin2024}, since they are mainly interested in the high $B$-field region close to the stellar surface. Due to how computationally demanding it is to resolve the full particle gyration in these high fields \footnote{In the QED regime, the particles are in discrete Landau states thus one does not have classical gyro orbits.} of pulsars, we would have to find new numerical techniques or optimisations to be able to start the simulations at the stellar surface while using realistic pulsar fields, which is a current problem for all PIC codes. Beyond future hardware improvements, we can find a better adaptive time step method to improve the stability of the 12th-order numerical scheme as discussed in Chapter~\ref{sec:Paper2}. The code can be rewritten implementing Single Instruction, Multiple Data (SIMD) operations\footnote{This allows the processing of multiple data with a single instruction but is quite advanced data handling and coding and would require most of the code be rewritten.}, and assessing using the Graphics Processing Unit (GPU) to do some of the calculations. Importantly, with any of these implementations the accuracy and stability of the results should be re-evaluated against the results in Chapter~\ref{sec:Paper2} and~\ref{sec:Paper3} due to the high requirements in these extreme regimes. The code can also be developed into a hybrid PIC-MHD code to take advantage of probing the different physics at the different scales where each scheme is required, similar to \citet{Soudais2024}.  

Due to the general particle dynamics and radiation reaction used in the code, the code can be used to model other sources that require the general dynamics and solving the particle's gyro-orbit. This could extend to the cataclysmic variable AE Aquarii with its propeller state accretion but the addition of the accretion disc needs to be accounted for, intermediate Polars, other accreting binaries, WD Fast Radio Burst candidates, and millisecond pulsars due to their lower fields.   

%\subsection{Personal Contribution}
  
\section*{Acknowledgements}
\addcontentsline{toc}{section}{Acknowledgements}

I want to start by thanking my wife who has encouraged and supported me along this journey. Who stood with me when facing challenging circumstances, unwavering in support. Heartfelt thanks to my mother, brother and wife's family for all their individual and collective support over the years. Thank you to my supervisor C. Venter and co-supervisors A.K. Harding and Z. Wadiasingh for all their support and guidance throughout this work. A special thank you to C. Venter for all the conversations, support and leadership over seven years through projects, a mini-thesis, and finally my PhD thesis. Furthermore, I thank C. Kalapotharakos, A. Kundu, and P. Els for all their helpful conversations and input into the completion of this work. Above all our Heavenly Father, for His grace, guidance, and blessing throughout my life's journey.  

I would like to acknowledge that this work is based on research supported partially by the National Research Foundation of South Africa under award numbers PMDS22060820024 and PMDS23042496534. The AR Sco results were also expedited thanks to using the FSK-GAMMA-1 computer cluster facility at the Centre for Space Research (North-West University).

\renewcommand{\theequation}{A.\arabic{equation}}
\renewcommand\thefigure{A.\arabic{figure}}
\chapter*{Appendix}
\addcontentsline{toc}{chapter}{Appendix}

\section*{Particle Dynamics Paper Additions}
\begin{figure}[!h]
\begin{minipage}{19pc}
\centering
\includegraphics[width=\textwidth]{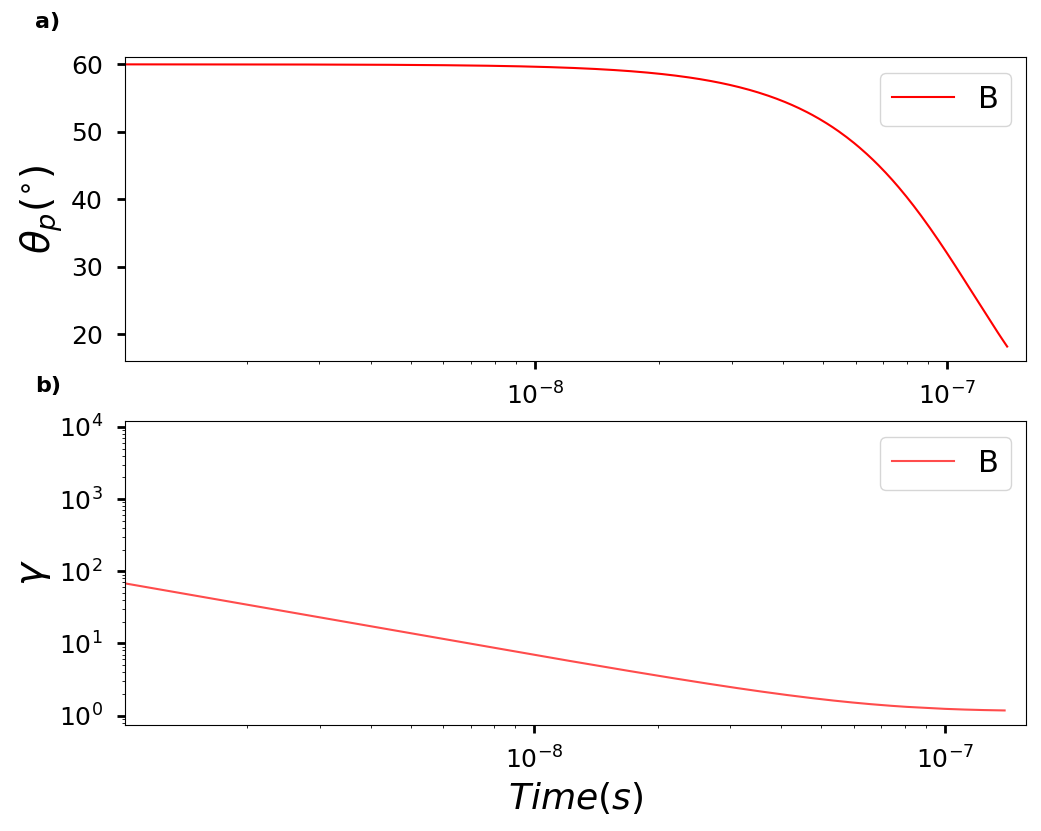} 
\caption[Large $\gamma$, Uniform $B$-field Longer Run]{Longer runtime results for the uniform $B$-field case in Chapter~\ref{sec:Paper2} Figure~18 panel a). Thus using $B = 10^{8} \, \rm{G}$ and $\gamma=10^{4}$ where panel a) shows the particle $\theta_{\rm p}$ and panel b) the particle $\gamma$.}
\label{AP2_long1}
\end{minipage}%\hspace{0pc}
\begin{minipage}{19pc}
\centering
\includegraphics[width=\textwidth]{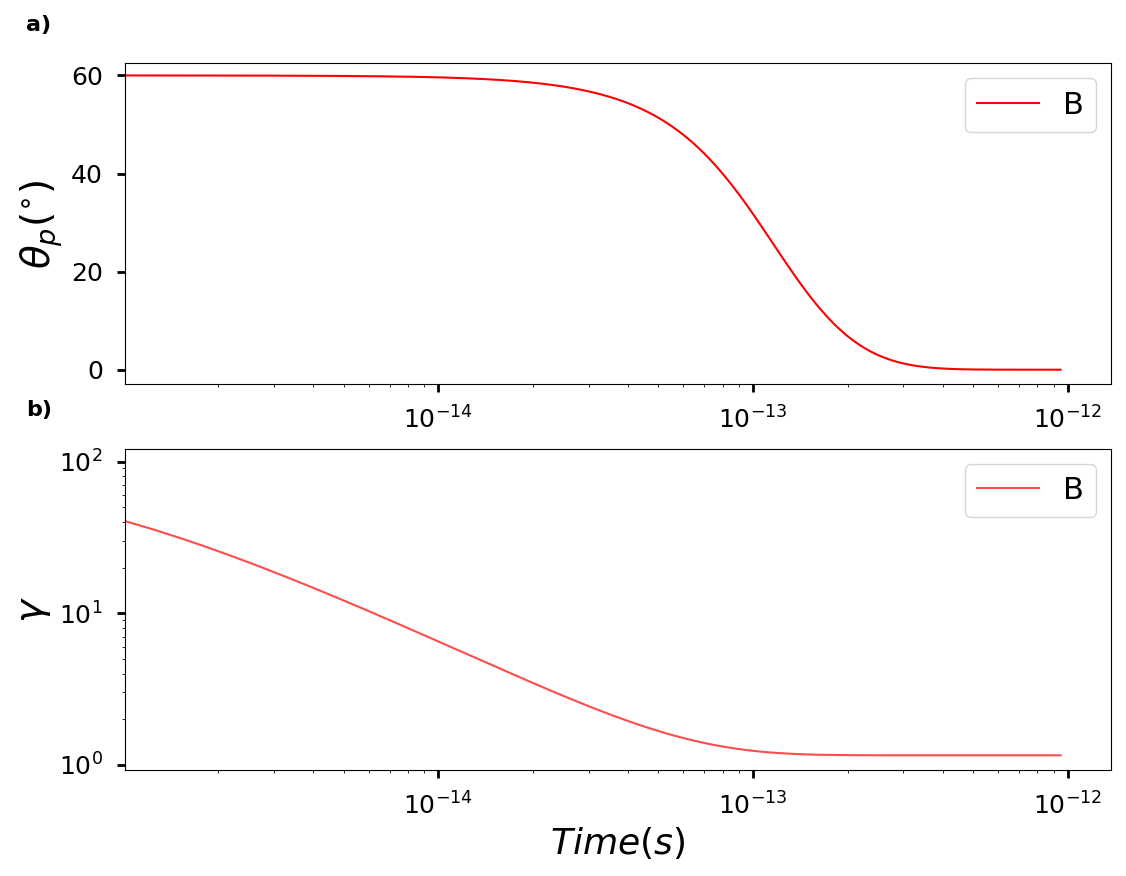} 
\caption[Large $B$-field, Uniform $B$-field Longer Run]{Longer runtime results for the uniform $B$-field case in Chapter~\ref{sec:Paper2} figure 18 panel b). Thus using $B = 10^{11} \, \rm{G}$ and $\gamma=10^{2}$ where panel a) shows the particle $\theta_{\rm p}$ and panel b) the particle $\gamma$.}
\label{AP2_long2}
\end{minipage}
\end{figure}

\addcontentsline{toc}{section}{Particle Dynamics Paper Additions}
Here I plot the extended run results for the two uniform $B$-field cases shown in Chapter~\ref{sec:Paper2}, Figure~18. Figure~\ref{AP2_long1} I use $B = 10^{8} \, \rm{G}$ and $\gamma=10^{4}$ where in Figure~\ref{AP2_long2} we use $B = 10^{11} \, \rm{G}$ and $\gamma=10^{2}$. These plots are added to show the effect the RRF has on the particle $\theta_{\rm p}$ for an extended simulation time, focusing on the point where $\theta_{\rm p}$ starts to decrease. In Figure~\ref{AP2_long1} one sees that $\theta_{\rm p}$ starts to slightly decrease at $\sim 10^{-8} \, \rm{s}$ where this corresponds to $\gamma \sim 10$. For Figure~\ref{AP2_long2} one sees $\theta_{\rm p}$ starting to slightly decrease at $\sim 10^{-14} \, \rm{s}$ corresponding to $\gamma \sim 7$. In both these cases, the $\gamma$ value is extremely low compared to standard pulsar parameters of $\gamma \sim 10^{7} - 10^{8}$ before $\theta_{\rm p}$ starts to be affected by the RRF. This shows the validity in the statement that the RRF does not affect $\theta_{\rm p}$ in the super-relativistic regime as discussed in Chapter~\ref{sec:Paper2}.

\section*{Calibration Paper Additions}
\addcontentsline{toc}{section}{Calibration Paper Additions}
\subsection*{Trajectories for Other Polar Cap Phases}
\begin{figure}[!h]
\centering
\includegraphics[width=\textwidth]{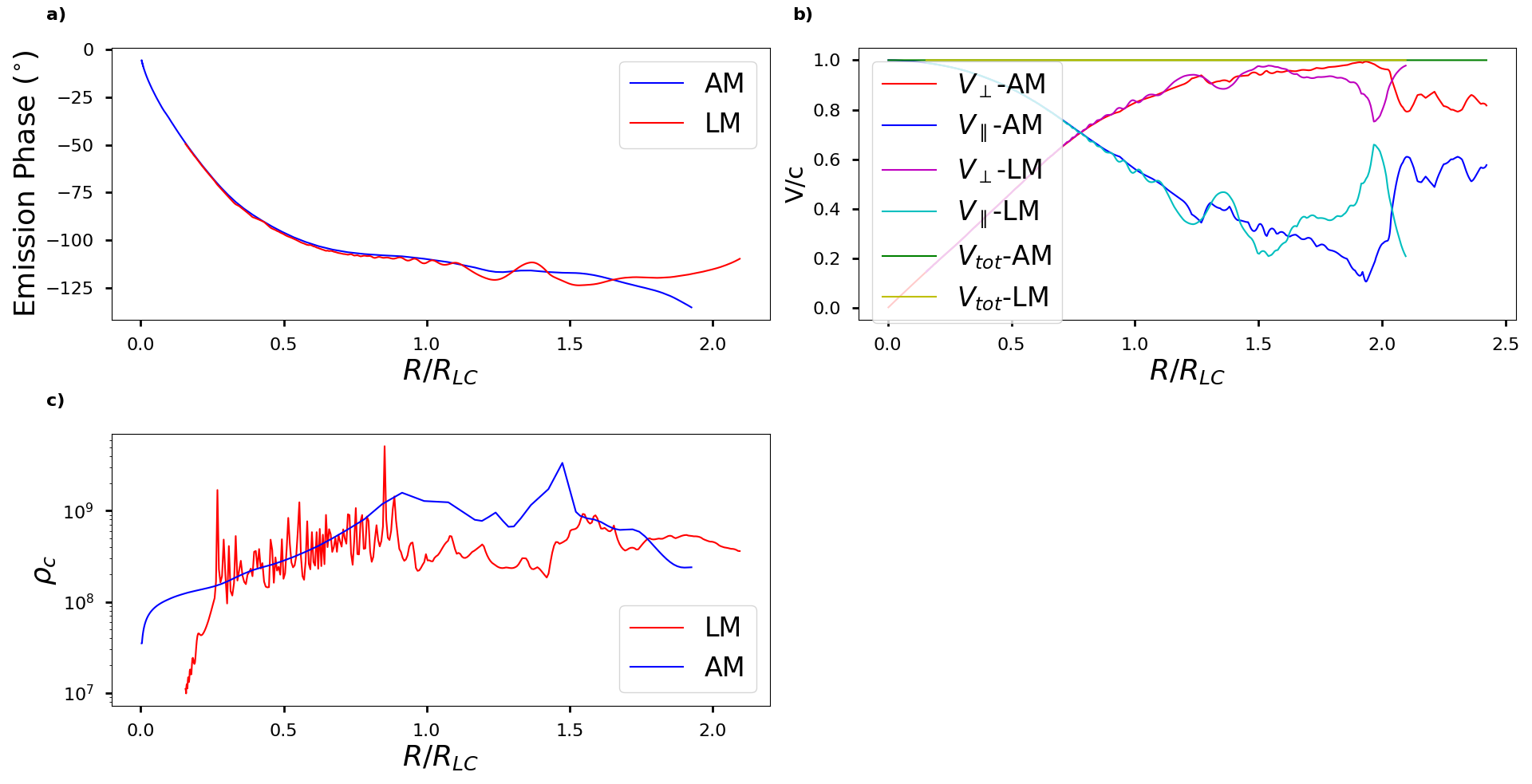} 
\caption[Trajectory Phase 90]{Vela calibration case following the field line at polar cap phase $90^{\circ}$ using $B_{\rm S} =8\times 10^{10} \, \rm{G}$. The corrected observer emission phase is shown in panel a), the normalised particle velocity components in panel b) and $\rho_{\rm c}$ in panel c). Here $LM$ represents my results and $AM$ the \citetalias{Barnard2022} model results. In panel b) we show the perpendicular velocity and parallel velocity to the local $B$-field as well as the total velocity normalised to $c$.}
\label{A_FF_traj_90}
\end{figure}

\begin{figure}[!h]
\centering
\includegraphics[width=\textwidth]{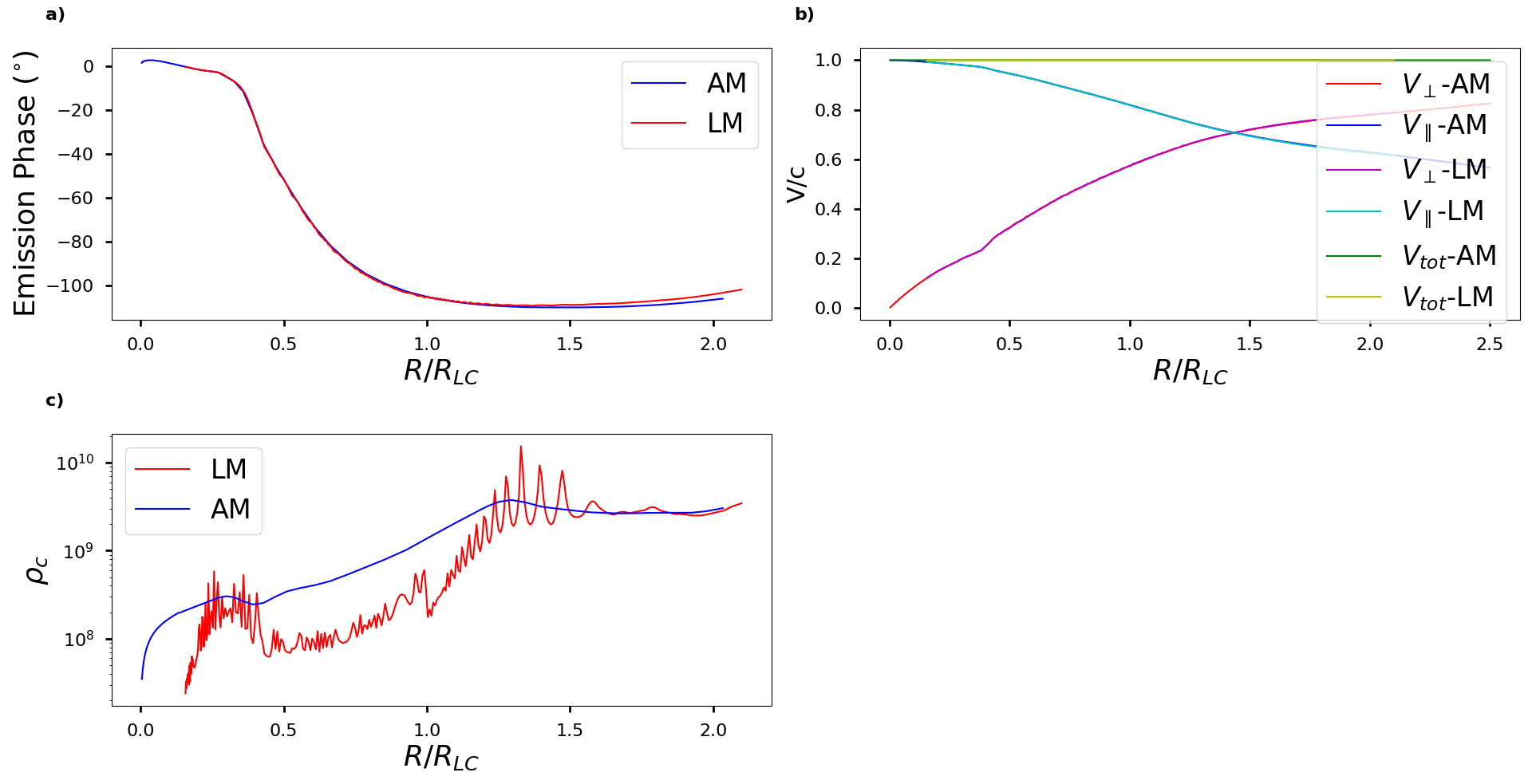} 
\caption[Trajectory Phase 180]{Vela calibration case following the field line at polar cap phase $180^{\circ}$ using $B_{\rm S} =8\times 10^{10} \, \rm{G}$. The corrected observer emission phase is shown in panel a), the normalised particle velocity components in panel b) and $\rho_{\rm c}$ in panel c). In panel b) I show the perpendicular velocity and parallel velocity to the local $B$-field as well as the total velocity normalised to $c$.}
\label{A_FF_traj_180}
\end{figure}

\begin{figure}[!h]
\centering
\includegraphics[width=\textwidth]{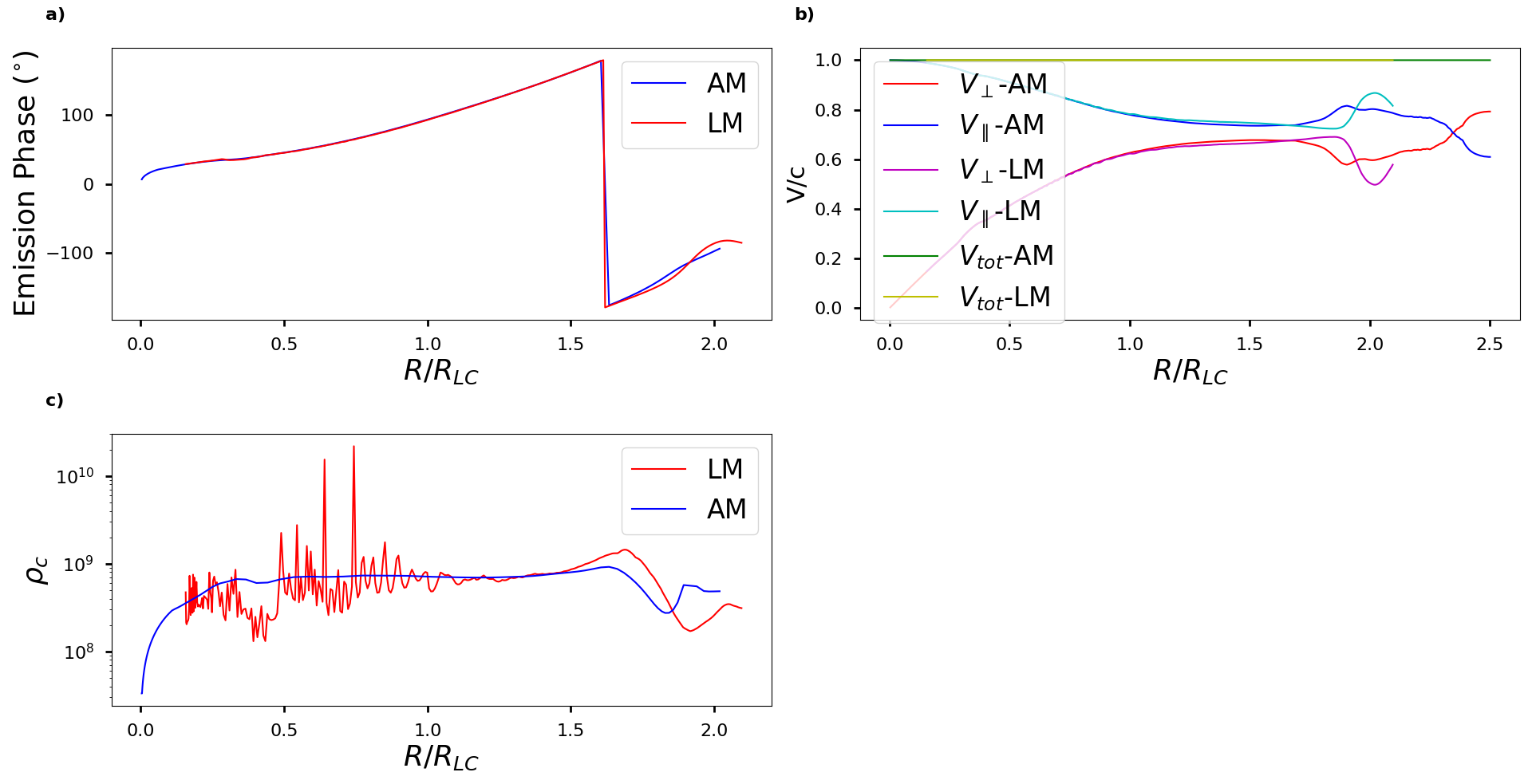} 
\caption[Trajectory Phase 270]{Vela calibration case following the field line at polar cap phase $270^{\circ}$ using $B_{\rm S} =8\times 10^{10} \, \rm{G}$. The corrected observer emission phase is shown in panel a), the normalised particle velocity components in panel b) and $\rho_{\rm c}$ in panel c). In panel b) I show the perpendicular velocity and parallel velocity to the local $B$-field as well as the total velocity normalised to $c$.}
\label{A_FF_traj_270}
\end{figure}

\addcontentsline{toc}{subsection}{Trajectories for Other Polar Cap Phases}
Here I show additional field lines that were followed for the $B_{\rm S} =8\times 10^{10} \, \rm{G}$ case discussed in Chapter~\ref{sec:Paper3}. Similar to  Chapter~\ref{sec:Paper3} the labels `LM' indicate my model results and `AM' the Harding model results generated with the \citetalias{Barnard2022} model. I have neglected the particle position and direction plots to avoid excessive plots in the Appendix, since one can infer from the observer phase plot that the particle directions are indeed the same. In Figure~\ref{A_FF_traj_90} I follow the field line for a particle initialised at polar cap phase $90^{\circ}$, in Figure~\ref{A_FF_traj_180} for polar cap phase $180^{\circ}$, and in Figure~\ref{A_FF_traj_270} for polar cap phase $270^{\circ}$. In all of the panel a) plots we see our model's corrected observer phase agrees well with that of the \citetalias{Barnard2022} model. In Figure~\ref{A_FF_traj_270} one sees both of our model results jump from $180^{\circ}$ to $-180^{\circ}$ at $\sim 1.6R_{\rm{LC}}$. This is due to the observer phase being defined between $[-\pi,\pi]$ and it is not a physical discontinuity. One has to simply add $2\pi$ to the negative values to make the curve continuous and remove the phase jump. In panel b) of Figures~\ref{A_FF_traj_180} and~\ref{A_FF_traj_270} one sees our velocity components agree well with those of the \citetalias{Barnard2022} model. In Figure~\ref{A_FF_traj_90} panel b) my model's velocity components mostly follow their model results but deviate close to $2.0R_{\rm LC}$ even though both models' total normalised total velocities conform to the condition of being $c$ as seen from the green and yellow curves. This could be due to the polar cap phase $90^{\circ}$ being where the polar cap notch is located and these field lines do not sweep out as smoothly as the rest of the field lines. In panel c) of all of the figures one sees my model $\rho_{\rm c}$ initially starts lower than the \citetalias{Barnard2022} model $\rho_{\rm c}$ but eventually oscillates around their $\rho_{\rm c}$ later in the magnetosphere/extended magnetosphere. As discussed in Chapter~\ref{sec:Paper3}, this is due to my model yielding $\rho_{\rm eff}$ since I am resolving the full particle gyration and the \citetalias{Barnard2022} model yielding the AE $\rho_{\rm c}$.      

\subsection*{Vela FF $B_{\rm S}= 8\times 10^{10}$~G-Case}
\begin{figure}
\centering
\includegraphics[width=\textwidth]{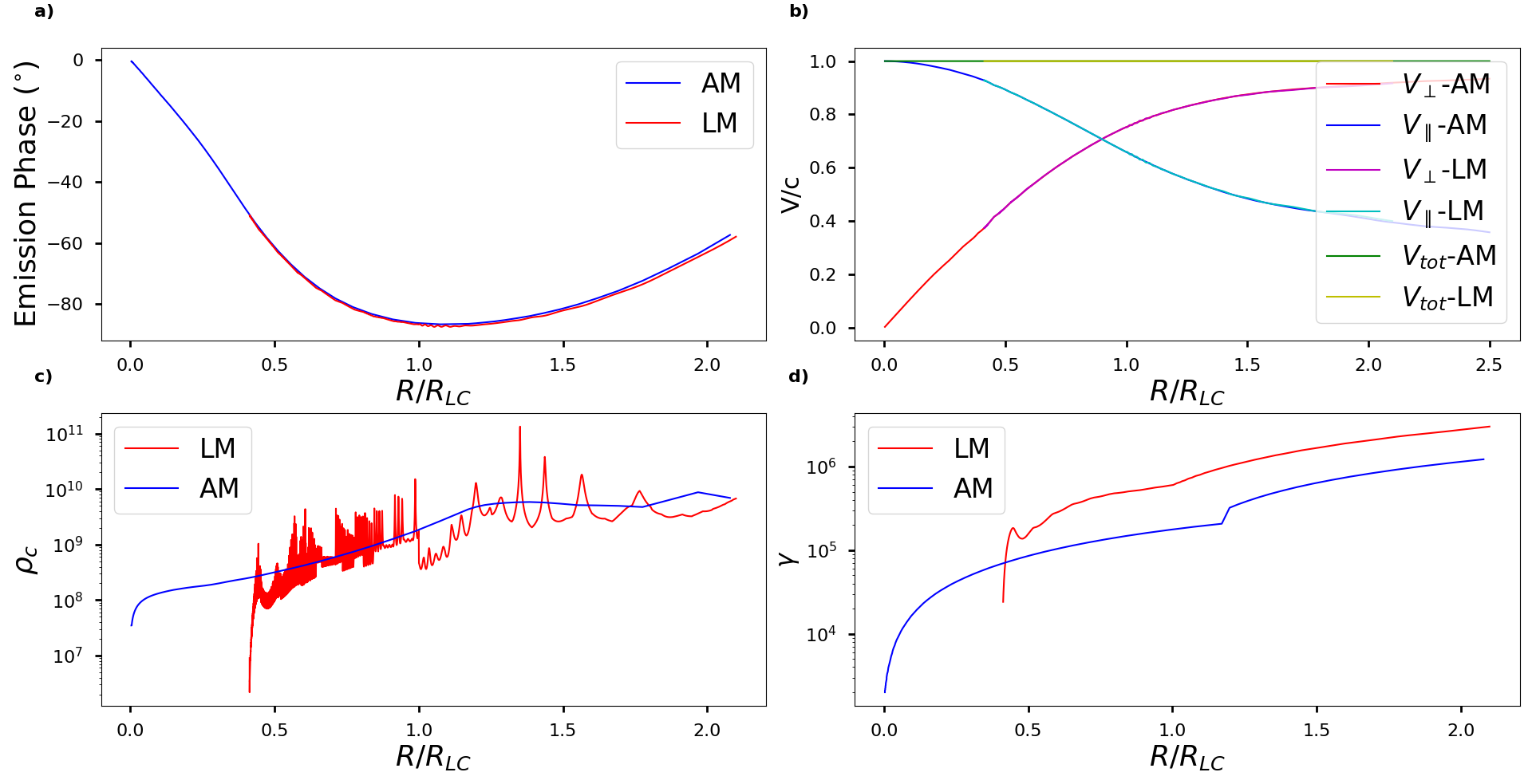}
\caption[Emission Phase Calibration Plot $B_{\rm S} =8\times 10^{10} \, \rm{G}$]{Similar to Figure~\ref{A_FF_traj_90} but for polar cap phase $0^{\circ}$, showing the corrected observer emission phase in panel a), normalised particle velocity components in panel b), $\rho_{\rm c}$ in panel c) and $\gamma$ in panel d). In panel b) I show the perpendicular velocity and parallel velocity components with respect to the local $B$-field as well as the total speed normalised to $c$.} 
\label{FF_Emission_phase_8e10}
\end{figure}

\addcontentsline{toc}{subsection}{Vela FF $B_{\rm S}= 8\times 10^{10}$~G-Case}
In Figure~\ref{FF_Emission_phase_8e10}, I use the $B_{\rm S} =8\times 10^{10} \, \rm{G}$ case and plot the observer-corrected phase in panel a), the velocity components in panel b), $\rho_{\rm c}$ in panel c), and $\gamma$ in panel d). In panels a) and b), one sees that my model results agree very well with those of the \citetalias{Barnard2022} model, with the total velocity converging to $c$ in panel b). In panel c), one sees that my model $\rho_{\rm c}$ oscillates around the $\rho_{\rm c}$ from the \citetalias{Barnard2022} model. As explained in Chapter~\ref{sec:Paper3}, this is due to my calculated $\rho_{\rm c}$ being equivalent to $\rho_{\rm eff}$, where the \citetalias{Harding2015, Harding2021, Barnard2022} model gives $\rho_{\rm c}$ for the gyro-centric trajectory. In panel d), I show the $\gamma$ from my model in red and the $\gamma$ from the \citetalias{Barnard2022} model in blue. I used a slightly lower $\gamma_{0} = 2\times 10^{4}$ to avoid initialising too close to the Schwinger limit for the RRF.

\begin{figure}
\centering
\includegraphics[width=\textwidth]{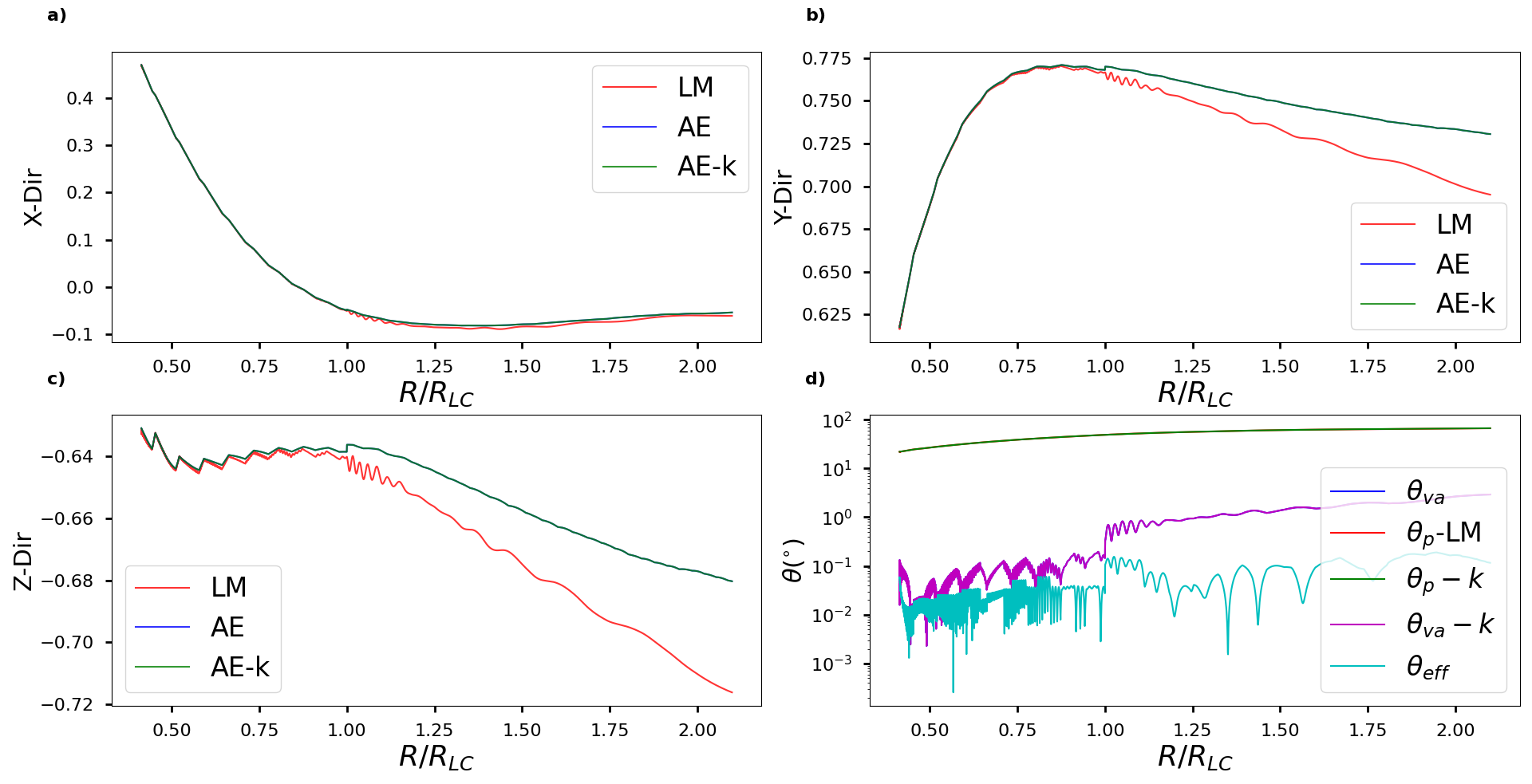}
\caption[AE Convergence Plot $B_{\rm S} =8\times 10^{10} \, \rm{G}$]{The AE convergence results Figure~\ref{FF_Emission_phase_8e10}. In this plot, AE labels the results of \citet{Gruzinov2012} and AE-k those of \citet{Kelner2015}. Panel a) shows the particle $x$-direction, panel b) the $y$-direction, and panel c) the $z$-direction. Here the blue curve is overlapped by the green. In panel d), I show the various angles discussed in Chapter~\ref{sec:Paper3}.}
\label{FF_AE_vel_8e10}
\end{figure}

\begin{figure}
\centering
\includegraphics[width=\textwidth]{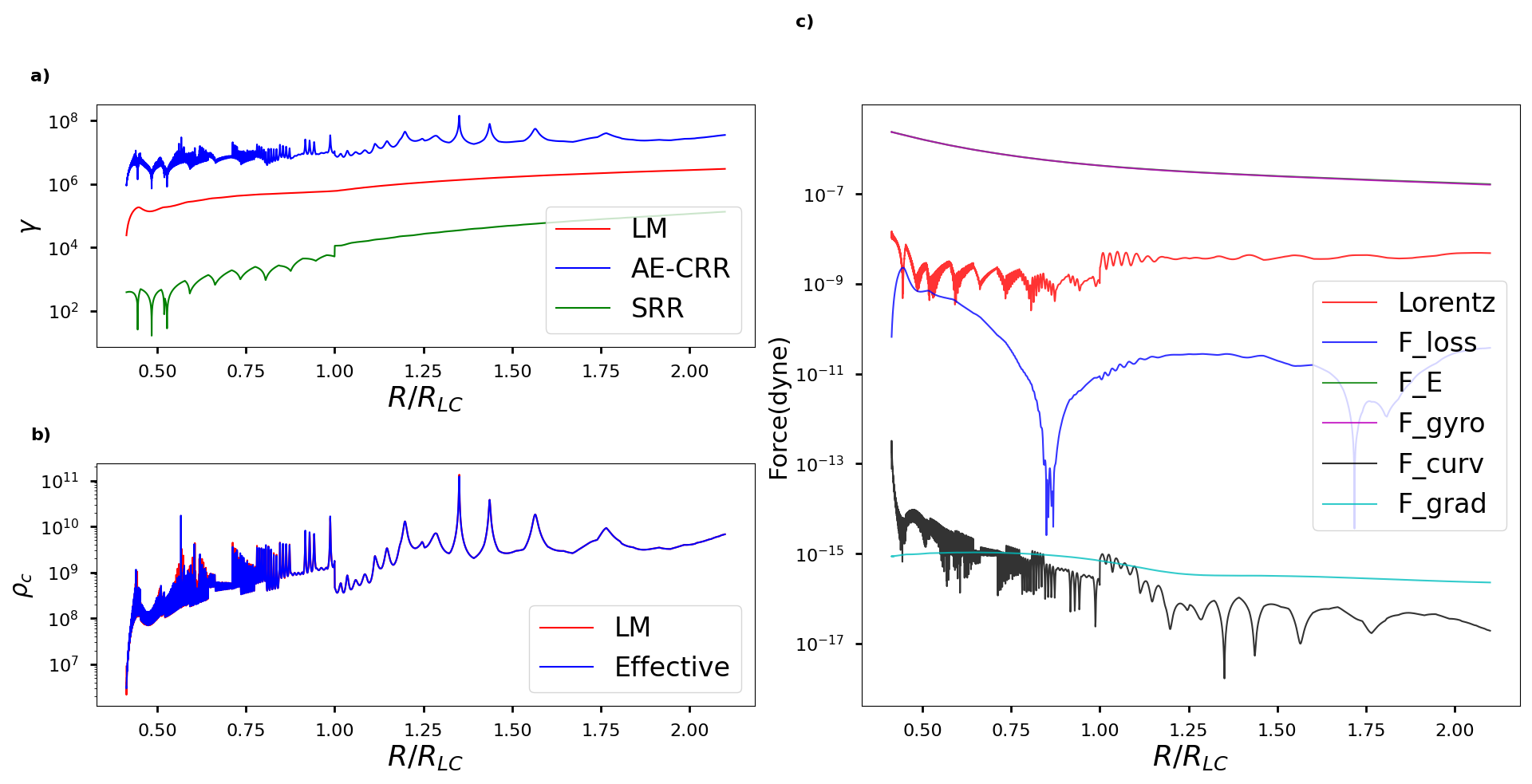}
\caption[Additional Parameter Convergence Plot $B_{\rm S} =8\times 10^{10} \, \rm{G}$]{Parameter results for the case in Figure~\ref{FF_Emission_phase_8e10}, with panel a) showing my model $\gamma$ results in red $\gamma_{\rm c}$ in blue, and $\gamma_{\rm SRR}$ in green. Panel b) shows my model particle $\rho_{\rm c}$ in red and the effective $\rho_{\rm c}$ from \citet{Kelner2015} in blue. Panel c) shows the different force components, namely the Lorentz force in red, the RRF in blue, the gyro-component of the Lorentz force in magenta, the $E$-field component of the Lorentz force in green, the curvature drift in cyan, and the gradient drift in black.}
\label{FF_AE_rho_c_8e10}
\end{figure}

In Figure~\ref{FF_AE_vel_8e10}, I plot the AE convergence results for the same parameters as used in Figure~\ref{FF_Emission_phase_8e10}. In panels a), b) and c), one sees that our results in red follow the AE results in the overlapping blue and green well until $1.0R_{\rm LC}$, where the jump in $E_{\parallel}$ occurs and the results start to diverge. In panel d), one sees the same feature where $\theta_{\rm VA}$ is small until $1.0R_{\rm LC}$ where it spikes to a higher value. The particle does not seem to be in equilibrium at $2.0R_{\rm LC}$, since $\theta_{\rm VA}$ does not seem to have settled to a relatively constant value. In Figure~\ref{FF_AE_rho_c_8e10}, I plot variables for the same case as in Figure~\ref{FF_Emission_phase_8e10}. In panel a), I plot my model $\gamma$ in red, $\gamma_{\rm c}$ in blue and $\gamma_{\rm SRR}$ in green. One sees that my model $\gamma$ lies well between these two limits. In panel b), one sees that my model $\rho_{\rm c}$ is very similar to $\rho_{\rm eff}$ from Equation~(25) from Chapter~\ref{sec:Paper3}. Panel c) shows the various forces acting on the particle, where one sees that the Lorentz force in red and RRF in blue are not yet close to equilibrium.    

\subsection*{Divergence for RD Fields}
\begin{figure}[!h]
\begin{minipage}{19pc}
\centering
\includegraphics[width=\textwidth]{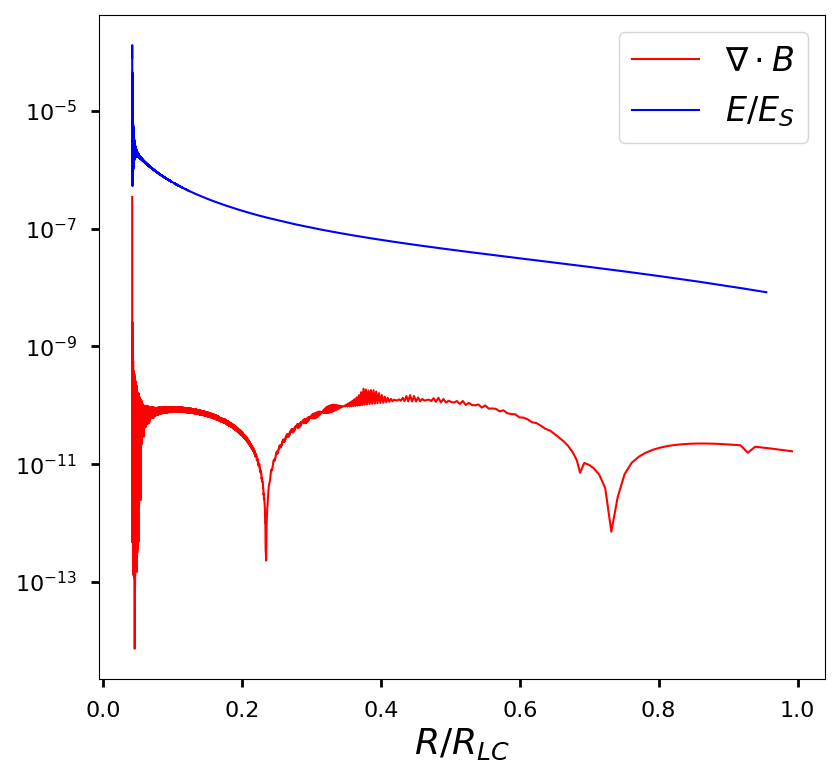} 
\caption[Divergence Plot RD Fields]{RD calibration case using $B_{\rm S} =8\times 10^{8} \, \rm{G}$, showing $\nabla\cdot\mathbf{B}$ in red and the field experienced by the particle normalised to the Schwinger field in blue.}
\label{Div_RD}
\end{minipage}%\hspace{0pc}
\begin{minipage}{19pc}
\centering
\includegraphics[width=\textwidth]{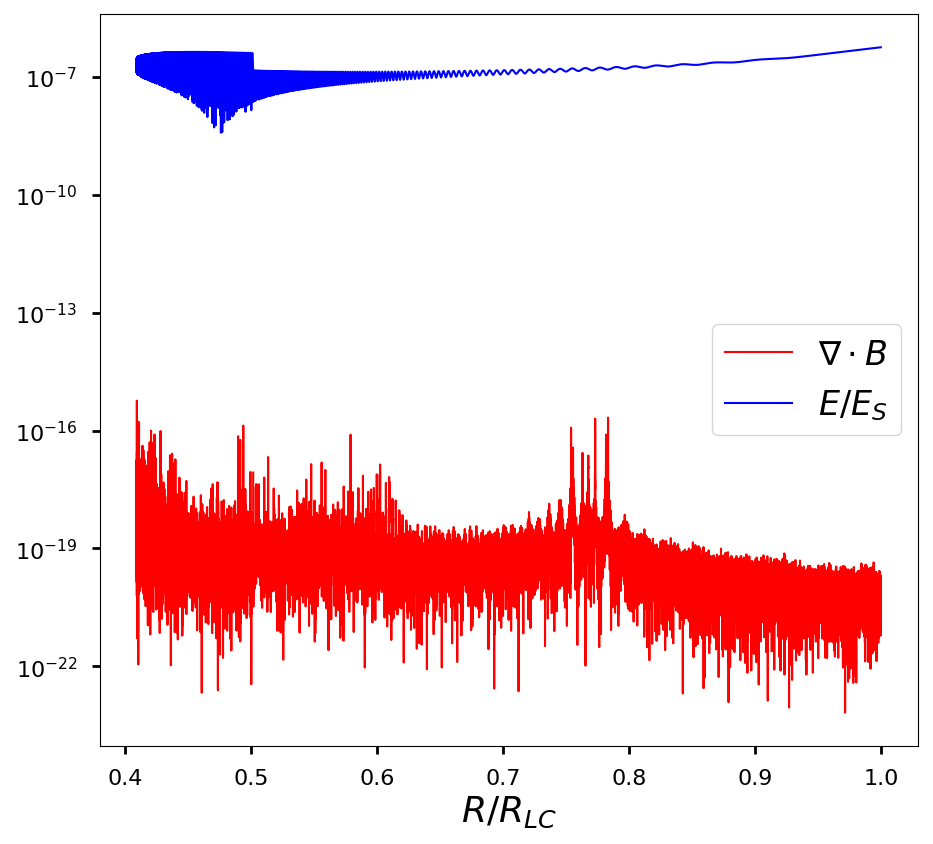} 
\caption[Divergence Plot RD Fields Mirror Scenario]{RD mirror case using $B_{\rm S} =8\times 10^{8} \, \rm{G}$, showing $\nabla\cdot\mathbf{B}$ in red and the field experienced by the particle normalised to the Schwinger field in blue.}
\label{Div_RD_mirror}
\end{minipage}
\end{figure}

\addcontentsline{toc}{subsection}{Divergence for RD Fields}
Here I add the $B$-field divergence and field experience by the particle plots for the RD case and RD mirroring case discussed in Chapter \ref{sec:Paper3}. For the RD fields I used $B_{S} =8\times 10^{8} \, \rm{G}$ and $R_{\rm acc} = 4.0\times 10^{-6} \,\rm{cm}^{-1}$ while starting at the stellar surface following the polar cap phase $0^{\circ}$ field line with a $\gamma_{0} = 10^{4}$ and $\theta_{\rm p} = 0.1^{\circ}$. For the mirror case, I started the particle at $0.5R_{\rm{LC}}$ and initial $\theta_{\rm p} = 160^{\circ}$ meaning the particle travels inward towards the polar cap. In these cases, I plot the divergence of the field $\nabla\cdot\mathbf{B}$ in red and the field experienced by the particle as defined in Chapter~\ref{sec:Paper3} in blue. For Figure \ref{Div_RD} one sees the divergence in the field has an initial spike but quickly drops to a low level where it fluctuates showing the field is divergence-free. In blue one also sees the field experienced is well below the Schwinger limit thus we are well within the classical RRF regime. Similarly in Figure \ref{Div_RD_mirror} one sees the divergence in the field is very low as well as the field experienced being well below the Schwinger limit.   

\newpage
%\section*{Article Permission}
\addcontentsline{toc}{section}{Article Permission}

\includepdf[pages=-,scale=0.9,offset=27.0mm -23.0mm,noautoscale]{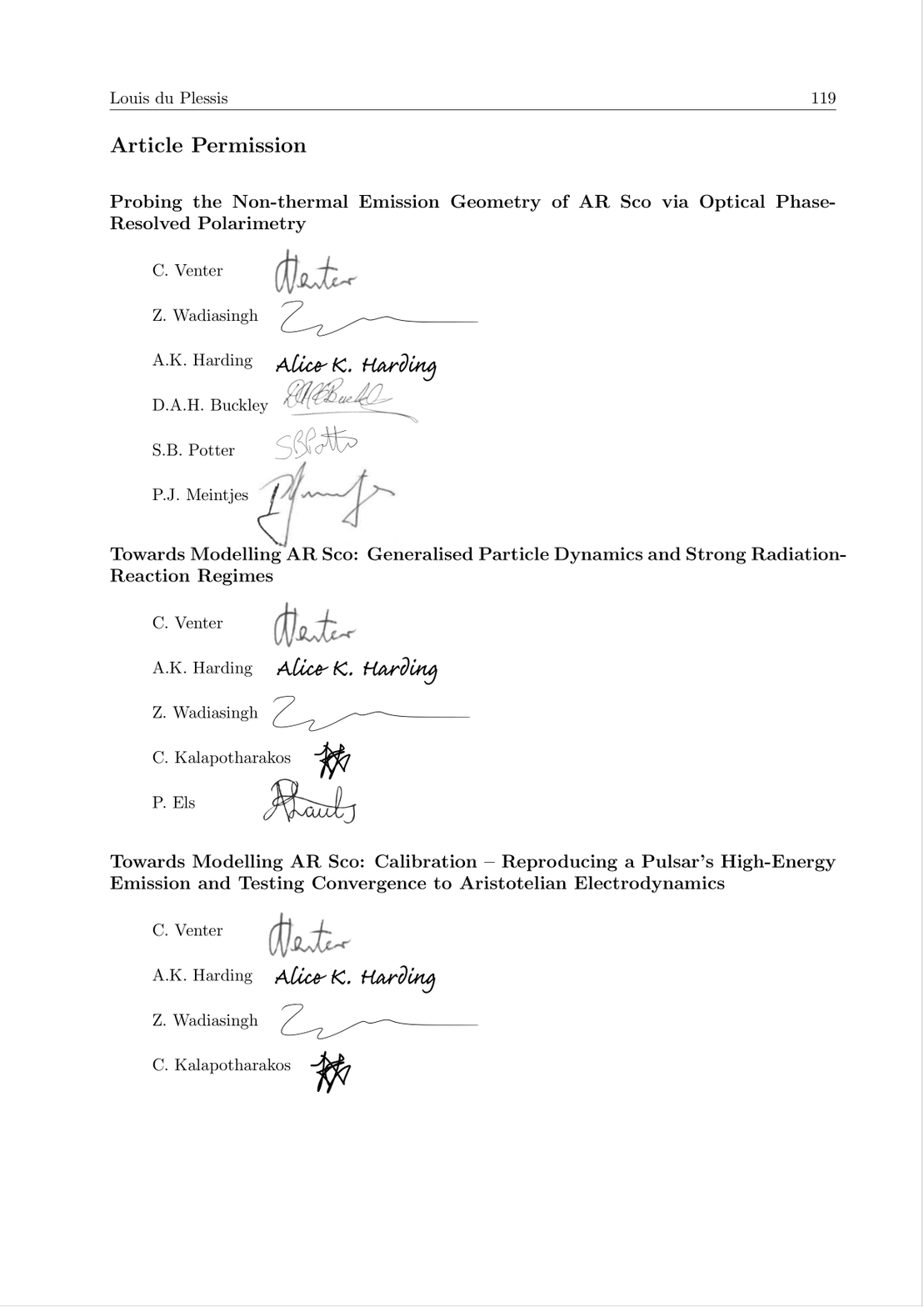}

%\subsection*{Probing the Non-thermal Emission Geometry of AR Sco via Optical 
%Phase-Resolved Polarimetry}
%\begin{itemize} 
%\setlength\itemsep{1em}
%\item[] C. Venter
%\item[] Z. Wadiasingh
%\item[] A.K. Harding
%\item[] D.A.H. Buckley
%\item[] S.B. Potter
%\item[] P.J. Meintjes 
%\end{itemize}
%
%\subsection*{Towards Modelling AR Sco: Generalised Particle Dynamics and Strong Radiation-Reaction Regimes}
%\begin{itemize} 
%\setlength\itemsep{1em}
%\item[] C. Venter
%\item[] A.K. Harding
%\item[] Z. Wadiasingh
%\item[] C. Kalapotharakos
%\item[] P. Els 
%\end{itemize}
%
%\subsection*{Towards Modelling AR Sco: Calibration -- Reproducing a Pulsar's High-Energy Emission and Testing Convergence to Aristotelian Electrodynamics}
%\begin{itemize} 
%\setlength\itemsep{1em}
%\item[] C. Venter
%\item[] A.K. Harding
%\item[] Z. Wadiasingh
%\item[] C. Kalapotharakos
%\end{itemize}

\bibliographystyle{aasjournal}                               
\bibliography{./References/References} 
\pagenumbering{gobble}                            
\pagebreak

\end{document}